\documentclass[prb,twocolumn,showpacs,floatfix]{revtex4}
\usepackage{epsfig}
\usepackage{mathrsfs}
\usepackage{amssymb}
\usepackage{color}
\newcommand{\bcen}{\begin{center}}
\newcommand{\ecen}{\end{center}}
\newcommand{\btab}{\begin{tabular}}
\newcommand{\etab}{\end{tabular}}
\newcommand{\bdes}{\begin{description}}
\newcommand{\edes}{\end{description}}

\newcommand{\beq}{\begin{equation}}
\newcommand{\eeq}{\end{equation}}
\newcommand{\bea}{\begin{eqnarray}}
\newcommand{\eea}{\end{eqnarray}}
\newcommand{\non}{\nonumber}

\newcommand{\half}{\frac{1}{2}}
\newcommand{\bary}{\begin{array}}
\newcommand{\eary}{\end{array}}
\newcommand{\benum}{\begin{enumerate}}
\newcommand{\eenum}{\end{enumerate}}
\newcommand{\bitem}{\begin{itemize}}
\newcommand{\eitem}{\end{itemize}}
\newcommand{\br} { \mbox{\boldmath $r$}}
\newcommand{\bs} { \mbox{\boldmath $s$}}

\newcommand{\bS} { \mbox{\boldmath $S$}}

\newcommand{\D}[1]{\mbox{d}{#1}}

\newcommand{\prn}[1] {(\ref{#1})}

\newcommand{\fig}[1]{FIG.~\ref{#1}}
\newcommand{\Fig}[1]{FIG.~\ref{#1}}
\renewcommand{\D}[1]{d #1}

\newcommand{\mean}[1]{\langle #1 \rangle}
\newcommand{\bra}[1]{{\langle #1 |}}
\newcommand{\ket}[1]{| #1 \rangle}

\newcommand{\dagg}[1]{#1^\dagger}

\newcommand{\figwidth}{9.0truecm}
\newcommand{\figtwowidth}{18.0truecm}
\newcommand{\Ejt}{E_{{JT}}}

\newcommand{\exic}{{\cal E}}

\begin{document}

%\preprint{}

%\preprint{}
\title{ Long Range Coulomb Interactions and Nanoscale Electronic Inhomogeneities in Correlated Oxides
}

%%Electronic Inhomogeneities in Systems with Coexisting
%%Polaronic and Delocalized States -- Influence of Long Range Coulomb Interactions\\~ \\
%%From `phase' separation to nanoscale electronic inhomogeneity due to (via*?) Coulomb interaction


\author{Vijay B.~Shenoy$^{1}$}\email[] {shenoy@mrc.iisc.ernet.in}
\author{Tribikram Gupta$^1$}\email[]{gupta@physics.iisc.ernet.in}
\author{H.~R.~Krishnamurthy$^{1,2}$}\email[]{hrkrish@physics.iisc.ernet.in}
\author{T.~V.~Ramakrishnan$^{3,1,2}$}\email[]{tvrama@bhu.ac.in}
\affiliation{$^1$Centre For Condensed Matter Theory, Indian Institute of Science, Bangalore 560 012, India\\$^2$ Jawaharlal Nehru Centre for
Advanced Scientific Research, Jakkur, Bangalore 560 064, India\\$^3$Department of Physics, Banaras Hindu University, Varanasi 221 005, UP, India }


\date{\today}

\begin{abstract}
Electronic, magnetic or structural inhomogeneities ranging in size from
nanoscopic to mesoscopic scales seem endemic, and are possibly generic, to
colossal magnetoresistance manganites and other transition metal oxides.
They are hence of great current interest and understanding them is of
fundamental importance. We show here that an extension, to include long range
Coulomb interactions, of a quantum two-fluid  $\ell-b$ model proposed recently for manganites
[Phys.~Rev.~Lett., {\bf 92}, 157203 (2004)]  leads to an excellent description of such inhomogeneities.
In the  $\ell-b$  model two very different kinds of electronic states, one
localized and polaronic ($\ell$), and the other extended or broad band
($b$) co-exist. For model parameters  appropriate to manganites, and even within a
simple dynamical mean-filed theory (DMFT) framework, it describes many of the unusual
phenomena seen in manganites, including colossal magnetoresistance (CMR),
qualitatively and quantitatively. However, in the
absence of long ranged Coulomb interaction, a system described by such a model
would actually phase separate, into macroscopic regions of $l$ and $b$ electrons
respectively. As we show in this paper, in the presence of Coulomb interactions,
the {\em macroscopic} phase separation gets suppressed, and instead nanometer scale
regions of polarons interspersed with band electron puddles appear,
constituting a new kind of quantum Coulomb glass.
We characterize the size scales and distribution of the inhomogeneity
using computer simulations. For realistic values of the long range Coulomb interaction parameter
$V_0$, our results for the thresholds for occupancy of the $b$ states are in agreement with, and hence
support, the earlier approach mentioned above based on a configuration averaged DMFT treatment
which neglects $V_0$; but the present work has new features that can not be addressed in the DMFT framework.
Our work
points to an interplay of strong correlations, long range Coulomb interaction
and dopant ion disorder, all inevitably present in transition metal oxides,   as the origin of nanoscale inhomogeneities
rather than disorder frustrated phase competition as is generally believed.
As regards manganites, it argues against explanations for CMR based on disorder
frustrated phase separation and for an intrinsic origin of CMR. Based on this, we argue that the
observed micrometer(meso)-scale inhomogeneities owe their existence to extrinsic causes, eg.~strain due to cracks
and defects. We  suggest possible experiments to validate our
speculation.
\end{abstract}


\maketitle

\section{Introduction}\label{Intro}

In the last decade or so, a number of experiments on several families
of transition metal oxides have provided
evidence\cite{Dagotto2005,Dagotto2005a} that these often
consist of patches of two different kinds of
electronic/structural/magnetic states. The patches range in size from
nanometers to micrometers, and can be static or dynamic. In
doped manganites (Re$_{(1-x)}$Ak$_x$MnO$_3$ with Re and Ak being rare-earth and
alkaline earth ions respectively) where this phenomenon seems most widespread, one can have
insulating, locally lattice distorted (in some cases charge ordered)
regions coexisting with metallic, lattice undistorted
ones.\cite{Uehara1999,Renner2002,Fath1999,Zhang2002,Sarma2004,Dagotto2001,Dagotto2003,Rao2004,Shenoy2006}
In cuprates, an antiferromagnetic insulating state and a metallic (or
superconducting) state seem close in energy and may coexist under some
conditions, eg.~as nanoscale stripes (see, for example, the references in \cite{Lee2006}). There is a
view that the proximity of two states with very different long range orders (LRO) is a
defining characteristic of these systems. For example, such a view has
been most forcefully put forth in the work of Dagotto and
collaborators\cite{Dagotto2003,Burgy2001,Dagotto2003a,Dagotto2007} on the basis of
numerical simulations of finite sized samples of simplified lattice models capturing one or
more key features of manganites. In Ref.~\onlinecite{Dagotto2007}, models
with Mn $e_g$ electrons having Jahn-Teller coupling to classical phonons and double exchange coupling
to Mn $t_{2g}$ core spins and antiferromagnetic interactions $J_{AF}$ between the core spins
are studied for x=0.25 on lattices of size up to $12 \times 12$. The clean system has a first order transition (ending at a bi-critical temperature from a ferromagnetic metallic (F-M)
phase for small $J_{AF}$ to an antiferromagnetic(AF-I), charge ordered, insulating phase when $J_{AF}$ crosses a threshhold $J_{AF-t}$. The magneto-resistance  exhibits CMR like features when  $J_{AF}$ is tuned to be in the vicinity of the transition coupling $J_{AF-t}$.
But the tuning becomes less of a requirement in the presence of disorder, which further generates nanoscopic
interpenetrating patches of the two regions. These are then argued to be
generic features responsible for (nanoscale) two `phase'
coexistence, colossal magnetoresistance (CMR) and other phenomena
observed in manganites. Larger length scale (micron scale) coexistence is
also attributed to disorder effects amplified by proximity to a
critical value of competing parameters.\cite{Burgy2001} However, the
existence of manganites with negligible frozen disorder but showing CMR and other
characteristic phenomena\cite{Mathieu2004} suggests that these effects
are intrinsic, and that nano- and micron-scale
inhomogeneities could arise from other, intrinsic, causes.

In particular, the Coulomb interaction which is inevitably present can play a significant role if the
two phases involved have different charge densities, and will suppress phase
separation, leading to nanoscopic electronic inhomogeneity or `phase'
coexistence. This is the ubiquitous effect we investigate here
quantitatively, for the first time, in conjunction with the two fluid
$\ell - b$ model\cite{Pai2003,Ramakrishnan2003,Ramakrishnan2004} proposed recently for manganites. This model
invokes two very different types of electronic degrees of freedom, one polaronic and localized
(called $\ell$) with associated Jahn-Teller lattice distortion, and the other
(called $b$) forming a broad band, with no lattice distortion, and moving around primarily on sites
not occupied by the polarons. It has been shown\cite{Ramakrishnan2003,Ramakrishnan2004} that this approach
explains much of the unusual behavior of manganites qualitatively and
quantitatively. The work here shows in detail how the inclusion of electrostatic Coulomb
interactions in the $\ell - b$ model leads to nanometer scale inhomogeneities,
and quantifies this effect. More generally, our results describe the
intrinsic emergence of such nanoscale electronic inhomogeneities in any many
electron system with two very different microscopic states of
comparable energy. The model is different from disorder based phase
separation or domain formation as seen in computer
simulations of simple models\cite{Dagotto2003} or in statistical
Imry-Ma\cite{Imry1975,Burgy2001} type arguments. The frequently
observed micron scale inhomogeneity in many
manganites\cite{Uehara1999,Zhang2002,Sarma2004} is probably related to
strain effects, which are long ranged and unscreened. There are no
calculations of this effect, though simulations of elastic strain
effects and inhomogeneities in Jahn-Teller distorted lattice systems
have been made\cite{Ahn2004} on simplified models.

Specifically, in the $\ell-b$ model, at each lattice site there can be two
types of electrons, namely a nearly localized polaronic one (called
$\ell$) with site energy $-\Ejt$ and a broad band one (called $b$,
with site energy zero (0) and hopping amplitude $t$). There is a
strong local repulsion $U$ (we consider $U \rightarrow \infty$ in this
work) present between the $\ell$ polaron and the $b$ electron. The physical
origin and parameters of this model for manganites are
mentioned below (Section \ref{ellb}) and have been described in
references [\onlinecite{Pai2003,Ramakrishnan2003,Ramakrishnan2004,Krishnamurthy2005}].
If long range Coulomb interactions are neglected, the
system described by the model phase separates, i.~e., the ground state consists of two
separate phases, one a macroscopic region entirely of $\ell$ polarons at every site,
and  the other consisting of occupied $b$ band states,  the $b$ density
being fixed by the requirement of uniformity of the chemical potential $\mu$, whence
$\mu$ must equal $-\Ejt$. However, the two phases have very different charge densities.
With respect to ReMnO$_3$, which has one $e_g$ electron per site,
as the reference state, the polaronic phase is neutral, while the $b$  phase
has a positive charge density of $1-\bar{n}_b$ per site.
Furthermore, the doped alkaline earth ions (most likely randomly distributed on the lattice)
supply unit negative charge at each alkaline earth site. The relevant
``extended $\ell b$ model'' which includes long range Coulomb interaction
involving these charges is described in Section \ref{extellb}. It
leads to the suppression of macroscopic phase separation and the formation of
nanoscopic  ``puddles'' of $b$ electrons surrounded by regions with only $\ell$
polarons. We first present a simple analytical estimate of the size of
these regions, in particular their dependence on $V_0$, the energy
parameter that determines the strength of the long range Coulomb
interaction (Section \ref{Anlt}).  Next, we describe simulations
that we have carried out on finite systems of size up to
$20\times20\times20$, with the energy levels of the
$b$ electron puddles calculated quantum mechanically, and the Coulomb
effects treated in the Hartree approximation. The method for the
ground state determination is detailed in Section \ref{method}. The
results of the simulations are described and discussed in Section
\ref{Results}. The $\ell$ polarons are shown to form a Coulomb
glass\cite{Efros1975}.  Electronic states in the $b$ electron regions
are occupied (up to the chemical potential $\mu$) for a critical hole
concentration $x > x_{c1}$; these puddles connect percolatively and
the system is a metal for $x > x_{c2} > x_{c1}$ . We analyze and
discuss in detail $x_{c1}$, $x_{c2}$ as well as $b$ clump size
distributions as a function of $V_0, \Ejt$ and the arrangement of the
alkaline earth ions. We exhibit several examples of real space
inhomogeneous structures. The spatial autocorrelation of holes
(absence of polarons), and of holes with alkaline earth ions is also
elaborated. We compare the effective $b$ bandwidth obtained from
simulations which have the intrinsic nanoscale inhomogeneity mentioned
above with results from single site DMFT\cite{Pai2003} neglecting long
range Coulomb interactions but forcing homogeneity ('annealed disorder'), and find very good agreement.

In the final part of this paper (Section \ref{Disc}) we discuss some of the
approximations made, eg.~the assumption of random distribution of
dopant ions, the effect of disorder and its modeling. We also mention
a number of implications of our results for manganites such as Coulomb
glass behavior and other signatures of nanoscale inhomogeneities on
transport properties. We place our results in the context of the
inhomogeneities observed in experiments, and suggest some new
experiments to investigate long range strain effects.
A short description of this work has been published.\cite{Shenoy2007}

\section{Model and Method}\label{Model}

This section contains four parts. In the first part
(Sec.~\ref{ellb}), the $\ell b$ model\cite{Pai2003,Ramakrishnan2003,Ramakrishnan2004}
is briefly described. The second part (Sec.~\ref{extellb}) contains a description
of the extended $\ell b$ Hamiltonian. A simplified analytical solution
for the ground state of the extended $\ell b$ Hamiltonian is presented
in the third part (Section \ref{Anlt}). Finally, the determination of the ground
state of the extended $\ell b$ Hamiltonian is described in the final
part (Sec.~\ref{method}).

\subsection{Summary of the $\ell b$ model}\label{ellb}

\begin{figure}
\centerline{\epsfxsize=\figwidth \epsfbox{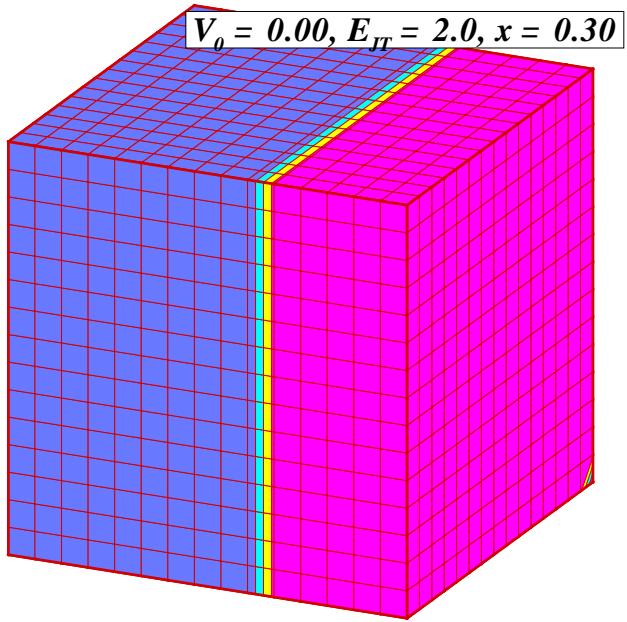}}
\caption{(color online) ``Macroscopic phase separation'' in the $\ell b$ model in  absence of long ranged Coulomb interaction. The lighter regions correspond to holes that form a large clump (with occupied $b$ states), the darker regions are occupied by $\ell$ polarons. This simulation is performed with a cube of size 16$\times$16$\times$16.}
\label{mphsep}
\end{figure}

The Mn$^{3+,4+}$ ions in Re$_{(1-x)}$Ak$_x$MnO$_3$ form a simple cubic
lattice (with lattice parameter taken as $a$, lattice sites indexed
by $i$). Each manganese ion experiences the octahedral environment of
oxygen atoms and thus the $d$ states are crystal field split into
$t_{2g}$ and $e_g$ orbitals.  For the purposes of describing the low energy physics of manganites
one can treat the $t_{2g}$ levels as always occupied by three electrons
at every site, with  parallel spin because of a strong Hund's rule,
and hence replace them by a  $S = \frac{3}{2}$,  ``$t_{2g}$ - core spin'',
and in the rest of this paper we approximate this as a classical spin, $S \hat {\Omega}$
where $ \hat {\Omega}$ is a unit vector. The remaining $(1-x)$
electrons per site move around among the $e_g$ orbitals with
an average hopping amplitude $t$. However, they have a strong Hund coupling $J_H$ with the core
spin, and there is a strong on-site Coulomb repulsion $U$ between the
$e_g$ electrons. Furthermore, they have a strong Jahn-Teller interaction
with the modes of distortion of the oxygen octahedron
surrounding the Mn$^{3+}$ ions.  As a consequence of this  and
phonon dynamics, two types of low energy, effective, $e_g$ electronic states
called $\ell$ and $b$ emerge. The polaronic $\ell$ state
is associated with a local lattice distortion in which an $e_g$ electron
can get self trapped, with a binding energy $\Ejt$. The hopping of the
$\ell$-polaron gets suppressed exponentially by the Huang-Rhys factor
$\eta$ ($= e^{-\Ejt/2 \hbar \omega_0} \approx 1/200$,
where $\omega_0$ is the frequency of the local lattice
distortion). The reduction arises
from the fact that the hop of the polaron from one site to a
neighboring one, in addition to the transfer of the electron,
involves the relaxation of the lattice distortion at the
original site and effecting a similar lattice distortion at the
neighboring site, and the exponentially small overlap of the corresponding
phonon wavefunctions. On the other hand, as $t >> \bar \omega_0$, $e_g$ electrons
can also hop fast among empty, undistorted or weakly distorted sites with
essentially the bare amplitude $t$, leading to the $b$ states.    The
greatly diminished hopping of the polarons allows for a useful
approximation: at temperatures much larger than $\eta t$, the polarons
may be considered as static. The effective $\ell b$ Hamiltonian can therefore be written as
\begin{widetext}
\bea
{\cal H}_{\ell b} & = & -\Ejt \sum_{i,\sigma} \ell^\dagger_{i \sigma}  \ell_{i \sigma} - t \sum_{<ij>} \left(b^\dagger_{i\sigma} b_{j \sigma} + \mbox{h.c.}\right) - J_H \sum_i (\bs_{i\ell} + \bs_{ib}) \cdot \bS_i  \label{lb} \non \\ && + U \sum_i n_{il} n_{ib} - \mu \sum_i (n_{il} + n_{ib}) - (J_{VDE} - J_{SE}) \sum_{<ij>} \bS_i \cdot \bS_j. \label{Hellb}
\eea
\end{widetext}
Here $\bs_{i\ell} (n_{i\ell})$ and $\bs_{ib} (n_{ib})$ are spin
(number) operators corresponding to the $\ell$ and $b$ degrees of freedom at site
$i$, and $\mu$ is the chemical potential determined from the condition
that $\mean{n_l + n_b} = 1 - x$. The last term contains the ``virtual
double exchange'' ferromagnetic coupling with strength $J_{VDE} \sim
x(1-x)\frac{t^2}{\Ejt}$. This coupling between
the neighboring core spins arises from fast virtual hopping
processes of the $\ell$-electron to a neighboring vacant
site. $J_{SE}$ is the antiferromagnetic superexchange.
The above Hamiltonian treats all the main energy scales that govern
manganite physics. The three largest local (strong correlation) energies are the on-site Coulomb or Mott-Hubbard repulsion $U \approx 5$ eV, the Hund's rule ferromagnetic exchange coupling $J_H \approx 2$eV and the Jahn-Teller energy $E_{JT}$ ($\approx$ 0.6 to 1 eV). The other important interactions are the nearest neighboring hopping $t \approx 0.2$ eV, the doping dependent virtual double exchange, and the superexchange $J_{SE} \approx 0.01$ eV.

Several simplifying approximations have been made in writing the Hamiltonian
\prn{Hellb}.  The first is that the kinetic energy in the Hamiltonian is
``orbitally averaged'', i.~e., the hopping amplitude $t$ represents an
average over the possible orbital configurations at the two pertinent
sites. Second, there are no cooperative/long range lattice effects,
i.~e., no intersite polaron correlations. Third, the virtual double
exchange has been approximated as a homogeneous interaction $J_{VDE}$
; in reality it acts only between pairs of sites where one is occupied by a
polaron and the other is empty. Fourth, the Jahn-Teller interaction has been
included only to the extent that it leads to the formation of polarons.
But its merit is that all the important energy scales governing  manganite physics are
included in \prn{Hellb}, at least in an approximate fashion.  Much of
the previous work on manganites is based on simplified
models\cite{Furukawa1995,Millis1996,Dagotto2003} that neglect one or more of
these energy scales.

The Hamiltonian \prn{Hellb} closely resembles the Falicov-Kimball
model (FKM)\cite{Freericks2003}; in fact, for $J_H \rightarrow \infty$, and
at $T=0$, when all the spins are completely ferromagnetically aligned,
it is the same as the FKM. The model was
solved\cite{Pai2003,Ramakrishnan2003,Ramakrishnan2004} using the
dynamical mean field theory\cite{Georges1996}. It is successful
in capturing the colossal magnetoresistance effect, the ferro-magnetic insulating
state found in low band width manganites, the systematics of the the
role of R and Ak ion radii,  etc.  The key point is that
for small doping $x$ the majority of the $e_g$ electrons get localized as $\ell$ polarons.
Because of the strong on-site Coulomb interaction $U$ (the largest energy scale in the
problem) the $b$ electrons avoid the polaronic sites,
and this reduces the effective half bandwidth
of the $b$ states from its bare value $D_0$ to $D_{eff}$  which is
strongly $(x)$ dependent.
At zero temperature with $U, J_H \rightarrow \infty$, and a
ferromagnetic ordering of the core spins, the effective half bandwidth
is given by\cite{Pai2003}
\bea
D_{eff} = \sqrt{x} D_0 \label{Deff}.
\eea
Thus, at low doping, the effective band bottom is above the
polaronic energy level $-\Ejt$ and therefore the $b$ band
is unoccupied and the chemical potential $\mu$ is
pinned at $-\Ejt$. Clearly, beyond a critical doping $x_c$ given by
\bea
 x_c = \left(\frac{\Ejt}{D_0}\right)^2 \label{xc}
\eea
the $b$-states begin to be occupied,
leading to a insulator to metal transition.
Furthermore, for $x > x_c$, starting from the metallic state at $T=0$,
the effective bandwidth reduces from its zero
temperature value with increase of temperature. This is because the
hopping of the $b$ electrons is strongly inhibited on account of their
Hund  coupling $J_H$ to the thermally disordered core spins. Hence one
gets a thermally induced ferro-metal to para-insulator
transition when the $b$ band moves above $\Ejt$.
For $T$ near $T_c$, the application of an external magnetic
field causes the core spins to align and thus increases the
effective bandwidth, leading to increase
in the number of thermally excited carriers by
orders of magnitude. This causes colossal magnetoresistance.

The work cited above did not address the issue of electronic
inhomogeneities. In fact, an effective homogeneous state was {\em
assumed} in the dynamical mean-field solution.
This is a drastic assumption, because from previous work on
the Falicov-Kimball model\cite{Freericks2002} (and as confirmed
by our computer simulations (see \Fig{mphsep})the ground state of the $\ell b$
Hamiltonian \prn{lb} (for the parameter ranges discussed above) is
known to be {\em a macroscopically phase separated state} due to the
strong on-site Coulomb correlation $U$. All the polarons cluster
on one side of the box; this allows band
states $(b)$ to optimize their kinetic energy by
moving  among the vacant `hole' sites in the other part of the box.
But the two portions have drastically different
electron densities. The motivation for the homogeneity assumption
made in the DMFT work  was that the phase separation will of course
be prevented by the long ranged Coulomb interactions which
are always present in the real system. In this work, we extend
the $\ell b$ model to include long ranged Coulomb interactions
to address the issue of electronic inhomogeneities, and
to check to what extent the DMFT homogeneity assumption is valid.

\subsection{The extended $\ell b$ Hamiltonian}\label{extellb}

The Hamiltonian is most conveniently developed in terms of `hole' operators $h^\dagger_i =
\ell_i$, which create vacant sites not occupied by polarons.
A hole carries a positive unit charge. The
charge is counterbalanced by the Ak$^{+2}$ ions which have
a negative unit charge. The Ak ions occupy the sites
$(\frac{l}{2},\frac{m}{2},\frac{n}{2})$ where $l,m,n$ are odd integers
(mimicking the perovskite type structure of manganites); the number of
Ak ions is equal to $xN$ ($N$ is the number of sites in the model) and they
are placed randomly on the lattice sites indicated. In the
orbital liquid regime\cite{Pai2003,Ramakrishnan2003,Ramakrishnan2004}
with large Hund coupling and on site Coulomb repulsion discussed above,
and at $T=0$, corresponding to a fully ferromagnetic alignment of
the core spins (spin drops out of the problem in this limit), the extended $\ell b$ Hamiltonian for fixed
particle number is given by
\begin{widetext}
\bea
{\cal H}_{\ell b}^{ext} = E_{\mbox{Ak}} + \sum_i \Phi_i \hat{q}_i +  V_0 \sum_{\{ij\}}  \frac{1}{r_{ij}} \hat{q}_i \hat{q_j} - t \sum_{<ij>} (b^\dagger_i b_j + b^\dagger_j b_i) + \Ejt \sum_{i} h^\dagger_i h_i \label{Hext}
\eea
\end{widetext}
where
\bea
\hat{q}_i =   \dagg{h}_i h_i - \dagg{b}_i b_i   \label{qop},
\eea
is the charge operator at hole site $i$ (in units of $|e|$), $E_{\mbox{Ak}}$ is the Ak-Ak
electrostatic interaction energy (a fixed number for a given realization of Ak
distribution in the lattice), $\Phi_i$ is the Coulomb potential at the hole site $i$ due to
Ak ions, $V_0$ is the strength of the Coulomb
interaction between nearest neighbor holes (of the order of 0.02eV,
$V_0 \approx 0.01t-0.1t$, see discussion below), $r_{ij}$ is the
distance between holes $i$ and $j$ in units of the lattice constant. Each hole has an energy penalty of
$\Ejt$. The sum $\{i,j\}$ is over all pairs of hole sites, while
$\langle ij \rangle$ denotes a sum only over nearest neighbor hole sites. The site charge
operators satisfy the constraint
\bea
\sum_i \mean{\hat{q}_i} = x N
\label{const}.
\eea
The $U \rightarrow \infty$ constarint implies that the $b$ electrons can only
move among `hole' sites, where $\dagg{h}_i h_i = 1$.

The Hamiltonian  \prn{Hext} embodies the competition between long
range Coulomb interaction which works to keep the holes as far apart
as possible, while the kinetic energy of band electrons promotes
the formation of ``clumps''. A clump is a collection of hole sites, each member
of which can be reached from any other member via a sequence of nearest neighbor
hops which visit only the members of the clump. Thus every configuration of holes
can be broken up into a set of clumps where each hole belongs to only one clump.
Site delocalized electronic $b$ states are possible in
a clump that has more than one member (a one member clump is called a
``singleton''). The kinetic energy gain can possibly promote an $\ell$
polaron to occupy the $b$ state in the clump leading to an electron ``puddle'',
which creates an additional hole in the system at the site from which
the $\ell$ polaron is removed, which changes the electrostatic energies,
etc. Thus the number of holes and number of band electrons is not
individually conserved, but only the constraint \prn{const} is
satisfied. The key question is: what is the ground state of the
Hamiltonian for a realization of the random distribution of the Ak
ions? The determination of this ground state requires an optimization of
the number of holes and their distribution [or equivalently, the configuration of the on site $\ell$ polarons,
which exist at sites where the holes are not there], as well as the clump structure, the
associated $b$ states and their occupancy (constituting the `b' puddles surrounded by the
$\ell$ plarons),  so as to achieve the lowest energy of the Hamiltonian
\prn{Hext}. An exact solution of this problem is beyond the available
techniques of correlated electron theory, so approximations have to be
resorted to. In this paper we explore two alternate ways; the first one is essentially
analytic (and very approximate), while the second one is a full scale numerical treatment.

\subsection{A simplified analytic treatment}\label{Anlt}

\begin{figure}
\centerline{\epsfxsize=\figwidth \epsfclipon \epsfbox[157 423 454 722]{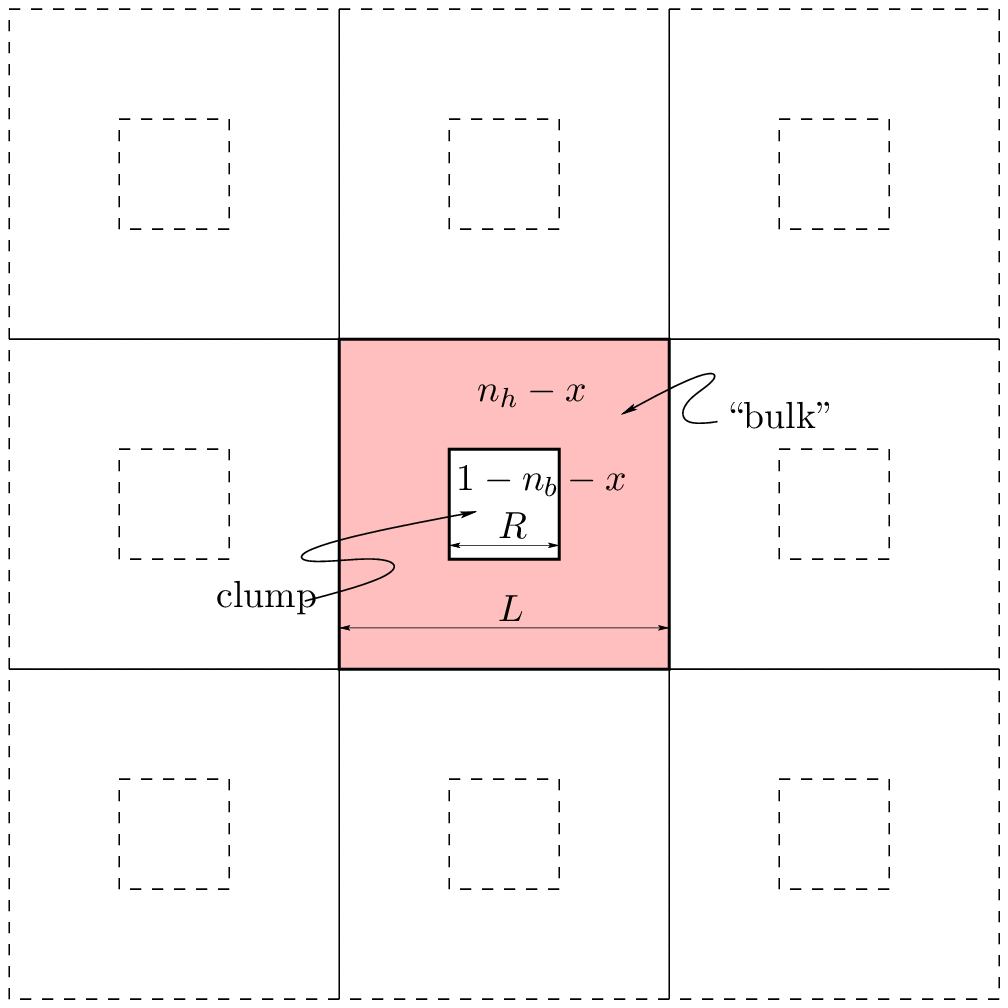}}
\caption{Schematic of the ground state used in the analytical calculation of Section \ref{Anlt}. ``Phase separation'' takes the system to two distinct type of regions called the ``bulk'' (regions with $\ell$ polarons) and the clump. The dashed lines indicate the assumed periodic nature of the clump distribution -- clumps of size $R$ (volume $R^3$) are assumed to be arranged in a periodic fashion (period $L$) with intervening ``bulk'' regions.  Site delocalized electronic states are found in the clumps. The initial hole density is $x$ (also equal to the background negative charge density). The charge density in the clump is $(1-n_b-x)$ and that in the ``bulk'' is $n_h -x$.  $n_b$ is the fraction of electrons that are promoted to delocalized states, and $n_h$ is the
fraction of holes created in the 'bulk'. The fractions $x$, $n_b$ and
$n_h$ are related via  charge balance.}
\label{scheme}
\end{figure}

A simple, analytical approximation for the ground state becomes possible if
we assume that the Ak ions are distributed
homogeneously in space, i.~e., model the charge of the Ak ions as
a ``jellium''. The approximation involved is a variational calculation with
an assumed ground state as shown in \fig{scheme}, where
clumps of size $R$ (volume $R^3$) are taken to be
spaced periodically with a spacing $L$. The regions between the clumps
-- the ``bulk'', is taken to have concentration of holes $n_h$ (the
concentration of holes in the clump is, of course, unity). Charge
conservation gives
\bea
(1 - n_b) R^3 + n_h (L^3 - R^3) = x L^3  \label{chargeconserve}
\eea
whence
\bea
n_h = \frac{ \alpha^3 x + (b -1 )} {\alpha^3 - 1} \label{heqn}
\eea
with $\alpha = L/R$. Clearly, the charge density $\rho_C$ in the clump is $(1 - n_b -x)$ and that in the bulk $\rho_B$ is $n_h -x$, i.~e.,
\bea
\rho_C = 1 - n_b - x, \;\;\;\;\; \rho_B = n_h -x = - (1 - n_b -x) \frac{1}{\alpha^3 - 1} \label{rhoeq}.
\eea

Associated with this charge distribution, there is an electrostatic energy (per ``unit cell'' of volume $L^3 = \alpha^3 R^3$) given by
\bea
E_{ES}(n_b,\alpha,R) = V_0 {\cal E}(n_b,\alpha,R/a)
\eea
where $a$ is the lattice parameter of the underlying atomic lattice, ${\cal E}(n_b, \alpha)$ is the ``Madelung function'' accounting for the net electrostatic energy in the system.
Since a total number of $n_b R^3$ polarons have been promoted to delocalized $b$ electrons, there is a loss of polaron energy proportional to this number, i.~e.,
\bea
E_{P}(n_b,\alpha,L) = E_{JT} n_b R^3
\eea
Finally, there is the delocalized kinetic energy of the electrons, which is expected to be of the form
\bea
E_{KE}(n_b,R) = t{\cal K}(n_b,R)
\eea
where ${\cal K}$ is a function of $n_b,R$. It can be evaluated by
finding the kinetic energy of $b_b R^3$ electrons in a clump of
size $R^3$. Note that we do not take $E_{KE}$ to go as $1/R^2$ as might
be expected from a free electron picture since the lowest energy that
can be attained a tight binding scenario is bounded below by $-6t$
(for a cubic lattice).

Putting everything together, we get
\bea
E_{tot}(n_b, \alpha, R) = E_{ES}(n_b,\alpha,R) + E_P(n_b,R) + E_{KE}(n_b,R).
\eea
We need to minimize this energy  as a function of $n_b$, $\alpha$ and $R$;
this will give us the clump size, spacing and charge distribution.

In the remainder of this  section $R$ stands for a dimensionless number (clump size normalized by lattice parameter $a$). Similarly charge densities are all dimensionless. We shall now proceed to estimate the different energy functions noted above.

We estimate the electrostatic energy by assuming the box is
``spherical''. In this case the total charge in the box is zero, and
hence the electric field {\em outside the sphere vanishes}, hence to a
very crude approximation one box does not affect the energy of the
neighbor. This is expected to be quite reasonable when the Ak ions are
distributed homogeneously (as we have assumed in this calculation),
since the electrostatic interaction is expected to be well screened.
With these assumptions, we find,
\bea
E_{ES}(n_b, \alpha, R) = \frac{(4 \pi)^2}{45} V_0 ( 1 - n_b - x)^2 R^5 f(\alpha) \label{Esapprox}
\eea
where $f(\alpha)$ is the function
\bea
f(\alpha) = 1 + \frac{1}{\alpha} \left( \frac{(\alpha^6 - 5 \alpha^3 +9 \alpha - 5)}{(\alpha^3 -1 )^2} -  \right. \non \\  \left. \frac{ 5 (\alpha - 1)^2 (\alpha + 2)}{(\alpha^3 - 1)} + 5 ( \alpha - 1) \right)
\eea
The function $f$ is such that $f(1) = 1$ and $f(\infty) = 6$, a very slowly
varying function of $\alpha$ and always of order unity.

The polaron energy (again for a spherical clump) is
\bea
E_{P}(n_b,\alpha,L) = \frac{4 \pi} {3} E_{JT} n_b R^3.
\eea

The estimation of the kinetic energy entails probably the
crudest approximation of this analysis. We assume that the the density of states is ``flat''
\bea
g(\epsilon) = \frac{1}{12t} \;\;\;\;\;\; -6t \le \epsilon \le 6t
\eea
With this assumption we  estimate the kinetic energy  (for a spherical clump) to be
\bea
E_{KE}(n_b,R) =  - \frac{4 \pi}{3} R^3 (6 t) n_b(1-n_b)
\eea
The total energy is given by
\bea
E(n_b,\alpha,R) =  \frac{4 \pi}{3} \left( \frac{4 \pi}{15} V_0 ( 1 - n_b - x)^2 R^5 f(\alpha) + \right. \non \\ \left.  E_{JT} n_b R^3  - 6 t   n_b(1-n_b) R^3 \right) \label{Enefunc}
\eea
The analysis of the minima of the above function shows that $E_{JT} < 6t$ {\em is a necessary condition} for the existence of a clump. If $E_{JT} > 6t$, then the coefficients of $R^5$ and $R^3$ are both positive (for any value of $b$) and hence $R$ will vanish for energy minimum. Inspecting \prn{chargeconserve} we see that this will result in a homogeneous distribution of the holes in the ``bulk'' equal to the cation charge density -- the ``bulk'' becomes neutral.

We now consider  $E_{JT} < 6t$. Furthermore,
\begin{enumerate}
\item The $b$ electron density in the clump is determined by ``chemical potential balance''. For the case of a flat band density of states, this will imply that the ``chemical potential'' is $E_{JT}$ and
\bea
\int_{-6t}^{-E_{JT}} \frac{1}{12t} \D{e} = n_b \;\;\;\;\;\; \Longrightarrow   \;\;\;\;\;\;\;\; n_b = \half  \left(1 - \frac{E_{JT}}{6t} \right).
\eea

\item The factor $\alpha$ is determined from the condition that the hole density in the bulk vanishes (charge conservation resulting in \prn{heqn})
\bea
\alpha^3 = \frac{1 - n_b}{x}
\eea
\end{enumerate}

Within this framework, we minimize \prn{Enefunc} w.~r.~t.~$R$ to obtain
\bea
R^2 =  \frac{9 n_b ( 6t (1 - n_b) - E_{JT} ) }{ 4 \pi f(\alpha) V_0 (1 - n_b - x)^2}
\eea
Note that the clump size varies as $1/\sqrt{V_0}$.

\begin{figure}
\centerline{\epsfxsize=\figwidth \epsfbox{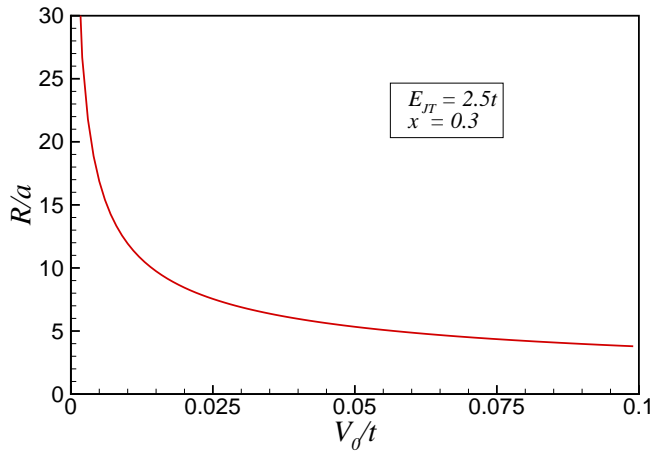}}
\caption{Electron puddle size (normalized by lattice parameter $a$ ) as a function of the Coulomb interaction parameter $V_0$ (normalized by $t$)  for $E_{JT} = 2.5 t$, $x = 0.3$.  }
\label{clumpsize}
\end{figure}

We take $E_{JT} \approx 2.5 t$, $x \approx 0.3$ to study the clump
size as a function of $V_0$. The result is shown in
\fig{clumpsize}. For these values, the centers of the clumps are
spaced at a distance of about two and a half times the clump size
($\alpha = 1.33$).  A reasonable estimate of $V_0$ in manganites is
about $0.1t$ ($V_0$ is likely to be between $0.01$ and $0.1$,
see discussion below). In this range we see that the clump size is
between 5 and 10 lattice spacings (taking $a = 5$ \AA, we get $R = 2 $ -- 5 nm),
in surprisingly good agreement with available experiments.
We must emphasize that we really do not have tight
control of the constants (such as $9/4 \pi$ appearing in the equation
for $R$). However, the main point that we learn is that the clump size
is a few lattice spacings, much as what the Coulomb interaction is
expected to do. We note here that there has been a recent calculation\cite{Kugel2005}
of ``phase separation'' in doped manganites adding in Coulomb effects
to the two fluid $\ell-b$ model of Ref. \onlinecite{Ramakrishnan2004} along lines similar to
those discussed above. However, that calculation does not uncover the dependence of the
clump size on the long ranged Coulomb interaction as we have done here.

We note again that the key assumption of the above analysis is the homogeneous
distribution of the Ak ions. A more realistic solution that takes into account
the inhomogeneous random distribution of the Ak ions requires a full
scale numerical treatment of \prn{Hext}, and we turn to this next.

\subsection{Approximate determination of the ground state}\label{method}

The key approximation that permits a numerical determination of the ground state of the
Hamiltonian \prn{Hext} is a Hartree-like treatment of the Coulomb
interactions between the $b$ electrons. This allows us to treat all the
electrostatic energy contributions in a classical fashion, i.~e., replace the
site charge operators $\hat{q}_i$ by their expectation
values $q_i$ in the ground state $\ket{~}$:
\bea
q_i =   \bra{~} \dagg{h}_i h_i - \dagg{b}_i b_i \ket{~}  \label{qh}.
\eea
In the absence of $b$ states the problem reduces to the Coulomb glass
problem\cite{Efros1975,Efros1976} -- a fully classical problem. The presence of the
$b$ states makes the system a new kind of quantum Coulomb glass --  the
$h-b$ glass -- with {\em coexisting localized and delocalized states},
in contrast to usual quantum Coulomb glasses\cite{Vojta2001}.
Since the Ak ions are randomly distributed, and in addition the density of $b$ electrons is very
low, a Hartree approximation is likely to be reasonably accurate.
We do indeed find (see below) that the interaction is well screened
and the charge distribution is almost homogeneous at scales larger than a few lattice spacings.

Our method of determination of the ground state is based on a
generalization of the method previously used in finding ground states
of the classical Coulomb glass\cite{Branovskii1979,Davies1984}. Here, the
ground state is obtained by starting from a trial configuration
(usually a random state), and ``performing'' transfers of electrons that lower
energy until  no transfer is possible that can lower the energy. In the
classical Coulomb glass each transfer involves moving an electron from its current position to
a vacant site, creating  a particle hole
``excitation''. To follow a strategy similar to the one above in the
present context, we investigate the energetics of possible
transfers in the $h-b$ glass.

The following definitions are useful to understand excitations in the
$h-b$ glass. Each clump $\alpha$ has many `delocalized' one particle
levels (obtained by diagonalizing the kinetic energy on this clump);
we label these one particle levels by $r$, and denote their kinetic
energy by $\epsilon_{\alpha,r} $ . The operator
\bea
\dagg{b}_{\alpha,r} = \sum_{i(\alpha)} \Psi^{\alpha, r}_i \dagg{b}_i
\eea
creates a $b$ electron in the $r$-th band state of the clump $\alpha$,
where $\Psi^{\alpha, r}_i$ is the associated single particle wave
function spread out over the clump ($i$ runs over all hole sites in the clump $\alpha$ as
indicated by $i(\alpha)$).  Clearly, for a hole at site $i$
\bea
q_i = 1 - \sum_{r} |\Psi^{\alpha, r}_i|^2 \theta (\mu -\epsilon_{\alpha r})
\eea
Here $\alpha$ is the clump to which the hole $i$ belongs, and
$r$ runs over all the {\em occupied} band states in the clump $\alpha$. The
potential $\phi_i$ at hole site $i$ is defined as
\bea
\phi_i = V_0 \sum_{j \ne i} \frac{1}{r_{ij}} \, q_j.
\eea

Consider, for a given $x$, a configuration containing $N_h$ holes
and $N_b$ band electrons (obviously, $\sum_i q_i = N_h - N_b = xN$),
with $N_c$ clumps.
The extra energy $E^h_i$ in the system due to the
addition of a hole at a hole-vacant site $i$ {\em without changing
the clump structure}  is,
\bea
E^h_i = \Phi_i + \phi_i + \Ejt. \label{hole}
\eea
The addition of a hole can, in general, change the clump structure since the new hole can modify existing clumps or create new clumps. If the modified clumps contain band electrons, then their energies will change. We ignore these effects in writing \prn{hole}.
A similar definition of $E^b_{\alpha,r}$, the energy increment for the addition of a band electron in the vacant band state ${\alpha,r}$,  is
\bea
E^b_{\alpha,r}  & = & \epsilon_{\alpha,r}  - \sum_i (\Phi_i + \phi_i) |\Psi^{\alpha,r}_i|^2  \non \\
& &  +  V_0 \sum_{\langle i, j(i \ne j) \rangle} \frac{1}{r_{ij}} |\Psi^{\alpha, r}_i|^2 |\Psi^{\alpha, r}_{j}|^2 , \label{band}
\eea
where $i,j$ run over all the hole sites in clump $\alpha$. The
definitions $E^h_i$ and $E^b_{\alpha,r}$ serve as the equivalent of
the ``single particle levels'' in the present context (Hartree approximation).

\begin{figure*}
\centerline{\epsfxsize=\figwidth \epsfbox{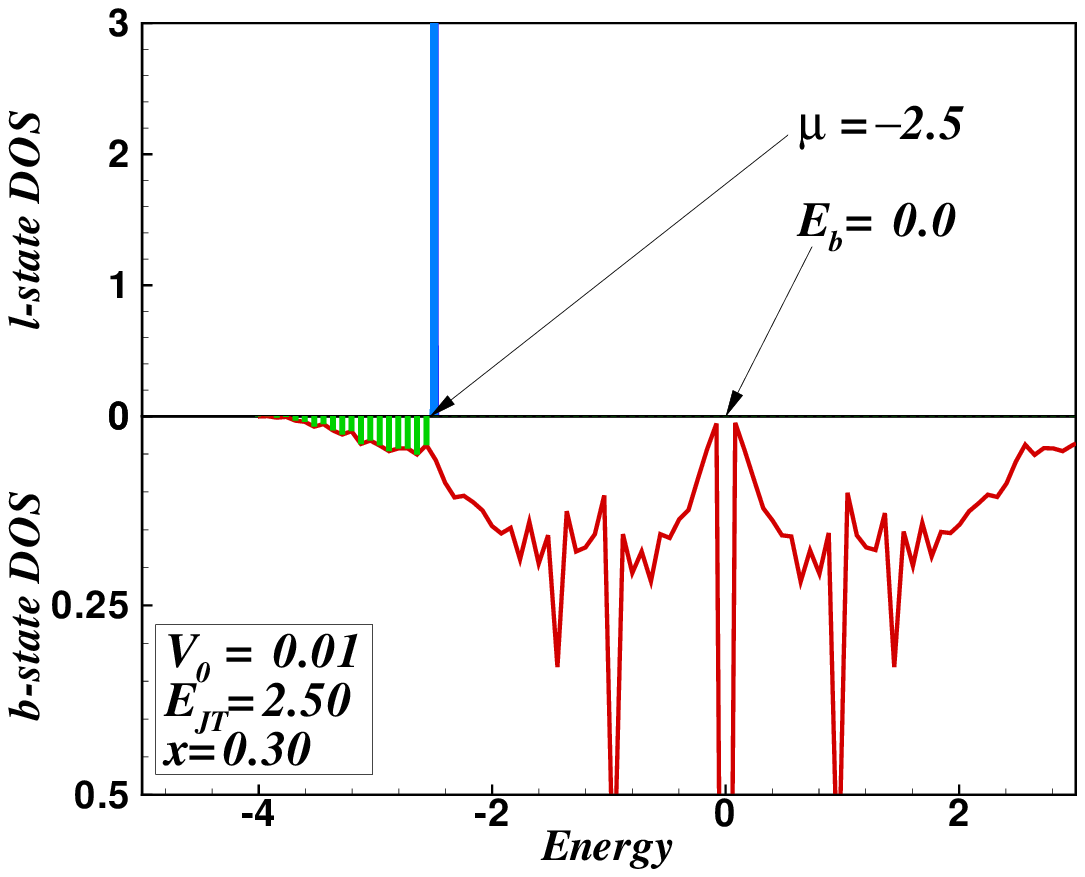}~~~~~\epsfxsize=\figwidth \epsfbox{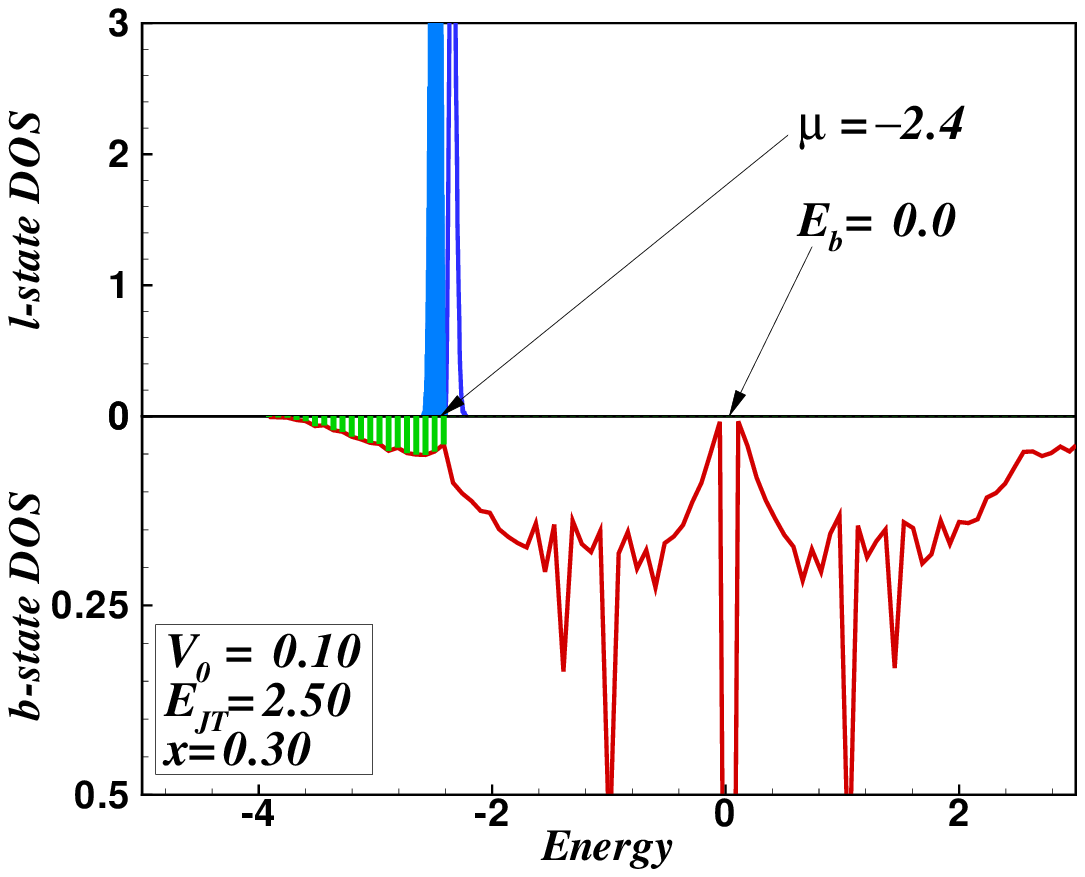}}
\centerline{(a)\hspace{9.0truecm}(b)}
\centerline{\epsfxsize=\figwidth \epsfbox{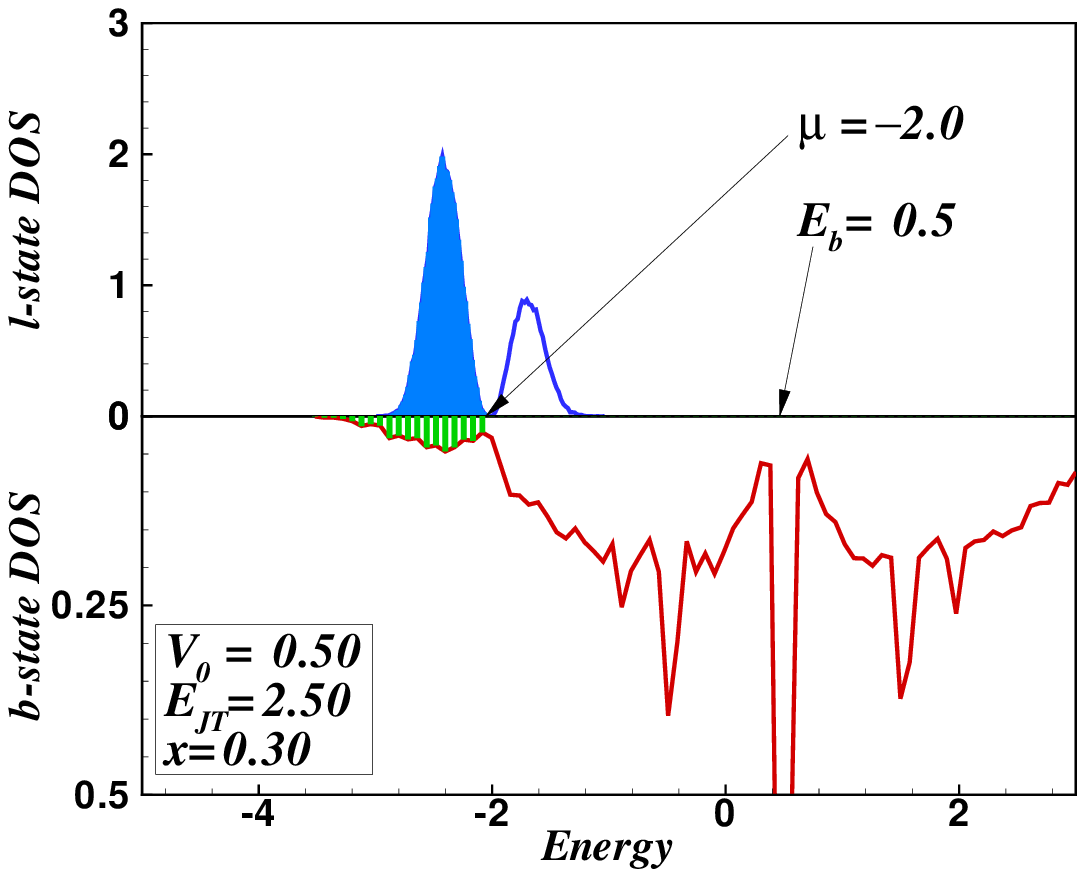}~~~~~\epsfxsize=\figwidth \epsfbox{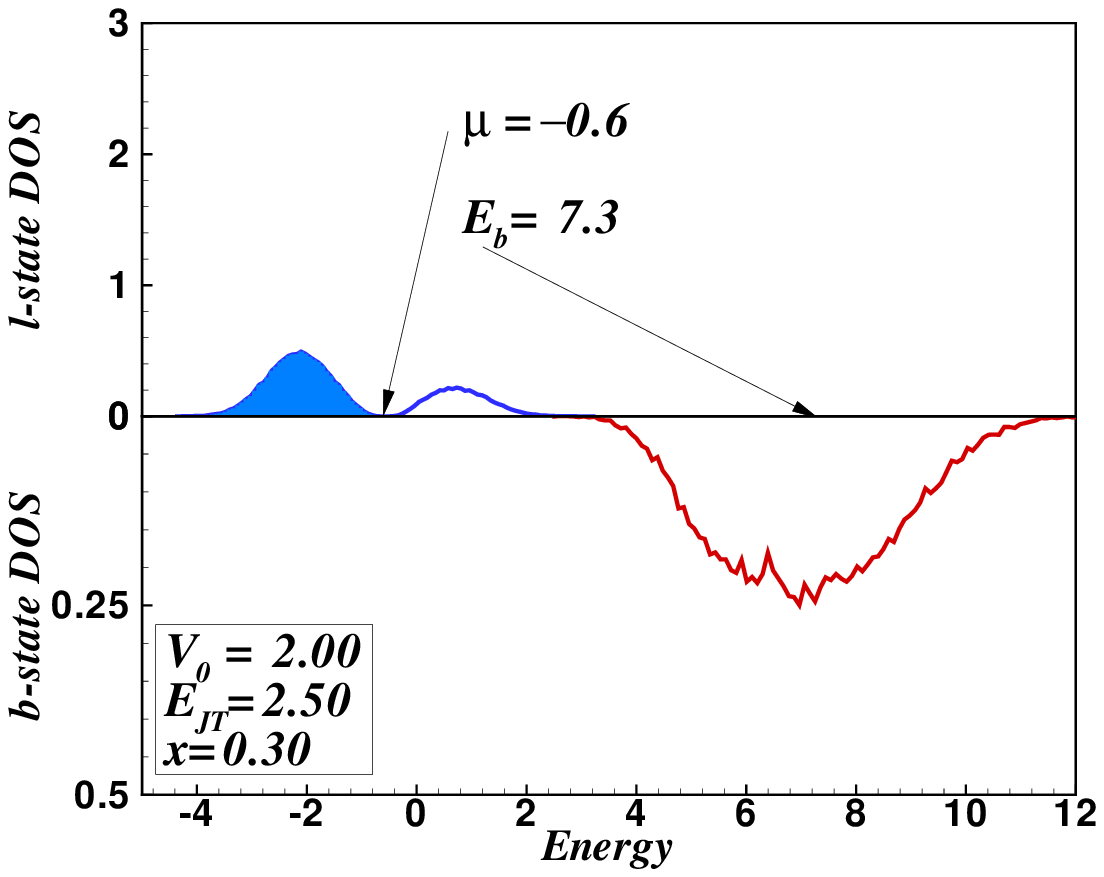}}
\centerline{(c)\hspace{9.0truecm}(d)}
\caption{(color online) Density of states (DOS) of $\ell$ polarons and $b$
electrons, where their respective energies are defined by \prn{hole}
and \prn{band}. The occupied states are shaded. The chemical
potential $\mu$ and the center of the $b$-band $E_b$ are indicated. The different panels (a)--(d) show the variation of the densities of states as a function of $V_0$ for $\Ejt = 1.0$ and $x=0.3$. The polaron density of states shows a
Coulomb gap at the chemical potential for all values of $V_0$
including $V_0 = 0.01$ (graph (a), top-left).
}
\label{dosV}
\end{figure*}

These definitions allow us to determine energies of ``excitations'' in the
$h-b$ glass corresponding to the
creation/annihilation of holes or
$b$/$\ell$-electrons. Strictly, the clump structure changes in an
excitation; as mentioned earlier we ignore this and our excitations are
``frozen-clump excitations''.

Three types of excitations are possible in the $h-b$ glass:

\begin{enumerate}
\item ``$\ell-h$'' excitations: Here a hole-occupied  site  $j$ obtains an electron from a hole-vacant site ($\ell$ occupied site) $i$. The energy for this excitation  is
\bea
\exic^{\ell-h}_{i;j} = E^h_i - E^h_j - \frac{V_0}{r_{ij}} . \label{ell-h}
\eea
This type of excitations can affect the clump structure. This arises from the fact that the clump structure is determined by the position of the holes.

\item ``$\ell-b$'' excitations: Here a hole-vacant site $i$ (containing an $\ell$ electron), donates an electron to a vacant band state $(\alpha,r)$. The energy of this type of excitation is
\bea
\exic^{\ell-b}_{\alpha,r;i} = E^b_{\alpha,r} + E^h_i - V_0 \sum_j \frac{1}{r_{ij}} |\Psi^{\alpha, r}_j|^2,  \label{ell-b}
\eea
where $j$ run over all the holes-sites in clump $\alpha$. Again, this
type of excitation can change the clump structure. It must be noted
that the reverse of this process is also a possible excitation (a
$b-\ell$ excitation), in that a band electron annihilates a hole, the
energy of which is negative of \prn{ell-b}. Again, since both $\ell-b$ and $b-\ell$ excitations change the position of the holes, they can affect the clump structure.

\item ``$b-b$'' excitations: Here a $b$-electron transfers from $(\alpha,r)$ $b$ state to the vacant $(\beta,s)$ $b$ state. The energy of this excitation is
\bea
\exic^{b-b}_{\beta,s;\alpha,r} = E^b_{\beta,s} - E^b_{\alpha,r} - V_0 \sum_{i,j} \frac{1}{r_{ij}} |\Psi^{\beta,s}_i|^2 |\Psi^{\alpha,r}_j|^2, \label{b-b}
\eea
where $i,j$ run over $\beta$ and $\alpha$ (the last term requires obvious modifications if $\alpha = \beta$). Note that this type of excitation will not affect the clump structure since it does not affect the position of the holes.
\end{enumerate}

For a given doping, the ground state is obtained in two stages. In the
first stage, $b$ states are not accounted for and the classical
Coulomb glass ground state of the holes in presence of the electrostatic potential
from the Ak ions is obtained. This is achieved by performing a series
of $\ell-h$ excitations until a minimum energy configuration is
achieved.  In the second stage, starting from the classical Coulomb
glass state, all possible excitations are performed iteratively. Each
iteration consists of the following steps:
\begin{enumerate}
\item {\em Find best excitation}: From the definition of the single
particle levels $E^h_i$, $E^b_{\alpha,r}$, and the excitation energies
defined above, the best possible excitation (the one that reduces the
energy the most) is determined.
\item {\em Perform the excitation:} If it is an $\ell-h$ excitation, then
one hole is removed and one is added. If it is an $\ell-b$ or $b-\ell$
excitation, then either a hole is removed or added. If it is $b-b$
excitation, holes do not change. Thus, $\ell-h$, and $\ell-b$
excitations change the clump structure.
\item {\em Update clump structure:} If the clumps with occupied band
states are not disturbed, then no update is required. If some of the
clumps have been disturbed, then the intersection of the old clumps
and new clumps is found, and electrons are distributed in the new
clumps so that the charge distribution is as close as possible to that of the
previous
distribution. After update, the number of band electrons must be same
as before, unless isolated holes in the new clump structure are
annihilated by band electrons. All energies are recalculated on
update.
\end{enumerate}
Iterations are carried out until an energy minimum is achieved. It
must be noted that the ground state that we obtain is based on single
particle excitations. In general, the stability of the ground state
must be checked for two (and multi) particle
excitations.\cite{Davies1984} This process is computationally
intensive. The correctness of the ground state obtained in our case is
affirmed by the energies of the highest occupied states of the $\ell$
electrons and the $b$ electrons -- if the ground state calculation is
correct, then the highest occupied states of both types of electrons
will correspond to the chemical potential $\mu$. We have never found
serious violation of this criterion in several ten thousand simulations;
we therefore believe that our scheme is robust enough to determine the
ground state of the $h-b$ glass.

The electrostatic energy is calculated using the Ewald
technique\cite{Kittel1996} using fast Fourier transform routines made
possible by the use of periodic boundary conditions. The other
computationally intensive step is the calculation of  energies and
wavefunctions of the $b$ states in a clump. This is the most CPU intensive
step which limits the size of the simulation cell. The size of the
cell that can be calculated depends on the doping. For $x=0.2$ we have
calculated with cells as large as 20$\times$20$\times$20.

Some other points regarding our simulations may be noted. First, the
strong correlation $U$ that induces many body effects is treated
almost exactly (since $U$ in the real system is very large, the $U =
\infty$ limit is accurate); the $b$-electron quantum dynamics is
treated almost exactly ($b$-electrons do not hop to sites with $\ell$
polarons). In our calculation, the long ranged Coulomb energy (within
the Hartree approximation) is treated accurately using Ewald
techniques (as mentioned above), and no further approximations are
made in the calculation of the long ranged electrostatic
energy. Further, we note that the Hamiltonian we study and the method we
develop here to find the ground state is a novel generalization (both
model and method) of the quantum Coulomb glass. The key point is that
our model consists of two types of electronic states, one localized
polaronic (classical) and other delocalized band-like (quantum), and
our treatment accounts for both of these on an equal footing.

\begin{figure*}
\centerline{\epsfxsize=\figwidth \epsfbox{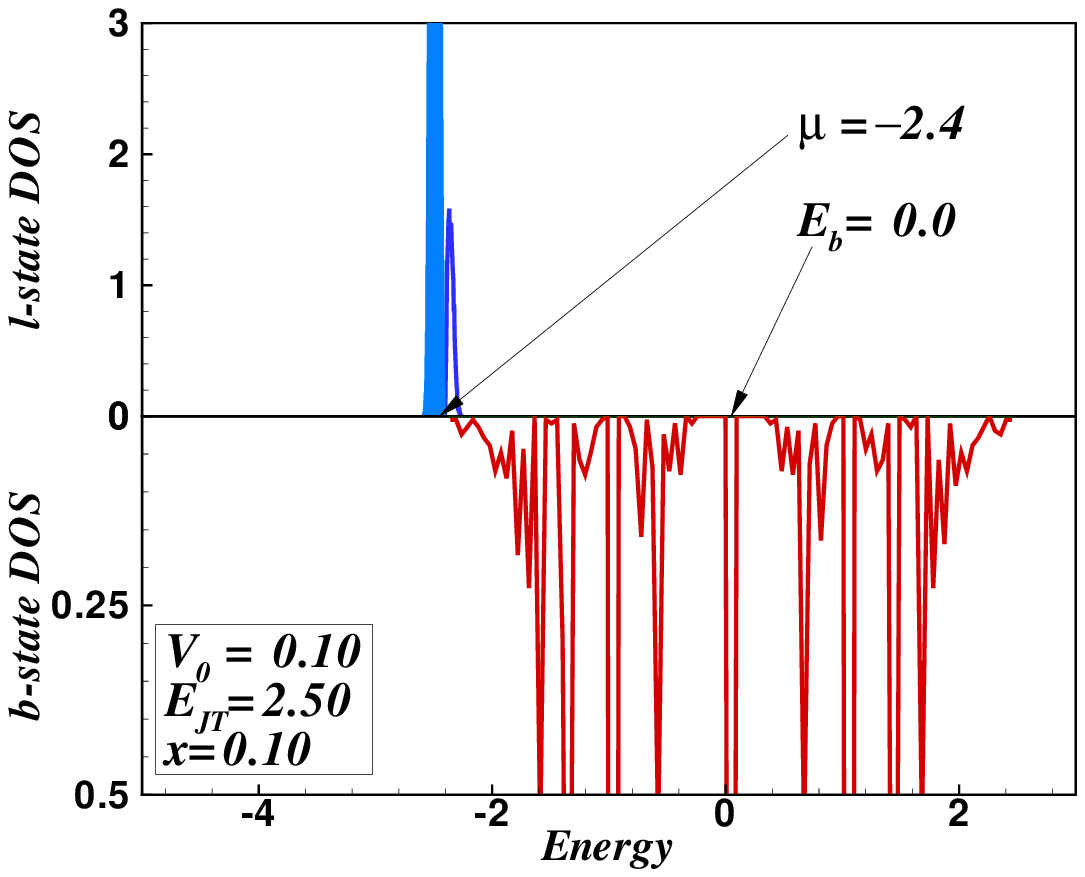}~~~~~\epsfxsize=\figwidth \epsfbox{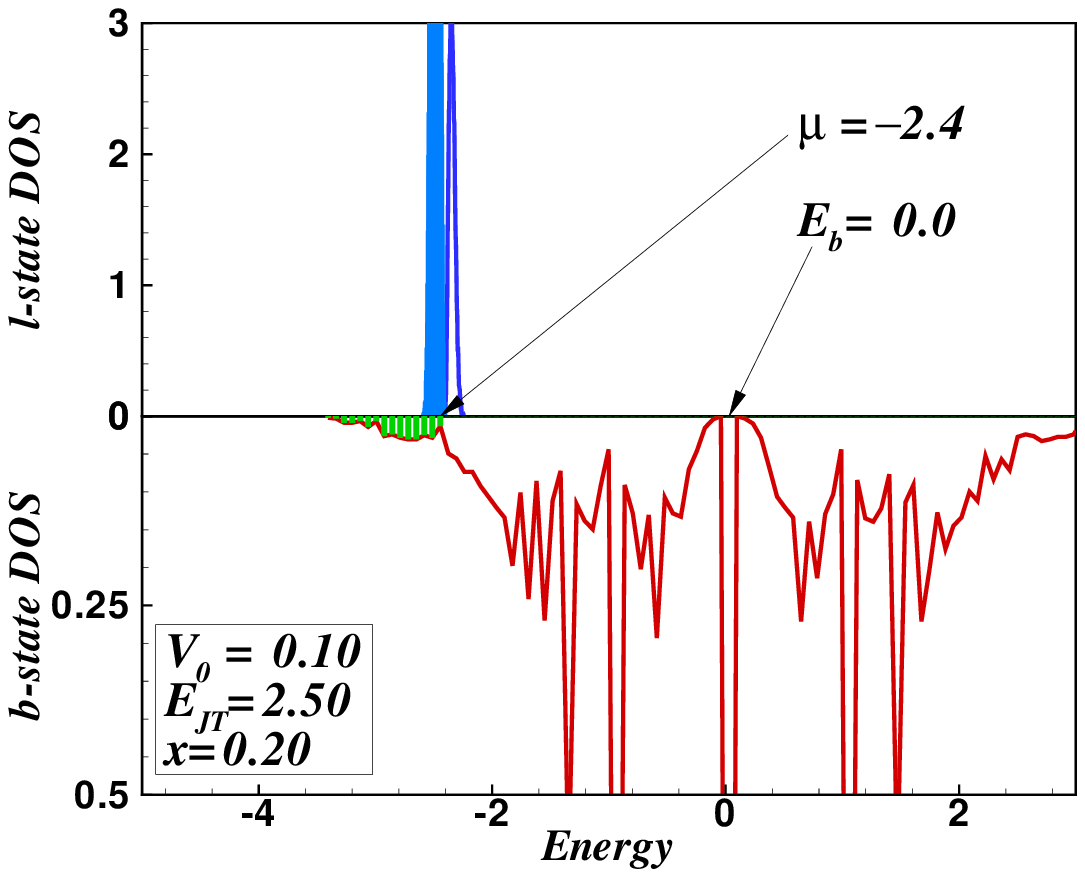}}
\centerline{(a)\hspace{9.0truecm}(b)}
\centerline{\epsfxsize=\figwidth \epsfbox{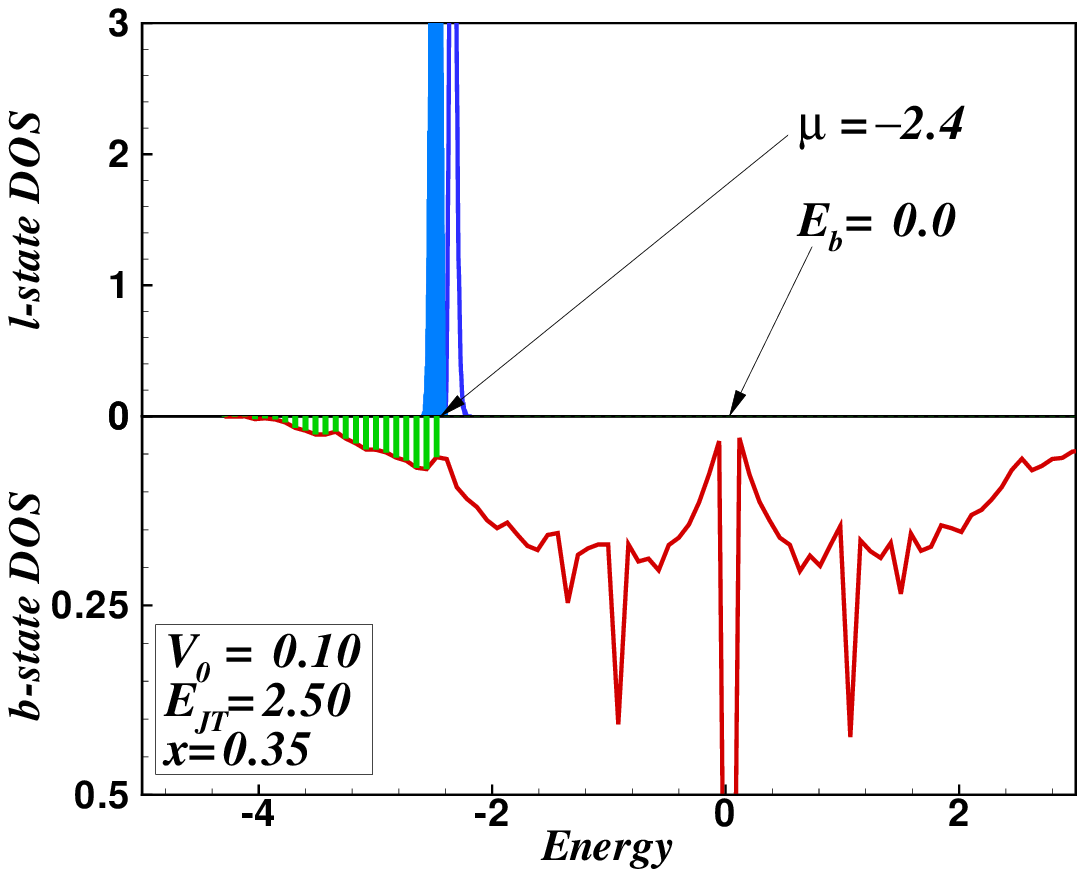}~~~~~\epsfxsize=\figwidth \epsfbox{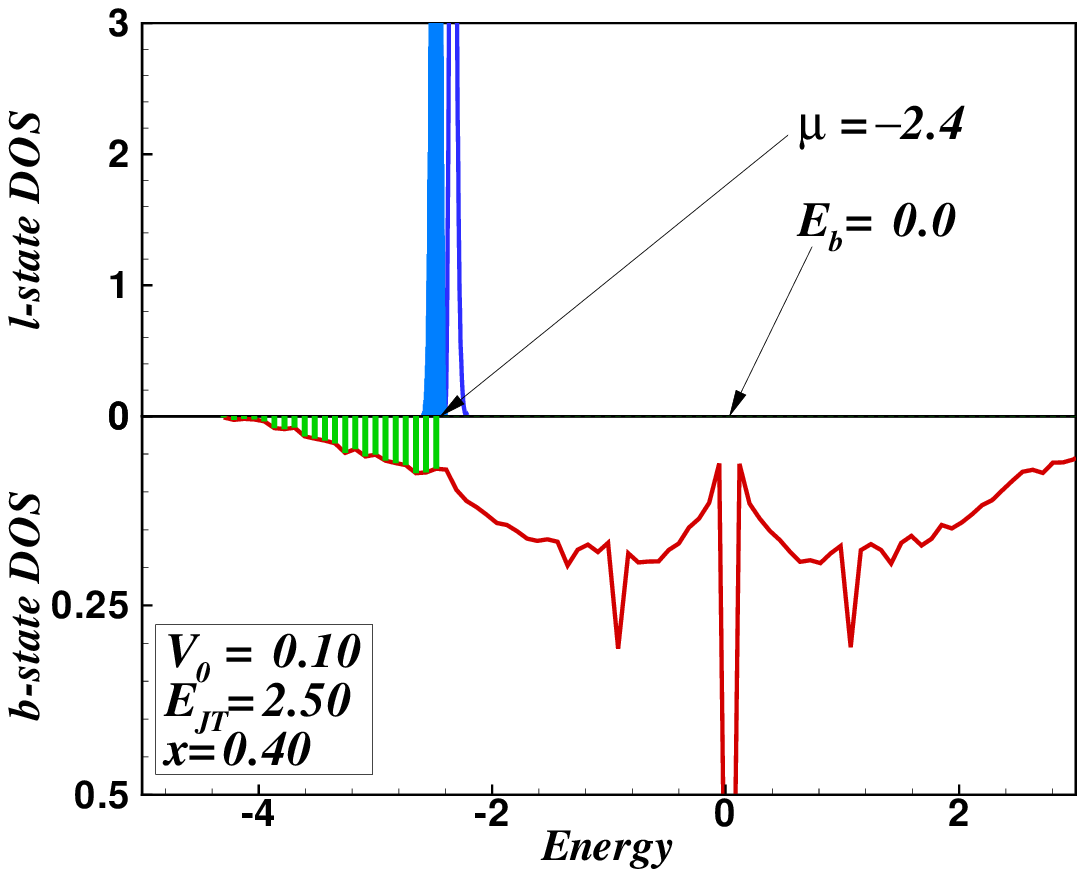}}
\centerline{(c)\hspace{9.0truecm}(d)}
\caption{(color online) Density of states (DOS) of $\ell$ polarons and $b$
electrons, where their respective energies are defined by \prn{hole}
and \prn{band}. The occupied states are shaded and the chemical
potential is marked by $\mu$. The center of the $b$-band $E_b$ is also indicated. The different panels (a)--(d) show the variation of the densities of states as a function of $x$ for $V_0=0.1$ and $\Ejt = 2.5$. It is evident that the $b$ bandwidth increases with increasing doping $x$.}
\label{dosx}
\end{figure*}

\begin{figure*}
\centerline{\epsfxsize=\figwidth \epsfbox{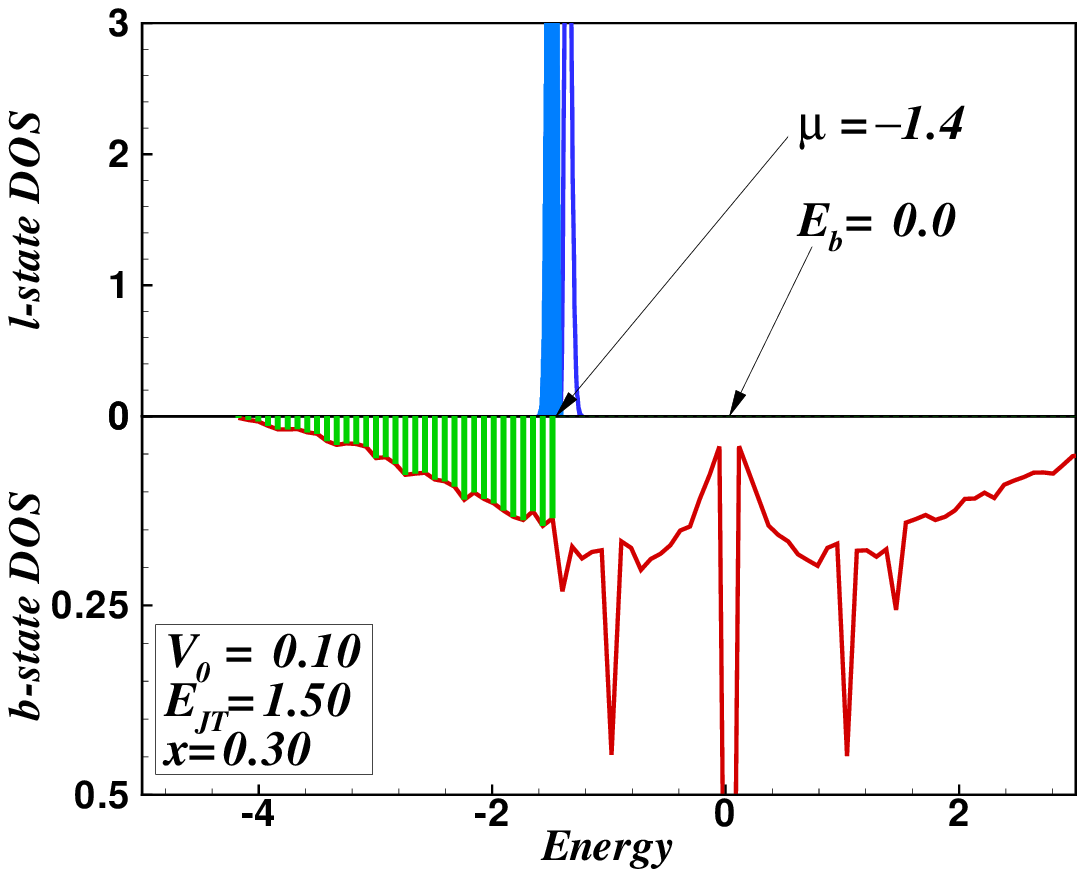}~~~~~\epsfxsize=\figwidth \epsfbox{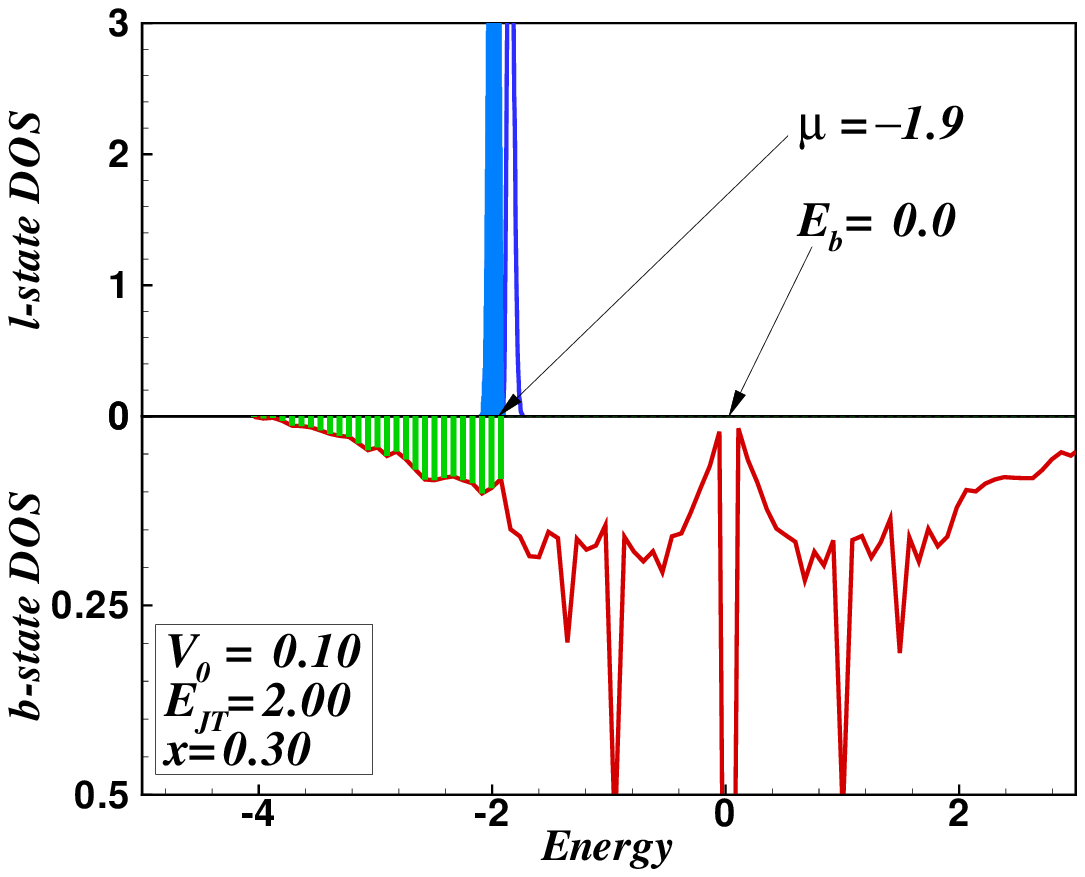}}
\centerline{(a)\hspace{9.0truecm}(b)}
\centerline{\epsfxsize=\figwidth \epsfbox{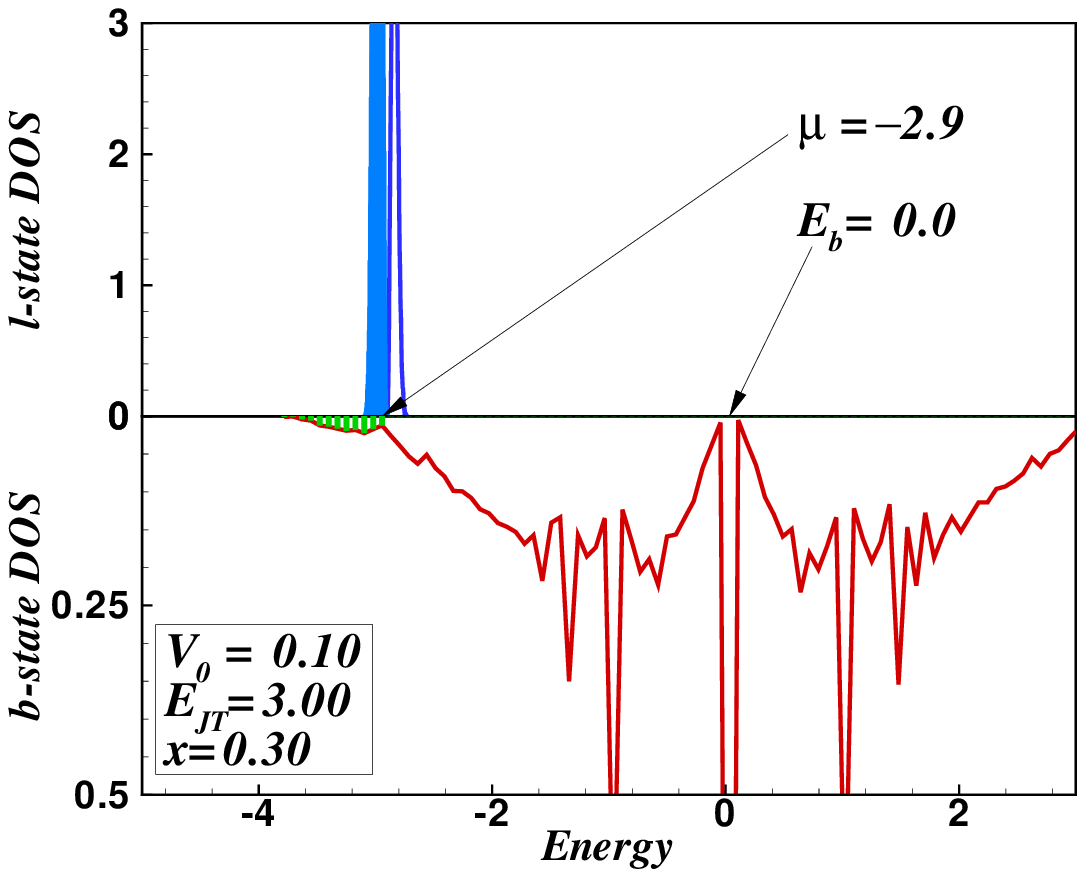}~~~~~\epsfxsize=\figwidth \epsfbox{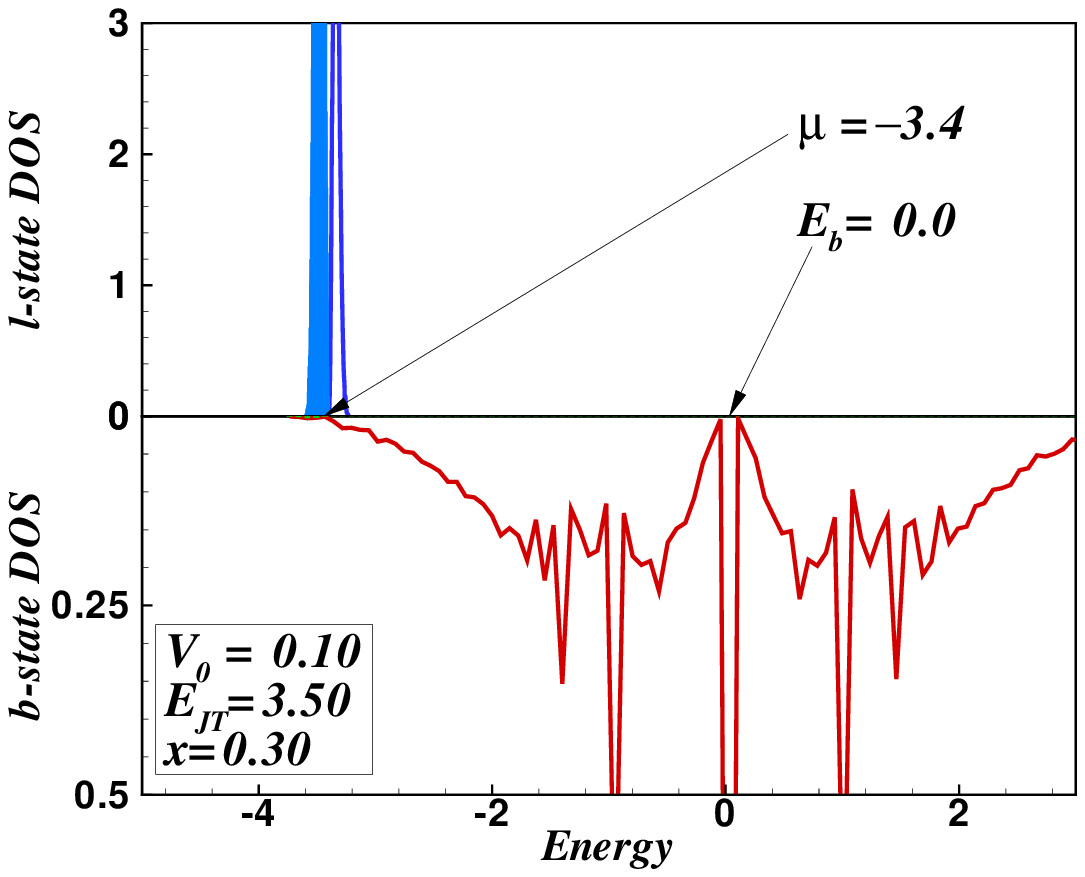}}
\centerline{(c)\hspace{9.0truecm}(d)}
\caption{(color online) Density of states of (DOS) $\ell$ polarons and $b$
electrons, where their respective energies are defined by \prn{hole}
and \prn{band}. The occupied states are shaded and the chemical
potential is marked by $\mu$. The center of the $b$-band $E_b$ is also indicated. The different panels (a)--(d) show the variation of the densities of states as a function of $\Ejt$ for $V_0=0.1$ and $x = 0.1$. The main effect of $\Ejt$ is to determine the chemical potential. $\Ejt$ does not affect width of the polaron density of states. The bandwidth of the $b$ band is not strongly affected by $\Ejt$, while the occupancy is strongly affected.}
\label{dosEjt}
\end{figure*}

\section{Results}\label{Results}

This section contains the results of our study of the
extended $\ell b$ Hamiltonian \prn{Hext}. The key parameters in the
Hamiltonian are energies $t,\Ejt,V_0$, and the doping $x$. The hopping
amplitude $t$ is taken as the basic energy scale, and $\Ejt$ and $V_0$
henceforth stand for the dimensionless values of the Jahn-Teller
energy and the long range Coulomb interaction strength parameter
normalized by $t$. Further all length scales are normalised by the
lattice parameter $a$.  We have checked for the size dependence of the
results and found that results for cubes larger than $8 \times 8
\times 8$ are essentially independent of the size. We checked
densities of states, clump size distribution etc., calculated in some

cases up to cubes of size $20\times20\times20$. We found that results
followed essentially  the same trends, independent of size. All the results
shown here are averages from calculations obtained with one
hundred random initial conditions using $10\times10\times10$ cells
unless stated otherwise.

\begin{figure*}
\centerline{\epsfxsize=\figtwowidth \epsfbox{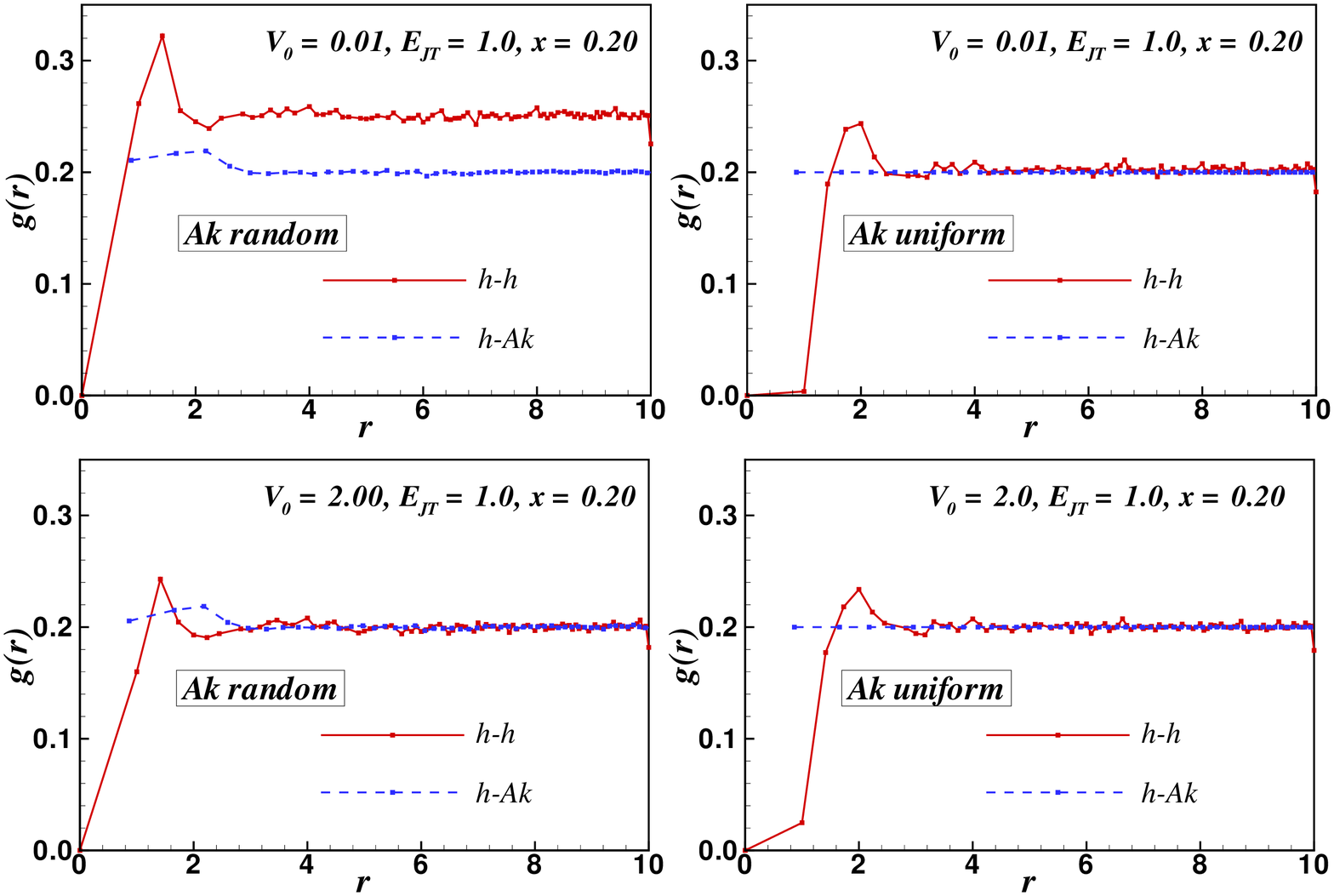}}
\caption{(color online) Position correlation function for $h - h$ and
$h-Ak$. The figures in the left column are for random distribution of
Ak ions, while those in the right column are for uniform distribution
of Ak ions (uniform distribution means that the total charge of the Ak
ions is distributed equally among the $Ak$ sites). The result is from
simulations with a $20 \times 20 \times 20$ cube (for a single realization of the random distribution of the Ak ions in the left-side graphs).}
\label{gofr}
\end{figure*}

The density of states obtained from simulations is discussed
first. The energies of the ``single particle states'' are defined in
\prn{hole} for a hole (with a similar expression for an $\ell$
polaron) and \prn{band} for the band electron. Figures \ref{dosV},
\ref{dosx} and \ref{dosEjt} show plots of densities of states for
various values of $V_0,\Ejt$ and $x$, with the chemical potential $\mu$ and
$b$ band center $E_b$ indicated in each case. \Fig{dosV} shows the variation
of the density of states with the parameter $V_0$, \Fig{dosEjt} shows
the variation in the density of states with the parameter $\Ejt$
($V_0$ and $x$ fixed) and \fig{dosx} shows the effect of doping on the
density of sates for fixed values of $V_0$ and $\Ejt$. From a study of
these figures we observe the following: The chemical potential is given by
\bea
\mu = -\Ejt + A(x) V_0 \label{chempot}
\eea
where $A(x)$ is a ``very weak'' function of $x$, with size of order unity; the physics behind this
will be discussed later.

We now discuss the $\ell$ polaron
states:- As discussed in Sec.~\ref{extellb}, the polarons with long
range Coulomb interaction form a Coulomb glass\cite{Efros1975}.
A Coulomb glass possesses a ``soft gap'' at the chemical
potential $\mu$, in that the density of states scales as
\bea
\rho_\ell(\epsilon - \mu) \sim (\epsilon - \mu)^2.
\eea
as is seen in \Fig{dosV}. It is also evident that the width (energy
spread) of the two lobes in the polaron density of states separated by
the chemical potential increases with increasing $V_0$. Thus the
polaron energies do not have a single value $-\Ejt$, but have a
distribution whose width scales with $V_0$. The energy spread arises
out of fluctuations in the local electrostatic potential. This is
the basic reason why though the polarons form a narrow band, there are
no heavy fermion like specific heat effects.  The occupied polaron states (indicated by shaded
portion below the chemical potential in \Fig{dosx}) per unit volume
are proportional to $(1-x)$ for large values of $V_0 (\gtrsim
1.0)$. For lower values of the $V_0$, the occupied spectral weight is
less, owing to the fact that some electrons are promoted to the $b$
states. The effect of $\Ejt$ on the polaron density of states is shown
in \Fig{dosEjt}. Since $\Ejt$ determines the chemical potential
via \prn{chempot},  it does not affect the width of
the distribution, but only affects the $\ell$ occupancy (amount of spectral
weight in each lobe).

Next, we turn to  the $b$ electron density of
states. Several points may be noted. First, it is clear from
\Fig{dosV} that for $V_0 \lesssim 1.0$, the $b$ band is centered
at $E_b = B(x) V_0$ ($B(x)$ is of order unity). The $b$ band center is
independent of $\Ejt$ (see \Fig{dosEjt}).  Second, as is evident from
\Fig{dosx}, the bandwidth of the $b$ states increases with increasing
$x$ (in fact, as $\sim \sqrt{x}$ as we shall discuss below), and
again essentially independent of $V_0$ (\Fig{dosV}) as well as of $\Ejt$
(\Fig{dosEjt}). Finally,  the number of occupied $b$
states (shown by shaded region in the figures), decreases with
increasing $V_0$ and $\Ejt$.

\begin{figure}
\centerline{\epsfxsize=\figwidth \epsfbox{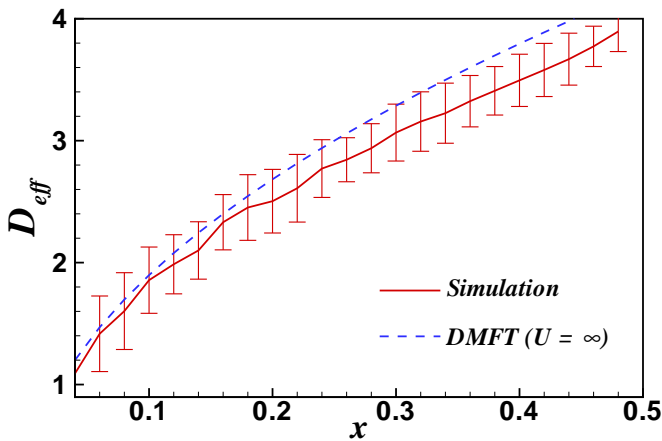}}
\caption{The effective half bandwidth $D_{eff}$ obtained from simulations compared
with the DMFT prediction \prn{Deff}. The half bandwidth in the
simulation is calculated by using the clump structure obtained by
minimizing the energy of the Coulomb glass. One hundred initial
configuration are averaged over to obtain the simulation curve with
the standard deviation bars indicated. }
\label{width}
\end{figure}

The physics underlying these observations may be understood by
studying the real space structure of the ground state, for example
by studying the positional correlation of the holes present in the system
(including the holes that appear due to promotion of the $\ell$ polarons to $b$
electrons as well as the ones already present due to doping). \Fig{gofr} shows
both the $h$-Ak and the $h-h$  correlation functions, which measure, respectively, the
probability of finding a hole at a relative distance $r$ from an Ak ion, or from another hole.
(Note that the Ak ions are placed randomly in the A sites of the
perovskite lattice.)  If the system were purely random, the
probability for any given value of $r$ will be equal to the doping
level $x$. When $V_0$ is small ($V_0 = 0.01$ in \Fig{gofr}), we see
the hole-hole correlation function reaches a plateau at a distance $r
\gtrsim 2$ with a value larger than $x$. This latter is because of the
increased number of holes in the system due to $\ell$ polaron to $b$
electron promotion. At larger $V_0$, the plateau in the hole-hole
correlation function appears at $x$. The hole-Ak ion correlation
function reaches a plateau of $x$, independent of $V_0$, as
expected. The most important feature of the correlation function
appears at $r \lesssim 2$. We see that the probability of finding a
hole in the neighboring shells is non-zero and is highest at $r =
\sqrt{3}$. Further, the probability of finding the $Ak$ ions near the
hole is also increased, though the increase is smaller than that of
finding a hole.  This suggests that there is a natural clustering
tendency in the $h-b$ glass where the holes tend to cluster around Ak
ions. Clearly, this is due to their opposite charge causing gain in
electrostatic energy. However, the competing, repulsive electrostatic
energy between the holes contributes to control this clustering
tendency. Since the distance between the Ak ions and
holes is smaller than that between two $\ell$ polarons, the clustering
of holes near Ak ions will  offset the energy penalty of $\ell$
polarons coming together -- the short (\AA ngstrom sized) clustering
scale is result of this competition. Strong screening effects
are evident in the correlation functions of \Fig{gofr} -- the
correlation functions reaches a plateau at about 2-3 lattice spacings.

\begin{figure*}
\centerline{\epsfxsize=\figwidth \epsfbox{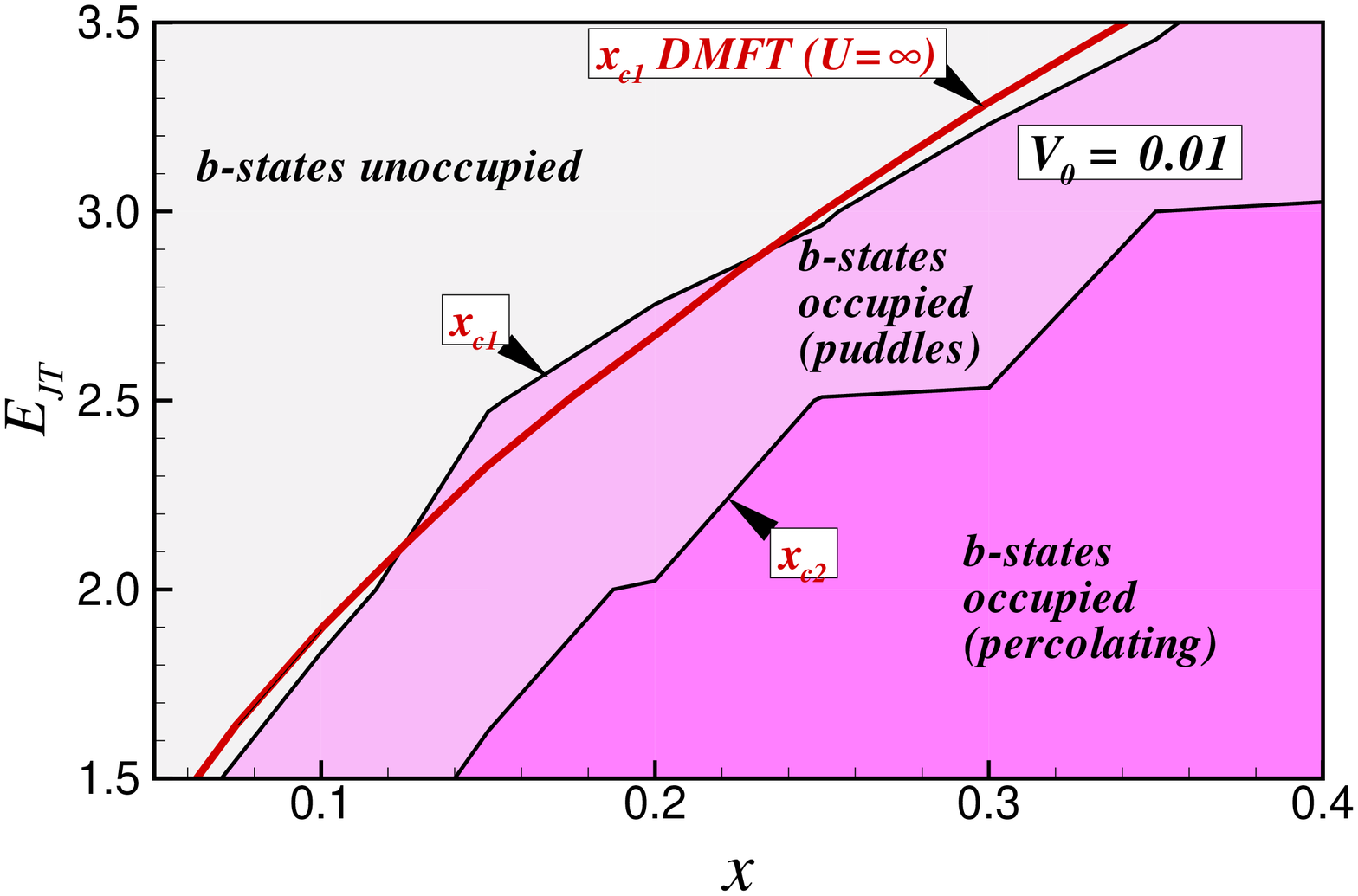}~~~~~\epsfxsize=\figwidth \epsfbox{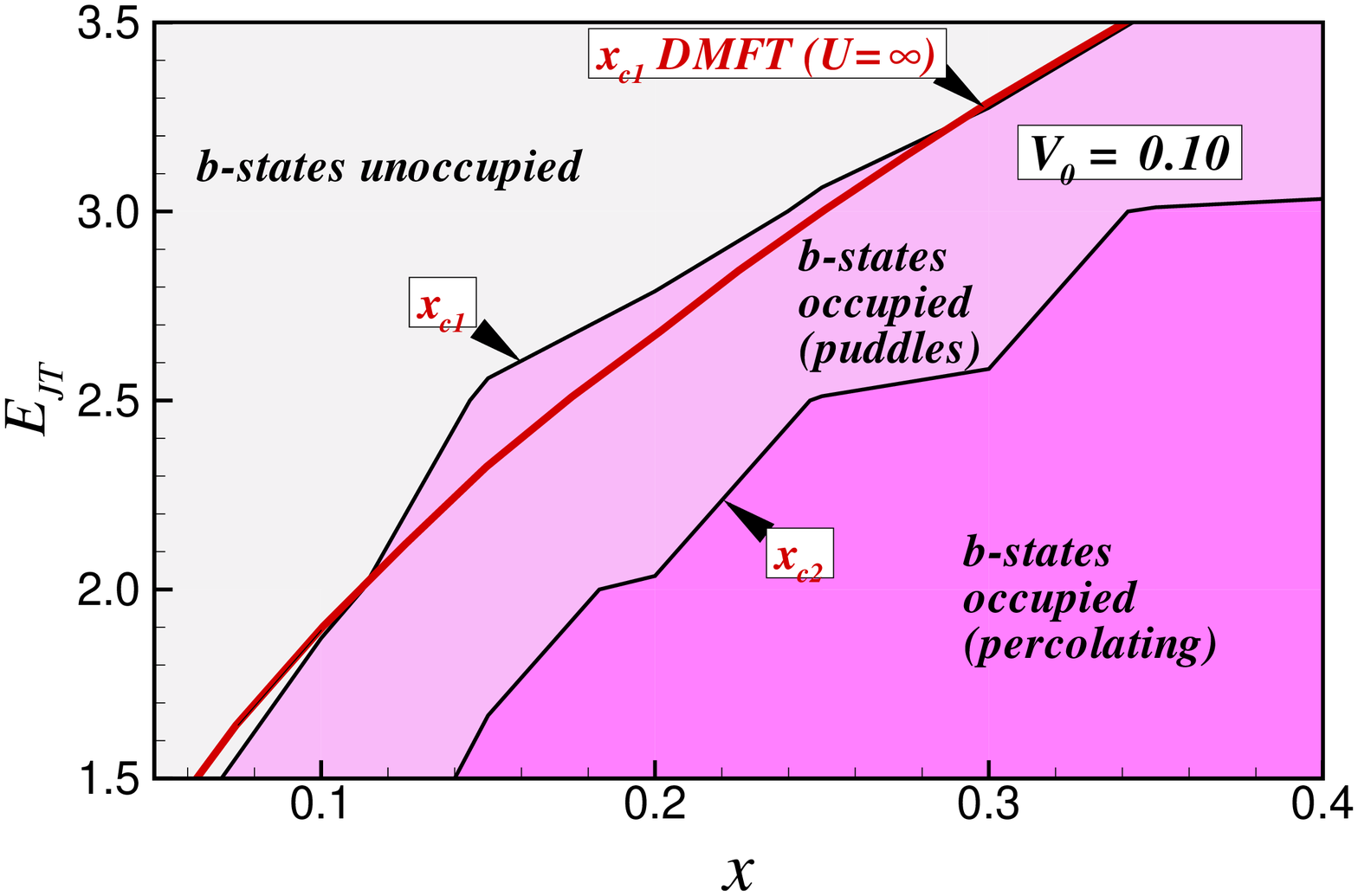}}
\centerline{(a)\hspace{9.0truecm}(b)}
\centerline{\epsfxsize=\figwidth \epsfbox{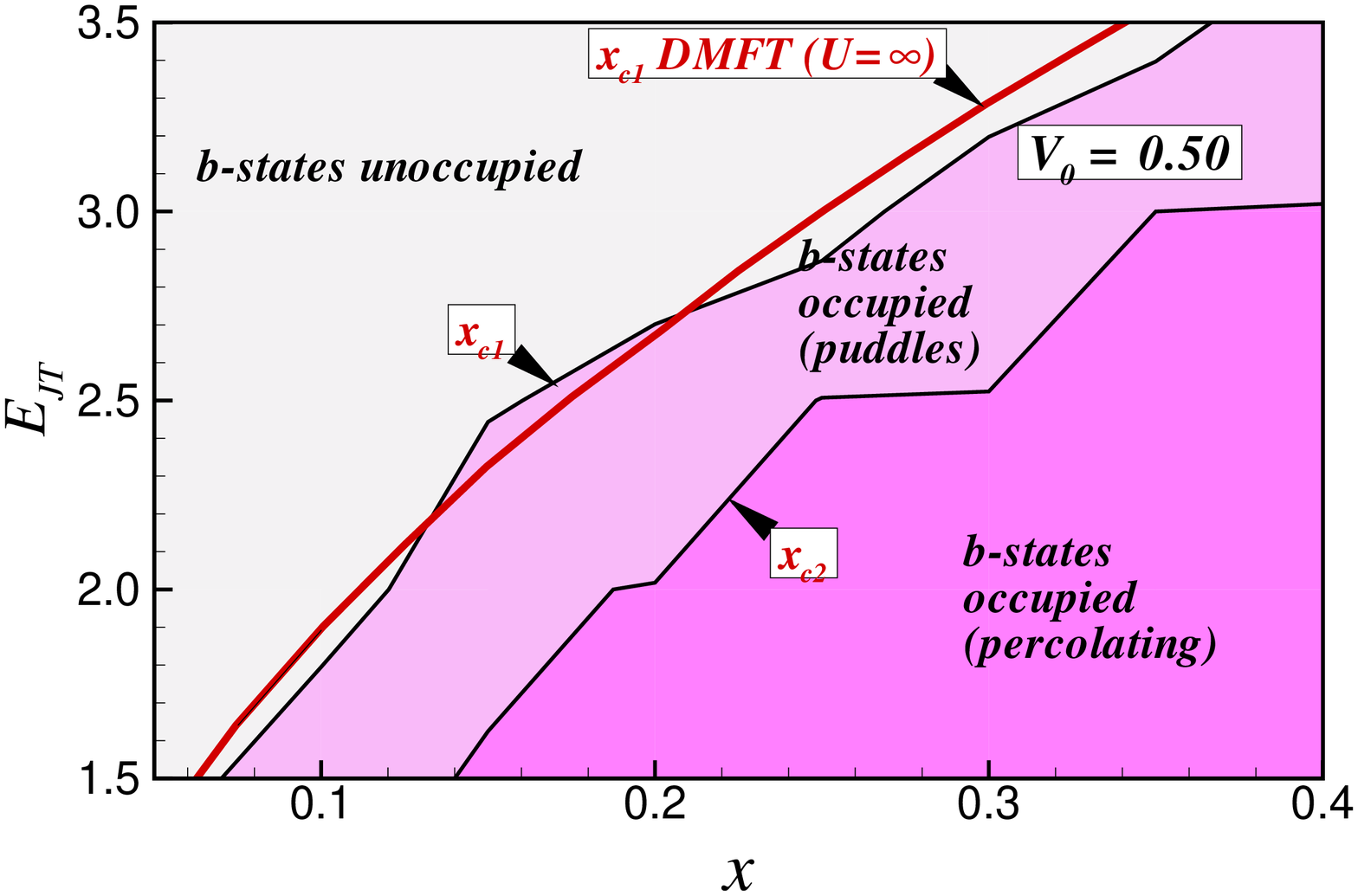}~~~~~\epsfxsize=\figwidth \epsfbox{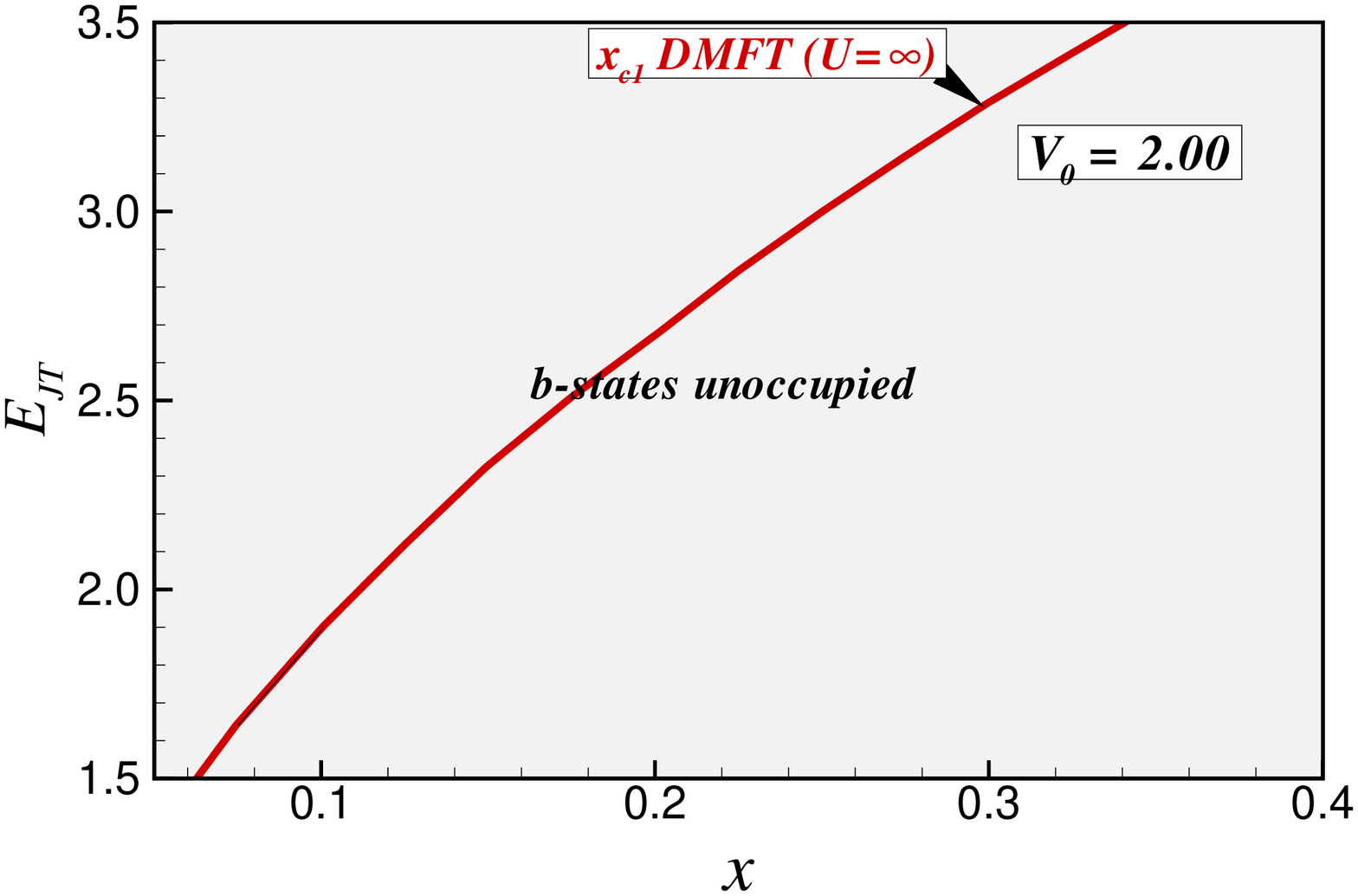}}
\centerline{(c)\hspace{9.0truecm}(d)}
\caption{(color online) Critical doping levels $x_{c1}$ and $x_{c2}$
obtained from simulations. The lightest region in the contour plot
contains no $b$ electrons, the intermediate shade has ``trapped'' $b$
states occupied, and the darkest regions corresponds to $b$ states
that percolate through the simulation box. The solid line corresponds
to the analytical DMFT result for $x_{c1}$.   }
\label{btype}
\end{figure*}

%% \begin{figure*}
%% \centerline{\epsfxsize=\figwidth \epsfbox{finrun_n010_V000.0100_sig.eps}~~~~~\epsfxsize=\figwidth \epsfbox{finrun_n010_V000.1000_sig.eps}}
%% \centerline{(a)\hspace{9.0truecm}(b)}
%% \centerline{\epsfxsize=\figwidth \epsfbox{finrun_n010_V000.5000_sig.eps}~~~~~\epsfxsize=\figwidth \epsfbox{finrun_n010_V002.0000_sig.eps}}
%% \centerline{(c)\hspace{9.0truecm}(d)}
%% \caption{(color online) DC conductivity $\sigma(\omega \approx 0)$ as a function of system parameters. For long long range Coulomb parameter $V_0$ greater than 0.4, the conductivity is zero for the indicated range of $\Ejt$ and $x$.}
%% \label{sig}
%% \end{figure*}

\begin{figure*}
\centerline{\epsfxsize=\figwidth \epsfbox{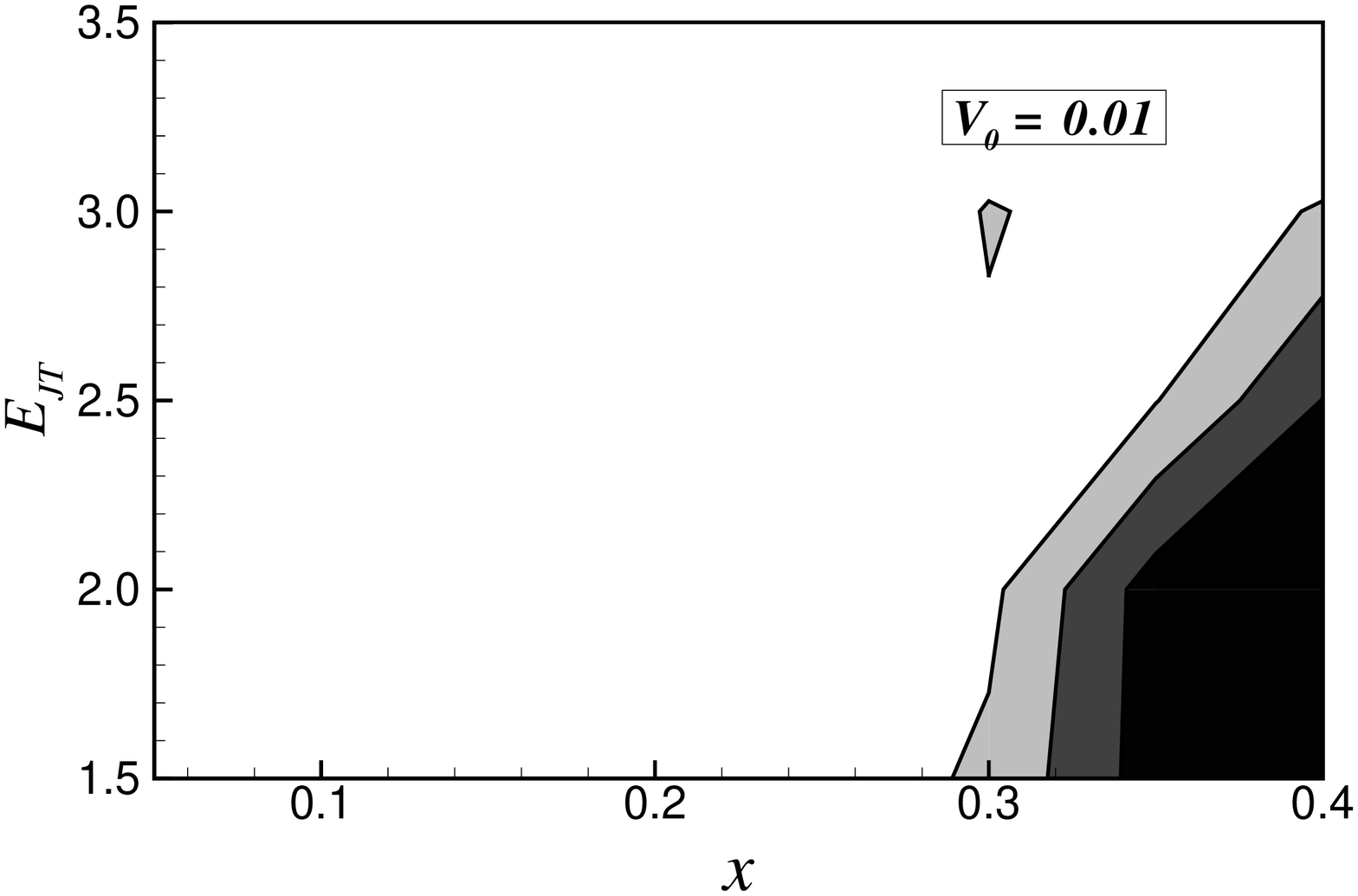}}
\centerline{(a)}
\centerline{\epsfxsize=\figwidth \epsfbox{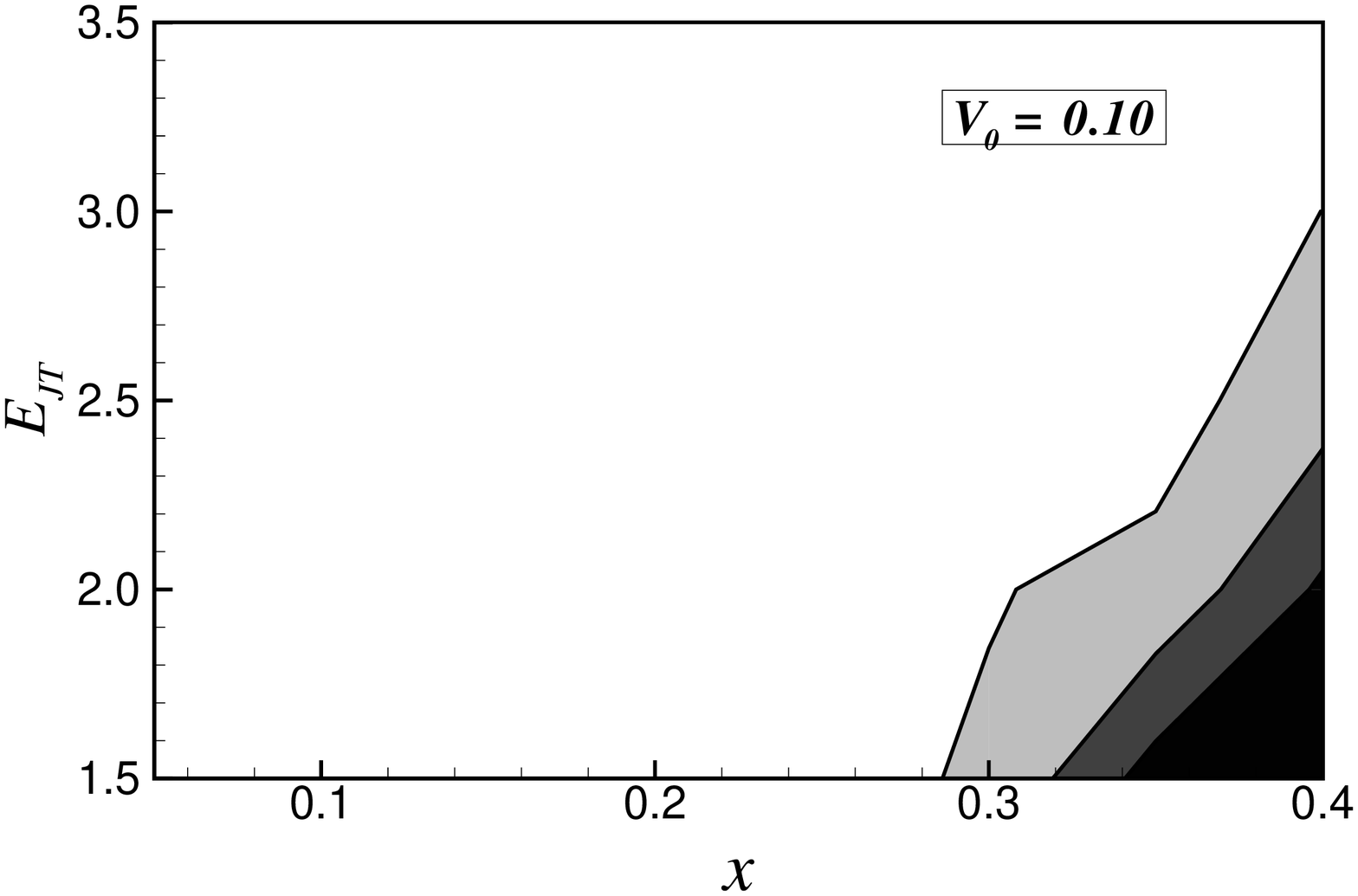}}
\centerline{(b)}
\centerline{\epsfxsize=\figwidth \epsfbox{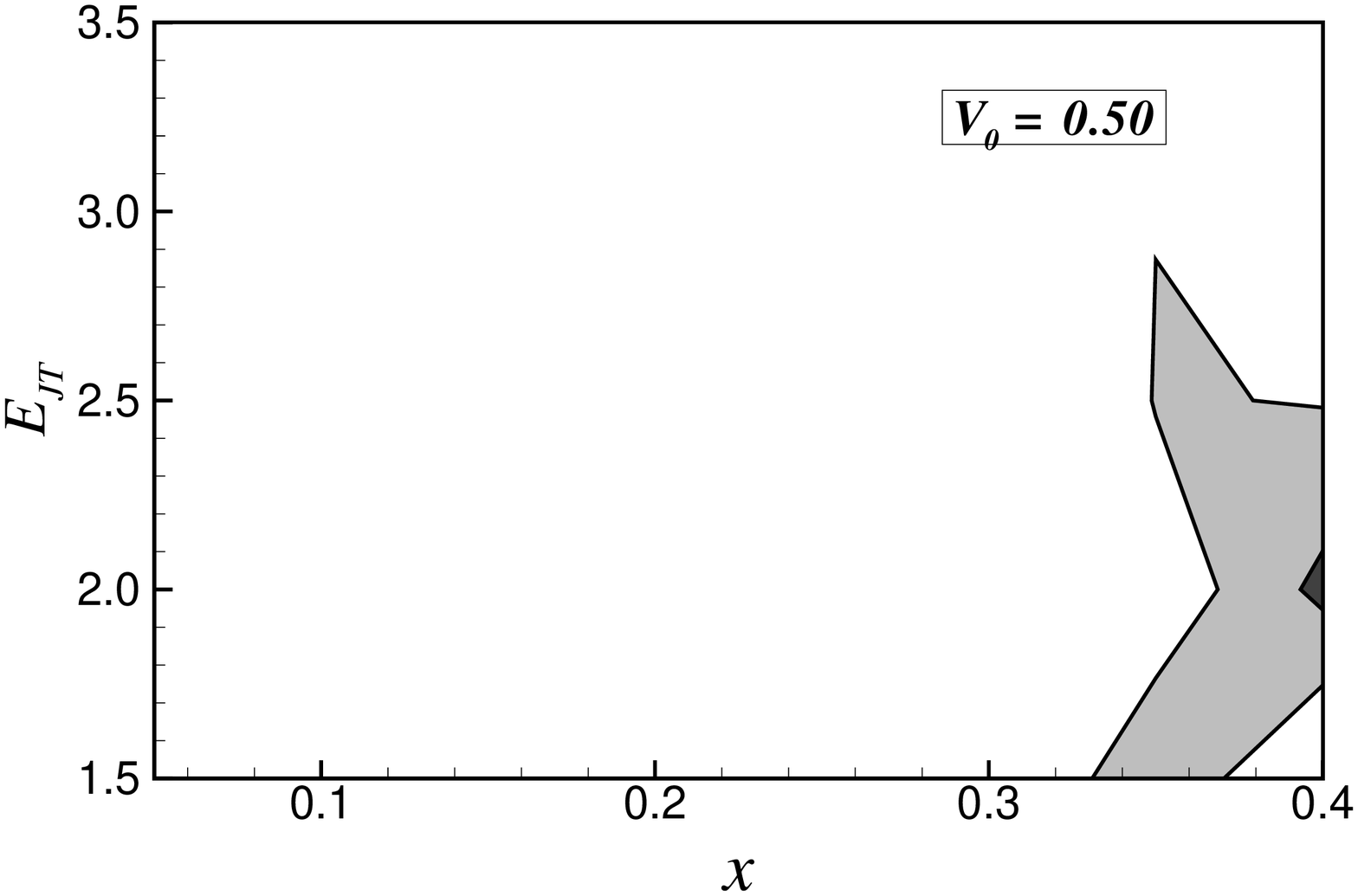}}
\centerline{(c)}
\caption{(color online) DC conductivity $\sigma(\omega \approx 0)$ as a function of system parameters. For long long range Coulomb parameter $V_0$ greater than 0.4, the conductivity is zero for the indicated range of $\Ejt$ and $x$. The darkest shade of the contour corresponds to a dimensionless conductivity $\sigma > 10$, the next darkest $5 > \sigma \ge 10$, the lightest $0 > \sigma > 5$. }
\label{sig}
\end{figure*}

The observations above allow us to obtain the chemical potential of
the system.  The hole clustering caused by the hole-Ak ion interaction
causes an additional effective electrostatic potential at an $\ell$
electron site which can be written as $A(x) V_0$ where $A(x)$ is a
(here undetermined) doping dependent factor of order unity. The
quantity $A(x) V_0$ can be interpreted as the average electrostatic
potential at $\ell$ polaron sites. The actual energies at the sites
fluctuates about the mean total energies of the $\ell$ polarons given
by $-\Ejt + A(x) V_0$. From the Coulomb glass problem it is known that
the chemical potential will be equal to the average
energy\cite{Efros1975,Efros1976} and \prn{chempot} follows.

\begin{figure*}
\centerline{\epsfxsize=\figwidth \epsfbox{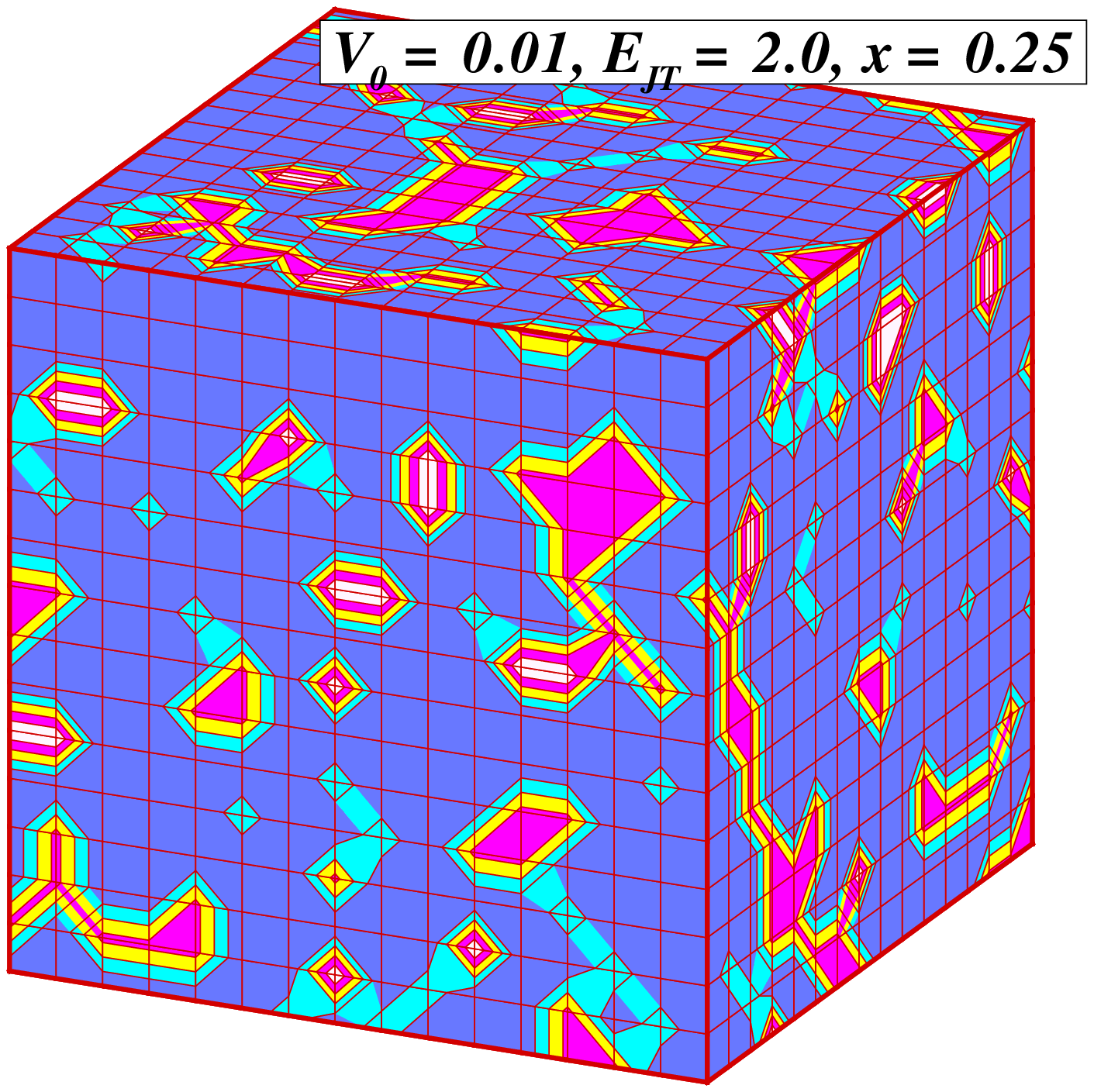}\epsfxsize=\figwidth \epsfbox{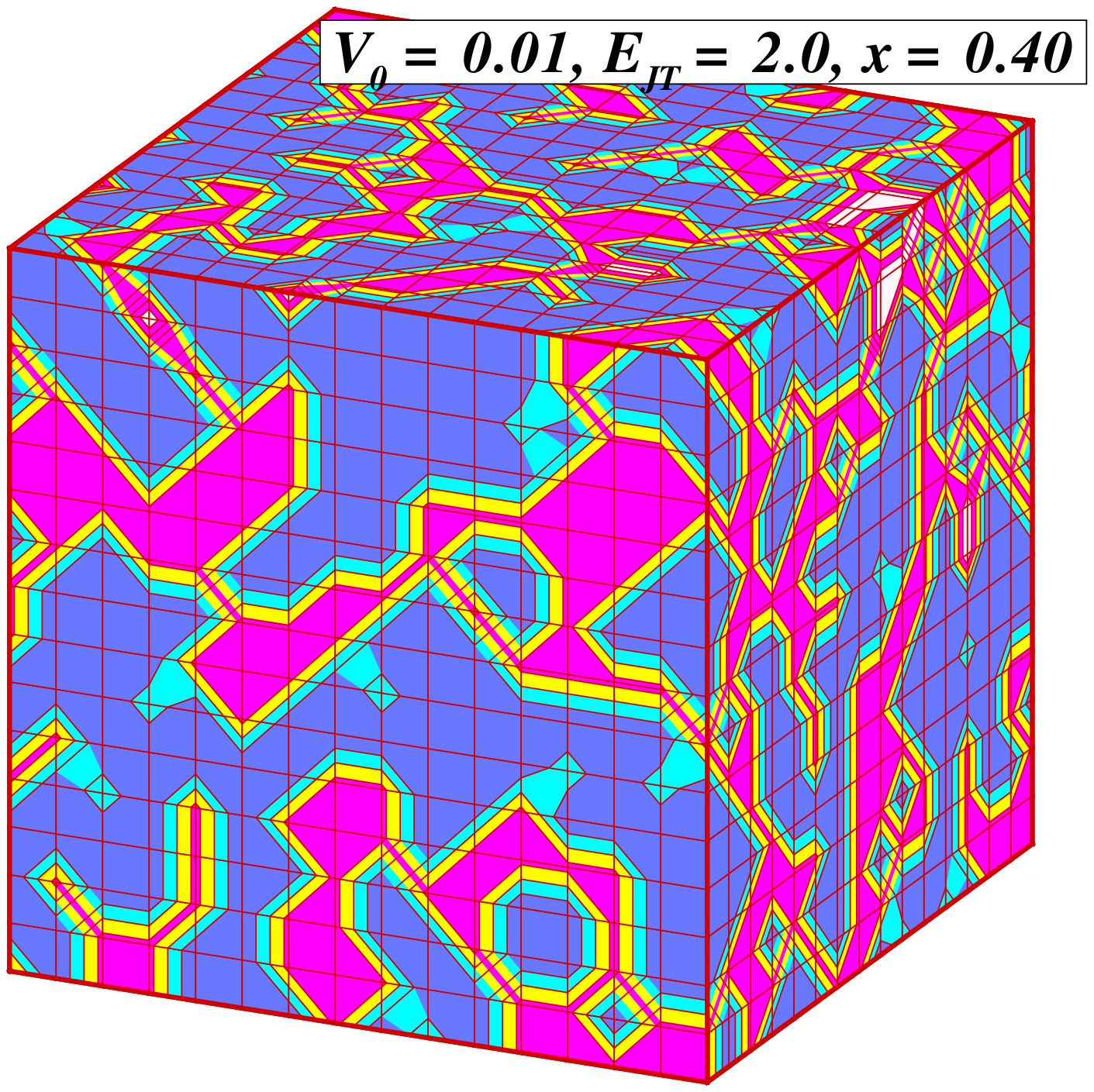}}
\centerline{(a)\hspace{9.0truecm}(b)}
\centerline{\epsfxsize=\figwidth \epsfbox{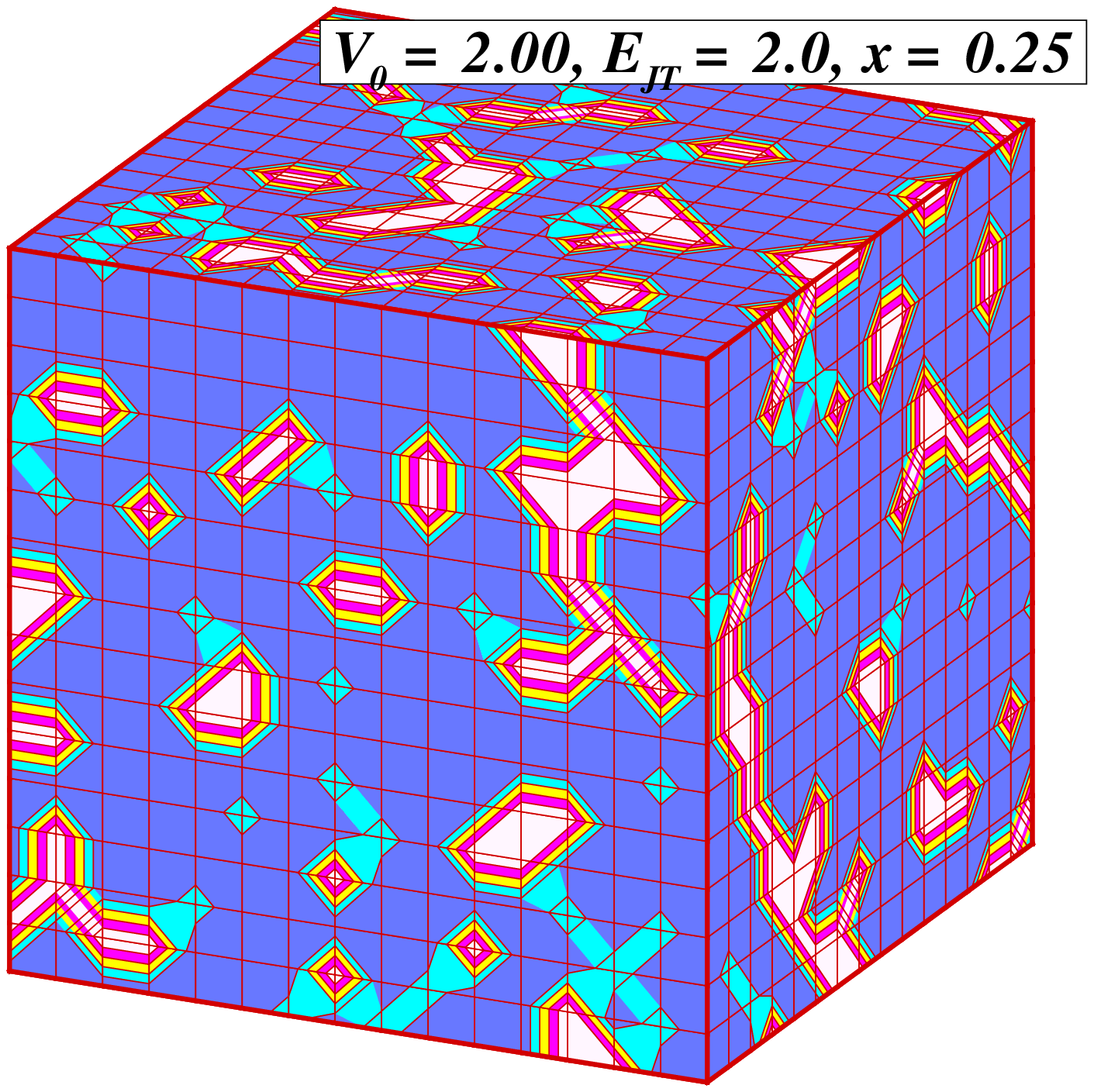}\epsfxsize=\figwidth \epsfbox{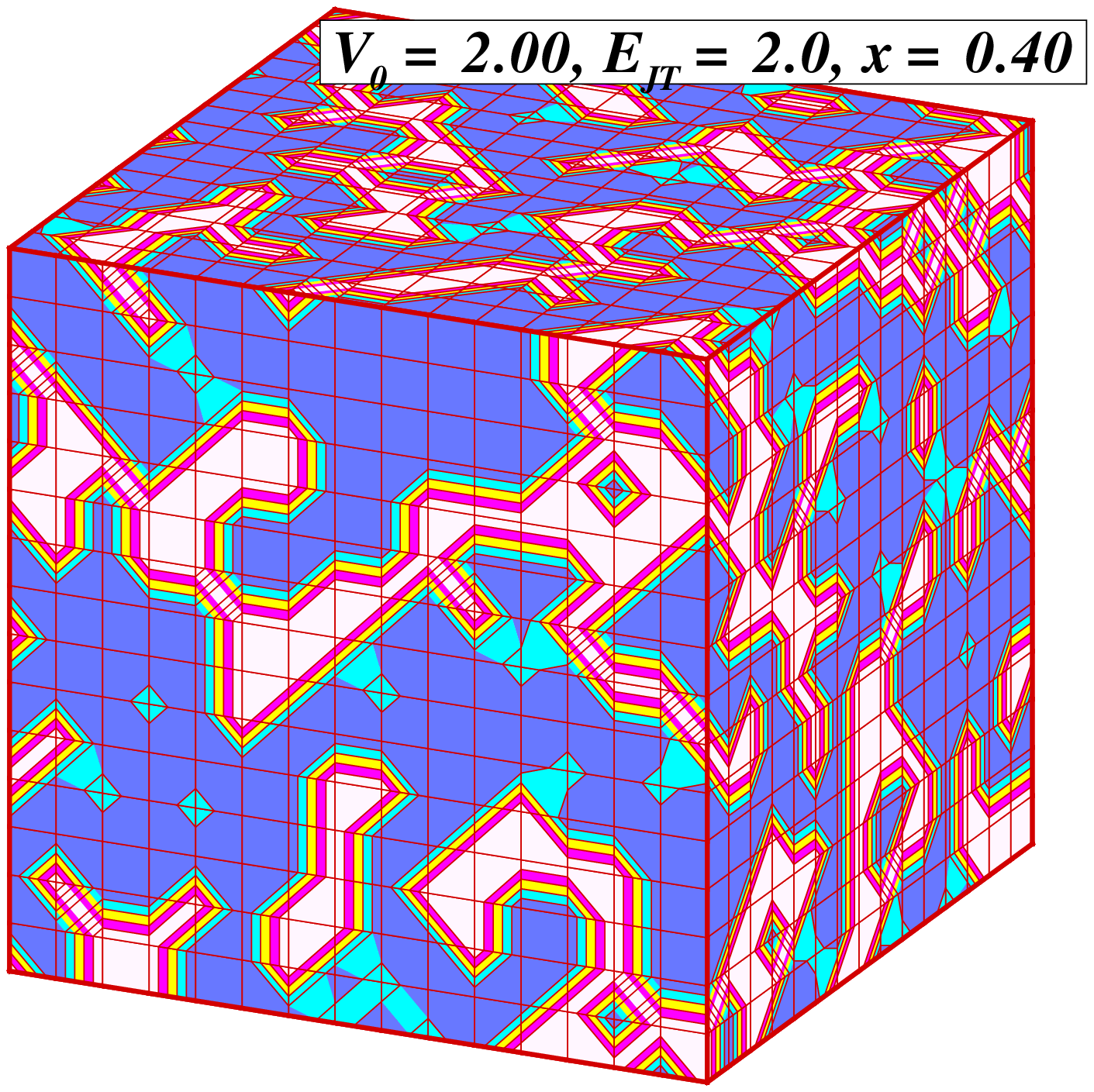}}
\centerline{(c)\hspace{9.0truecm}(d)}
\caption{(color online) Real space structure of the electronic
state. The darkest regions (magenta in online version) denote hole
clumps with occupied $b$ electrons, the lightest (white) denote hole
clumps with no $b$ electrons, the second lighter shade (cyan) denote
singleton holes, and the second darkest shade (light blue) represents
regions with $\ell$ polarons. The simulations for each $V_0$ are for the same realization of the random distribution of Ak ions. The cell size is $16\times16\times16$. }
\label{clump}
\end{figure*}

\begin{figure*}
\centerline{\epsfxsize=\figwidth \epsfbox{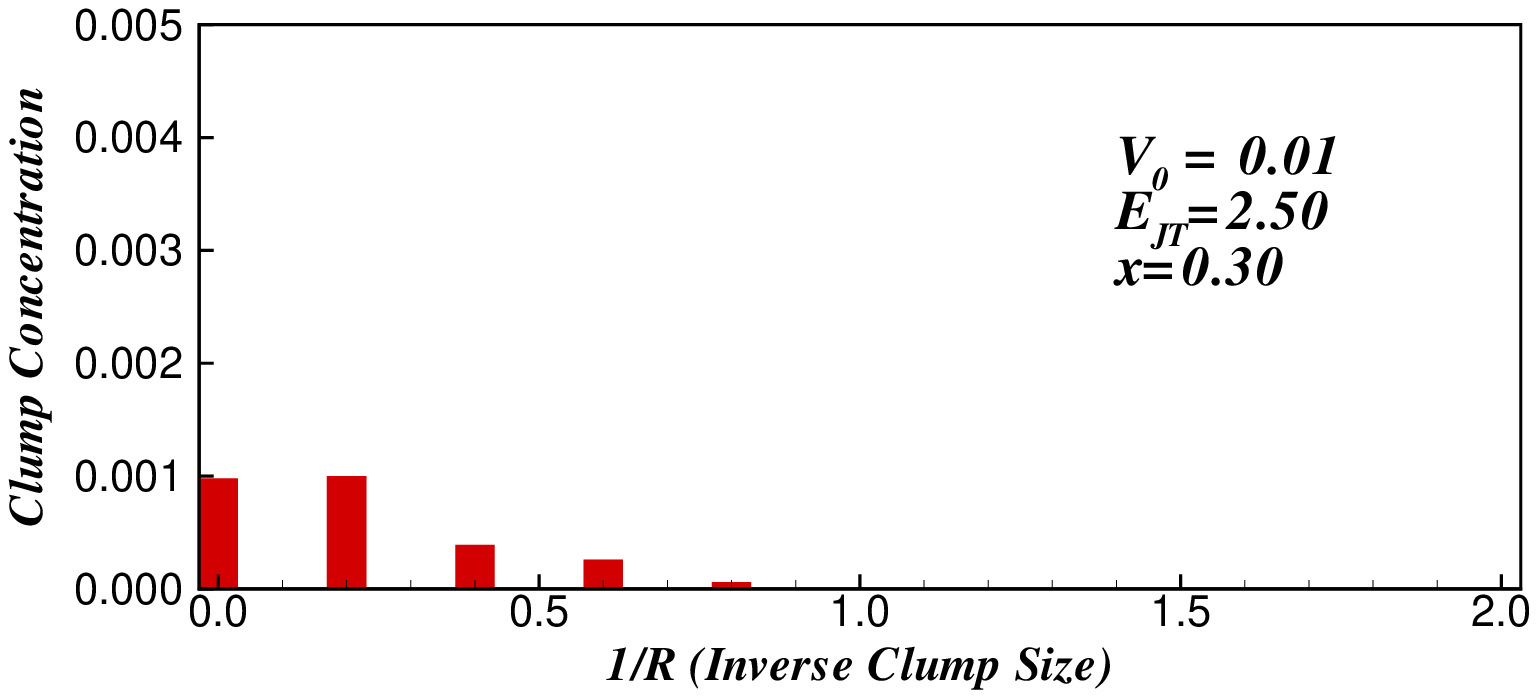}}
\centerline{(a)}
\centerline{\epsfxsize=\figwidth \epsfbox{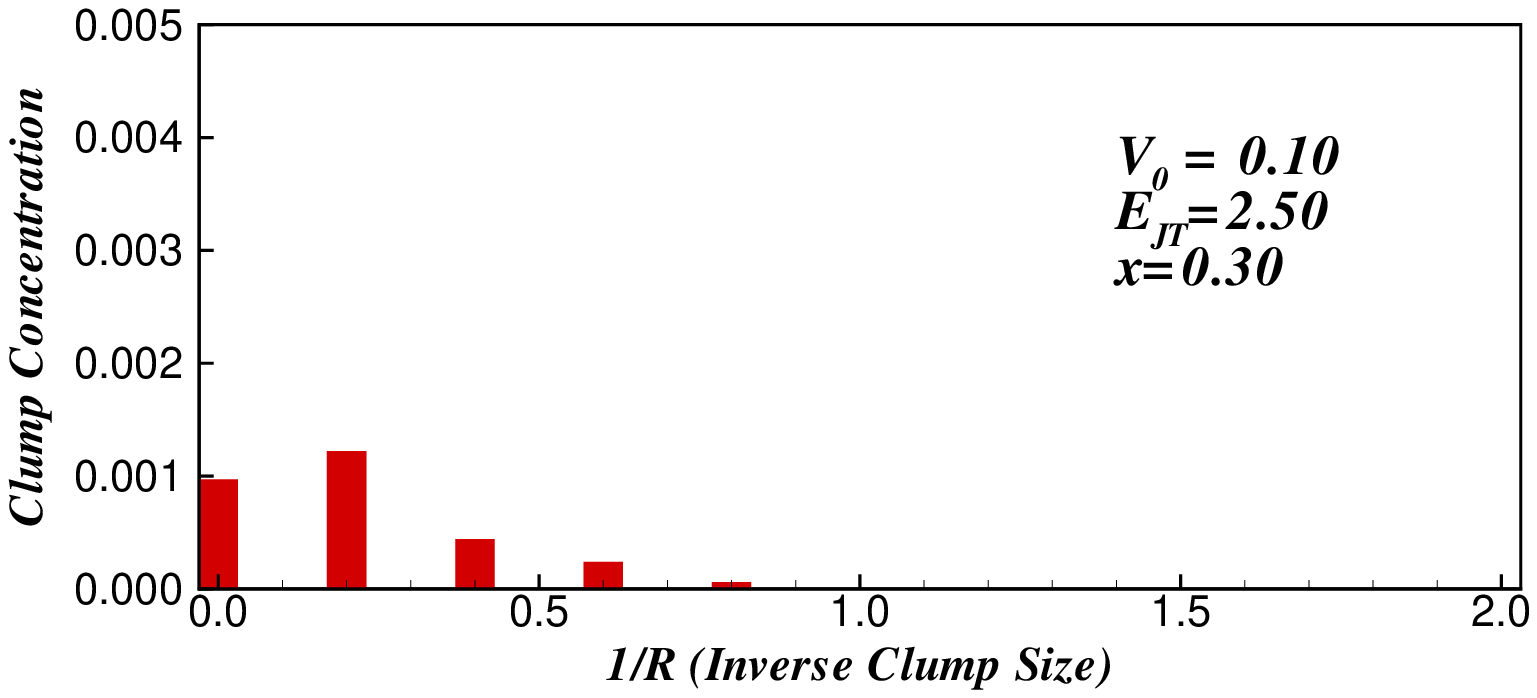}}
\centerline{(b)}
\centerline{\epsfxsize=\figwidth \epsfbox{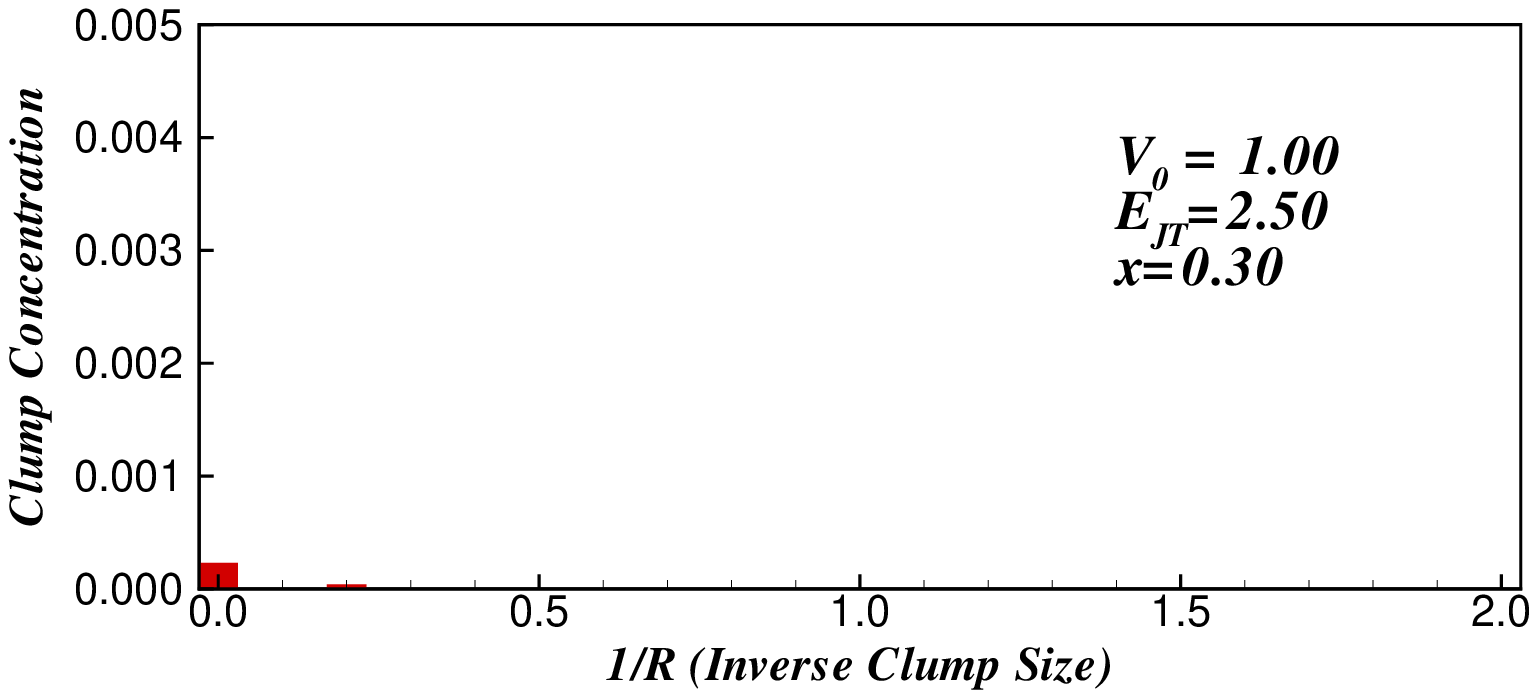}}
\centerline{(c)}
\caption{(color online) Size distribution of clumps with occupied $b$ electrons. Clump concentration is number of clumps per lattice site (simulated with a $10\times10\times10$ box). Size of the clump $R$ is the ``radius of gyration'' of the clump (see text), ordinate is inverse clump size. $1/R=0$ corresponds to a percolating clump, while $1/R = 2$ corresponds to the smallest clump with two sites. The plots show the effect of the Coulomb interaction $V_0$ on the distribution for $x=0.3$ and $\Ejt=2.5$. For $V_0 = 2.00$ there are no clumps with occupied $b$ electrons. }
\label{clzV}
\end{figure*}

\begin{figure*}
\centerline{\epsfxsize=\figwidth \epsfbox{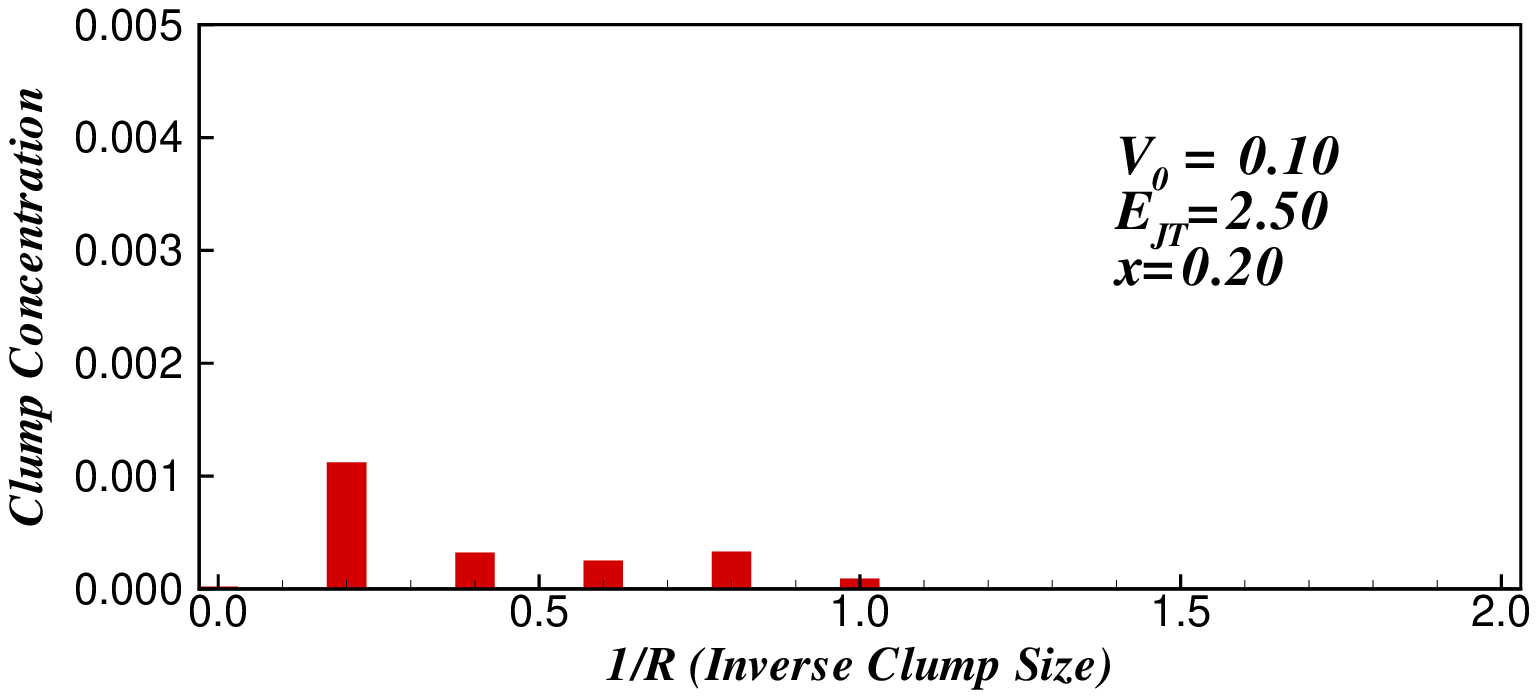}}
\centerline{(a)}
\centerline{\epsfxsize=\figwidth \epsfbox{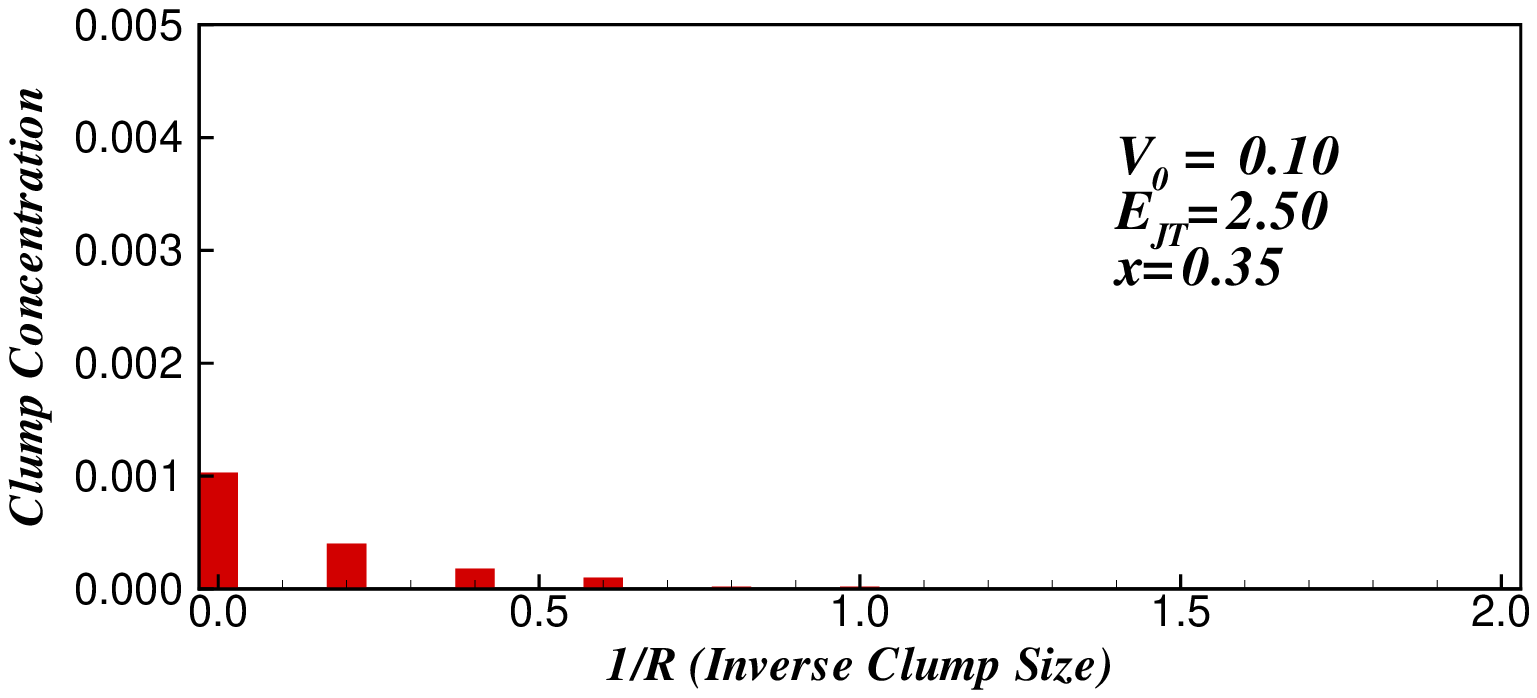}}
\centerline{(b)}
\centerline{\epsfxsize=\figwidth \epsfbox{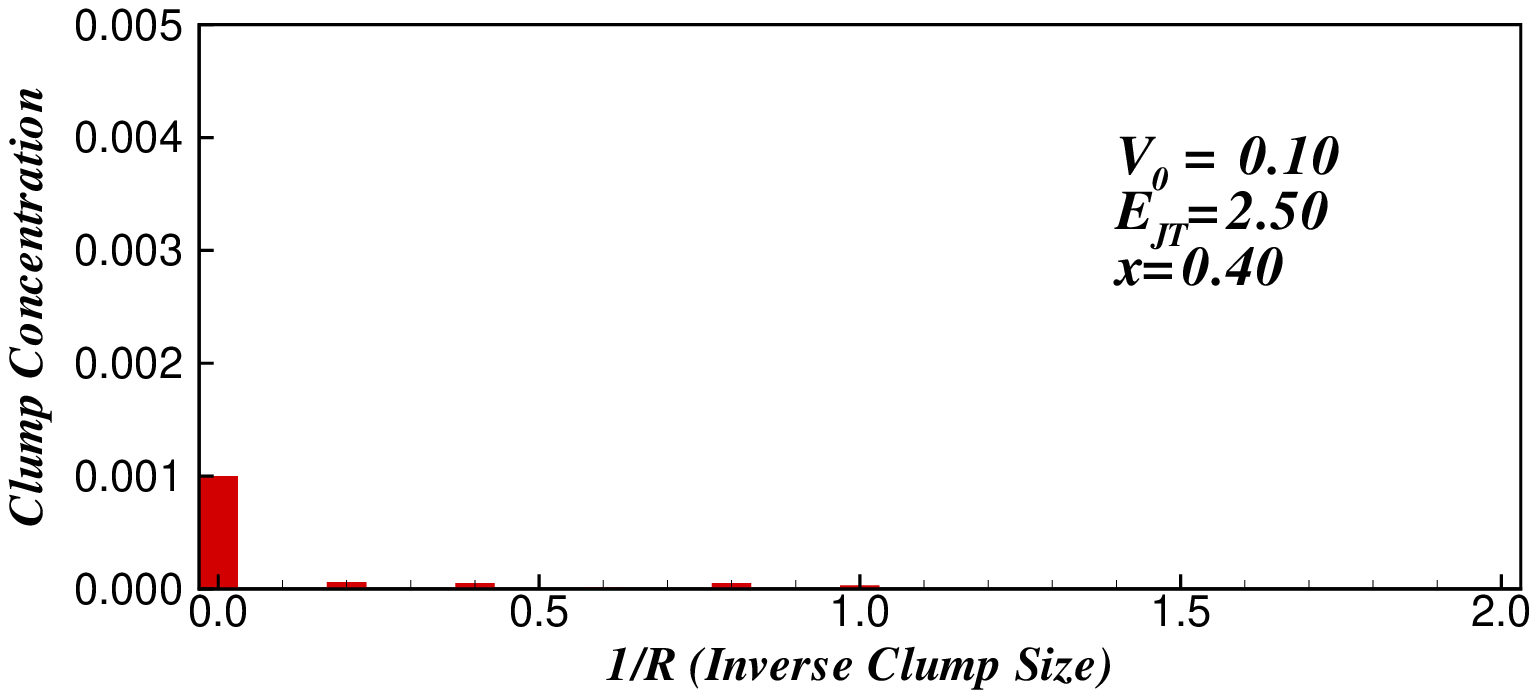}}
\centerline{(c)}
\caption{ (color online) Size distribution of clumps with occupied $b$
electrons. Clump concentration is number of clumps per lattice site
(simulated with a $10\times10\times10$ box). Size of the clump $R$ is the
``radius of gyration'' of the clump (see text), ordinate is inverse
clump size. $1/R=0$ corresponds to a percolating clump, while $1/R = 2$
corresponds to the smallest clump with two sites. The plots show the
effect of the doping $x$ on the distribution for
$V_0=0.1$ and $\Ejt=1.0$. For $x \ge 0.4$ there is only one percolating clump.
}
\label{clzx}
\end{figure*}

\begin{figure*}
\centerline{\epsfxsize=\figwidth \epsfbox{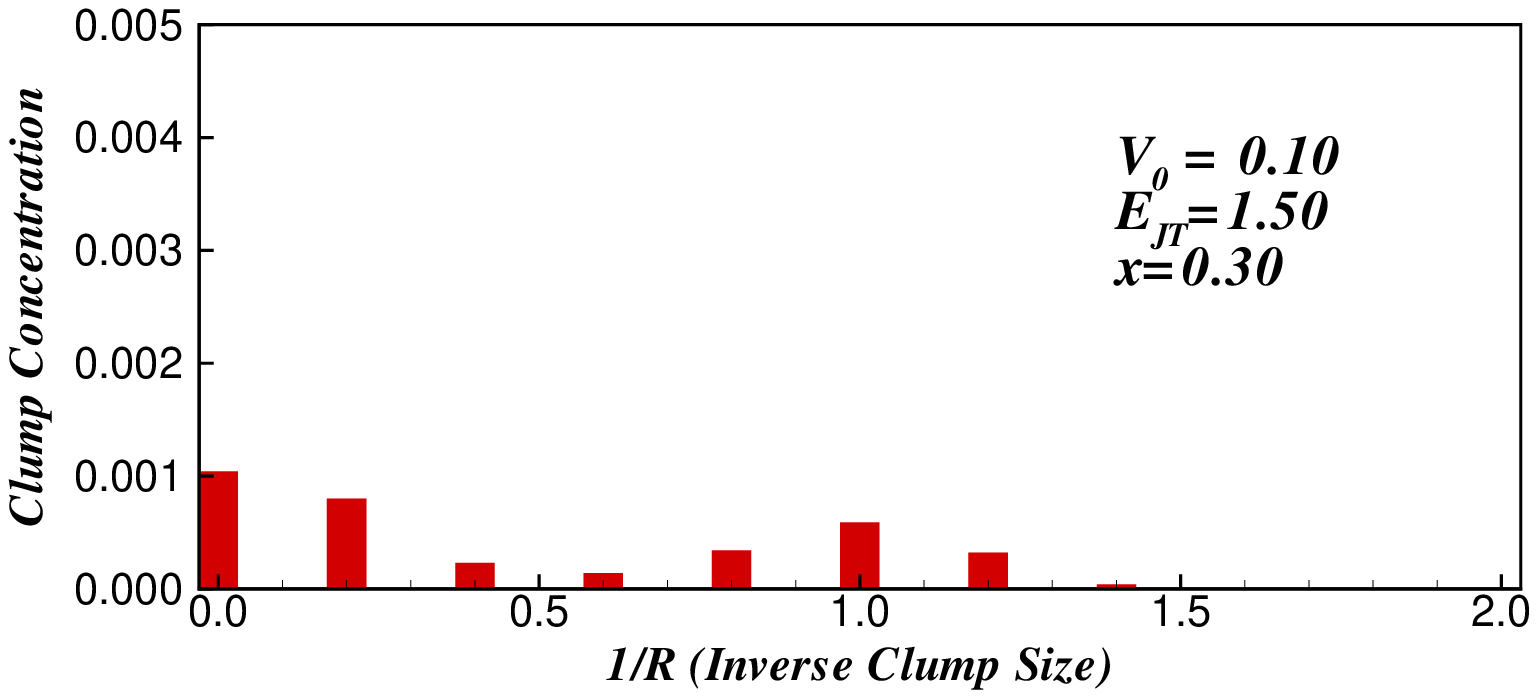}}
\centerline{(a)}
\centerline{\epsfxsize=\figwidth \epsfbox{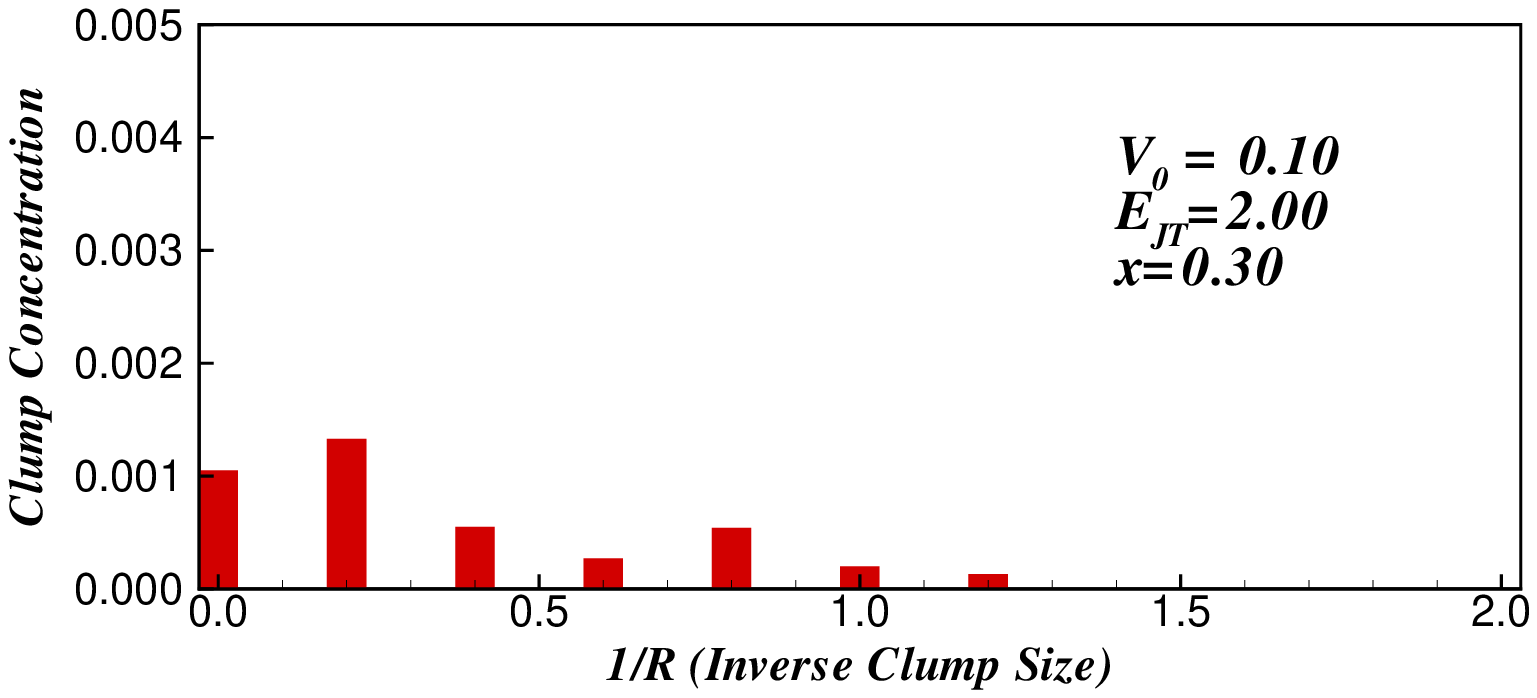}}
\centerline{(b)}
\centerline{\epsfxsize=\figwidth \epsfbox{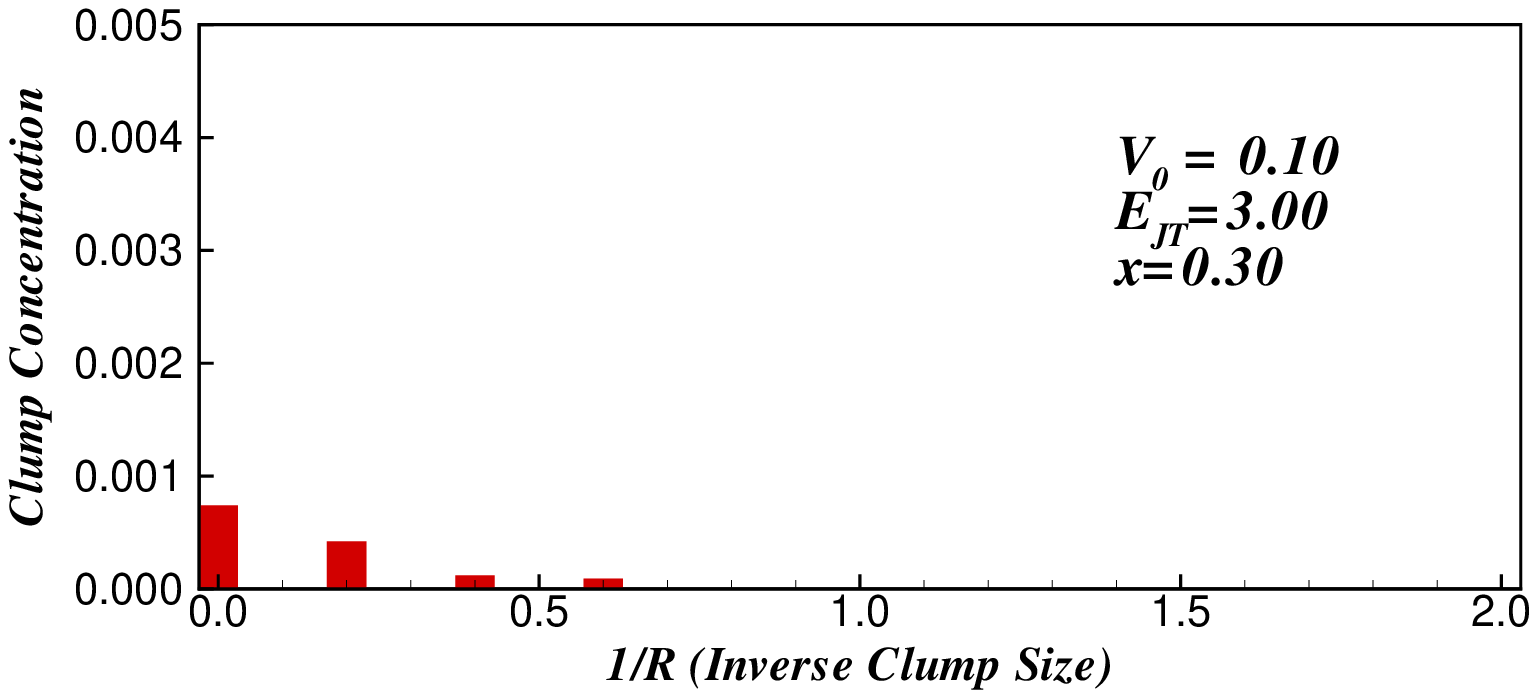}}
\centerline{(c)}
\caption{(color online) Size distribution of clumps with occupied $b$
electrons. Clump concentration is number of clumps per lattice site
(simulated with a $10\times10\times10$ box). Size of the clump $R$ is the
``radius of gyration'' of the clump (see text), ordinate is inverse
clump size. $1/R=0$ corresponds to a percolating clump, while $1/R = 2$
corresponds to the smallest clump with two sites. The plots show the
effect of $\Ejt$ on the distribution for
$V_0=0.1$ and $x=0.3$. For $\Ejt = 3.5$ there are no occupied clumps.}
\label{clzEjt}
\end{figure*}

We now discuss the energetics and the nature of the $b$ states, since
the transport properties depend crucially on the nature of the $b$
states and whether they are occupied. In case they are not occupied or
if all the occupied $b$ states are localized, the system is an insulator; on the other hand
if extended $b$ states are occupied one has a metal. We summarize our
results in this regard, and compare them where ever possible with those
obtained in single site DMFT.\cite{Pai2003,Ramakrishnan2003,Ramakrishnan2004}

The hole site has an average electrostatic potential energy $B(x) V_0$
where $B(x) \approx 1$; this follows from arguments very similar to
those used earlier for the electrostatic (Hartree) energy of the
$\ell$-polaron sites. Since the $b$ electrons occupy clumps consisting
of hole sites, they too sense this average repulsive potential $B(x)
V_0$ and thus the band center $E_b$ shifts to this value (as is seen
in the simulation, see FIGS.\ref{dosV}, \ref{dosx}, \ref{dosEjt}). The
simulations indicate that the effective bandwidth $D_{eff}$ of the $b$ states depends only
on the doping $x$. To understand this result, we calculated $D_{eff}$
as a function of $x$ using clumps obtained from a Coulomb glass calculation
of the holes including the random distribution of the $Ak$ ions
(thus the calculation was performed with only $\ell-h$
excitations). The result of this calculation is shown in \fig{width}
along the the DMFT result for the bandwidth. It is evident that the
agreement is excellent for a wide range of $x$.  Note that the half
bandwidth is provided by the clump structure determined by long ranged
Coulomb interactions and the {\em random distribution} of the Ak ions
which induce the clustering tendency of holes around them (as
discussed above). Thus the result for the effective bandwidth
\prn{Deff} seems to arise out of two key factors: (a) The large $U$
limit which disallows simultaneous $b$ and $\ell$ occupancy, and (b)
The clumps of holes induced by the random distribution of the Ak ions.
This is interesting since it implies that the single site DMFT which
neglects long range Coulomb interactions and consequent nanoscale
inhomogeneities or clumping, and replaces it with a self consistent
annealed random medium in which all sites are equivalent, is accurate
for such purposes. We also note here that in the DMFT, the $\ell$ polaron
is a single level while in the simulations the occupied levels, though
localized, have a distribution of energies.

We now discuss the doping/hole density $x_{c1}$ at which $b$ states
are occupied. In the DMFT, for $U= \infty$ and at $T=0$, the critical
concentration $x_c$ at which the lowest energy $b$ state is occupied
is $x_{c1} = \displaystyle{\left(\frac{\Ejt}{D_0}\right)^2}$ where
$D_0$ is the bare half bandwidth\cite{Pai2003}. From our simulations,
the $\ell$ chemical potential $\mu$ (i.~e., the energy of the last
occupied $\ell$ state) is $-\Ejt + A(x) V_0$. At the onset of the
occupation of the $b$ states, equilibrium demands that the bottom of
the $b$ band has to just equal the chemical potential $\mu$. Noting
that the effective half bandwidth of the $b$ band is $D_0 \sqrt{x}$
and the band center $E_b = B(x) V_0$, one obtains on equating the $b$
band bottom to the chemical potential, a result for the critical
doping
\bea
\sqrt{x_{c1}} = \frac{\Ejt}{D_0} + \left(A(x_c) - B(x_c)
\right) \frac{V_0}{D_0}.
\eea
Since $A(x) \approx B(x) \approx 1$, and $V_0/D_0 \le 0.05$
(typically) $x_{c1}$ is expected to be very close to the DMFT
value. In \fig{btype}, we notice that as expected, for small/realistic
values of $V_0$, there is good agreement with DMFT and the simulation
results. For larger values of $V_0$ (of the order of $D_0$), $b$ state
occupancy needs much larger hole density $x$ than implied in the
DMFT. Indeed, there is critical value of $V_0 (V_0^c \approx 1.2)$
above which there is no band occupancy for $\Ejt \ge 1$; the system
becomes a Coulomb glass of $\ell$ polarons.

Furthermore, unlike in the DMFT, in the present context the fact that $b$ states
are occupied does not in itself make the system a
metal. Even when $x > x_{c1}$, the hole sites where $b$ states exist form compact clumps
in our simulations. The $b$ levels within the bigger clumps which are below
the chemical potential and are occupied are hence still localized.
\Fig{clump} shows the `clumps' with both the
occupied and unoccupied $b$ state regions indicated. We notice that
the clumps, and hence the $b$ puddles are generally isolated.
Hence the system is still an insulator. The system becomes
metallic for $x > x_{c2}$, when the $b$ {\em puddles} percolate, which
requires as a necessary condition that the clumps with the $b$ states
occupied percolate. $ x_{c2}$ can be estimated by calculating the inverse
participation ratio\cite{Thouless1974} and by checking for the
geometric percolation of the clump. The differences among $x_{c1}$
(DMFT), $x_{c1}$ (simulations), $x_{c2}$ (simulations) or in the
nature of the $b$ states and their occupancy, are due to the fact that
the single site DMFT of the $\ell b$ model \prn{Hellb} and the
numerical simulation of the extended $\ell b$ model \prn{Hext} are
very different approximations. In the former, the static $\ell$
polarons are repulsive (potential $U$) scattering centers for $b$
electrons, present randomly and independently at each site with a self
consistently determined probability close to $(1-x)$. The scattering
is treated in CPA (coherent potential approximation), so that there is
no Anderson localization of the $b$ electron states; there are diffusive
and extended. By contrast, in the simulation, the $b$ electrons
are completely excluded from $\ell$ polaron sites, and can hop freely from
site to site only among contiguous hole sites (clumps). So unless the clumps are
are connected percolatively,  the system is an insulator. Furthermore, even when
the clumps percolate, if the $\ell$ polaron
distribution and the $b$ electron propagation in such an $\ell$ polaron
medium could be treated realistically, one will find that in contrast
to single site DMFT, a fraction of the occupied $b$ states are Anderson localized
due to the $\ell$ polaron disorder and the repulsion $U n_{\ell i}
n_{bi}$. Hence $x_{c2}$ is larger than the value of $x$ at which
the clumps just percolate, as the highest occupied $b$ state could still
be Anderson localized within the percolating clump; hence $x_{c2}$ signifies the
value of $x$ at which the {\em mobility edge} in the $b$ band crosses the chemical
potential.

We have confirmed this by calculating the contribution of the $b$ states to the DC
conductivity using the standard Kubo formalism\cite{Mahan2000} -- the
results, in the form of contours of constant dc conductivity in the $\Ejt$, $x$ plane,
are shown in \Fig{sig}. Again, it is clear that the
metal-insulator boundary (the $x_{c2}$) curve is independent of the
long range Coulomb parameter $V_0$ when $V_0$ is small. However, the
DC conductivity completely vanishes for larger values of $V_0$ greater
than $V^m_0$ of about 0.6 and the system is an insulator (due to both
strong local disorder potential from the Ak ions and Coulomb repulsion). Not
unexpectedly, this critical value $V^m_0$ is much smaller than the
value $V^c_0$ that forbids $b$ state occupancy.

\begin{figure*}
\centerline{\epsfxsize=\figwidth \epsfbox{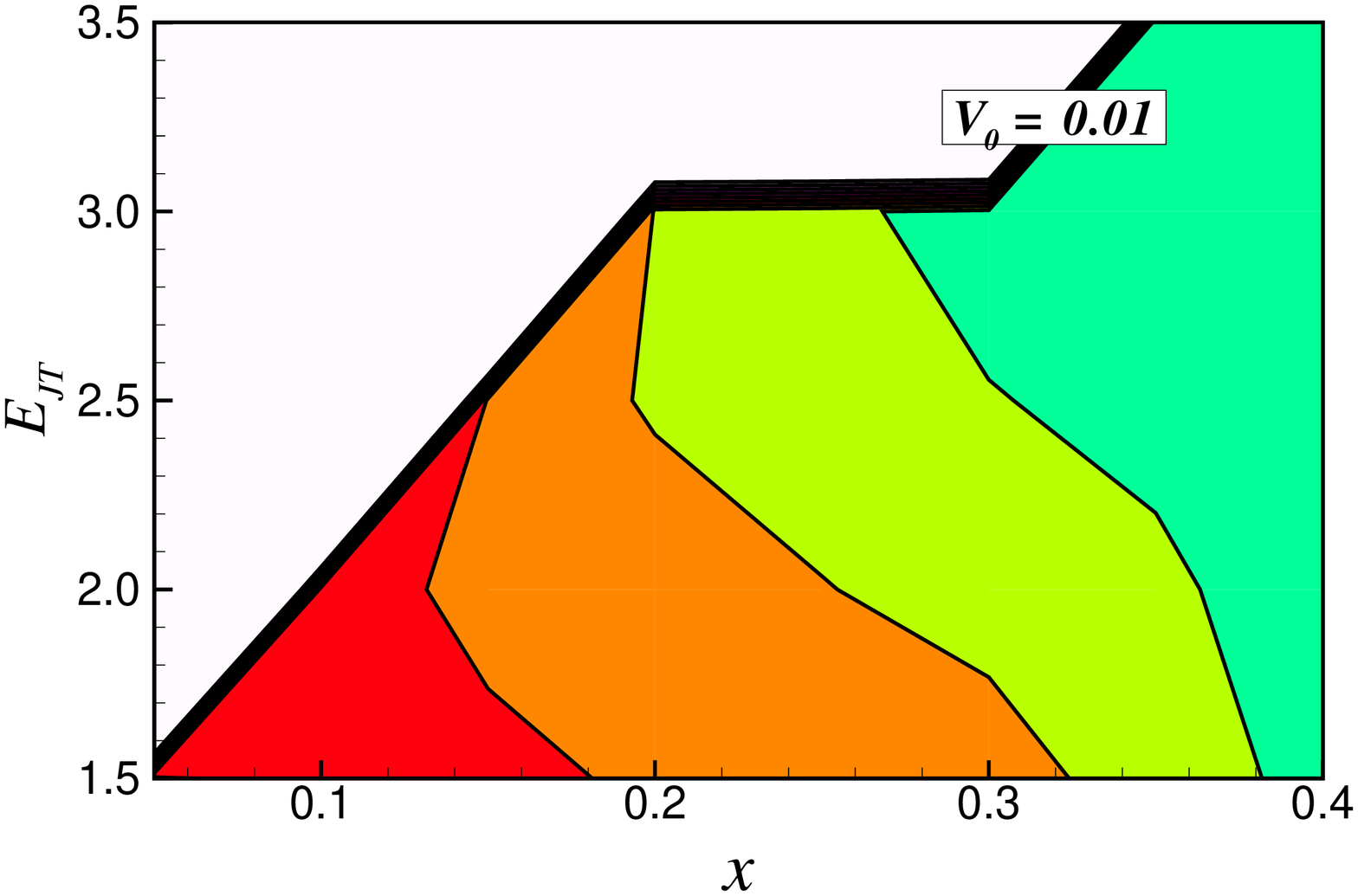}}
\centerline{(a)}
\centerline{\epsfxsize=\figwidth \epsfbox{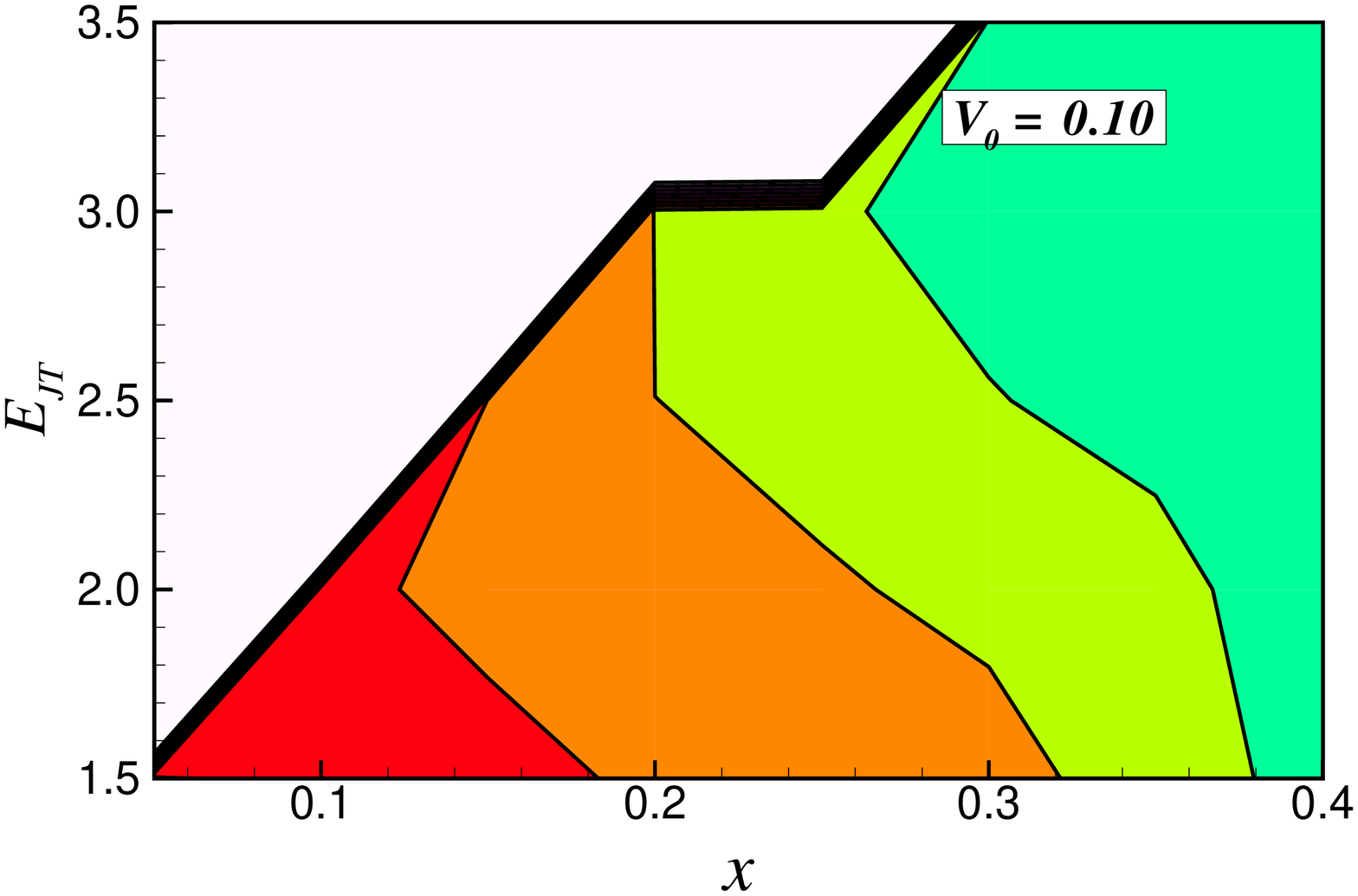}}
\centerline{(b)}
\centerline{\epsfxsize=\figwidth \epsfbox{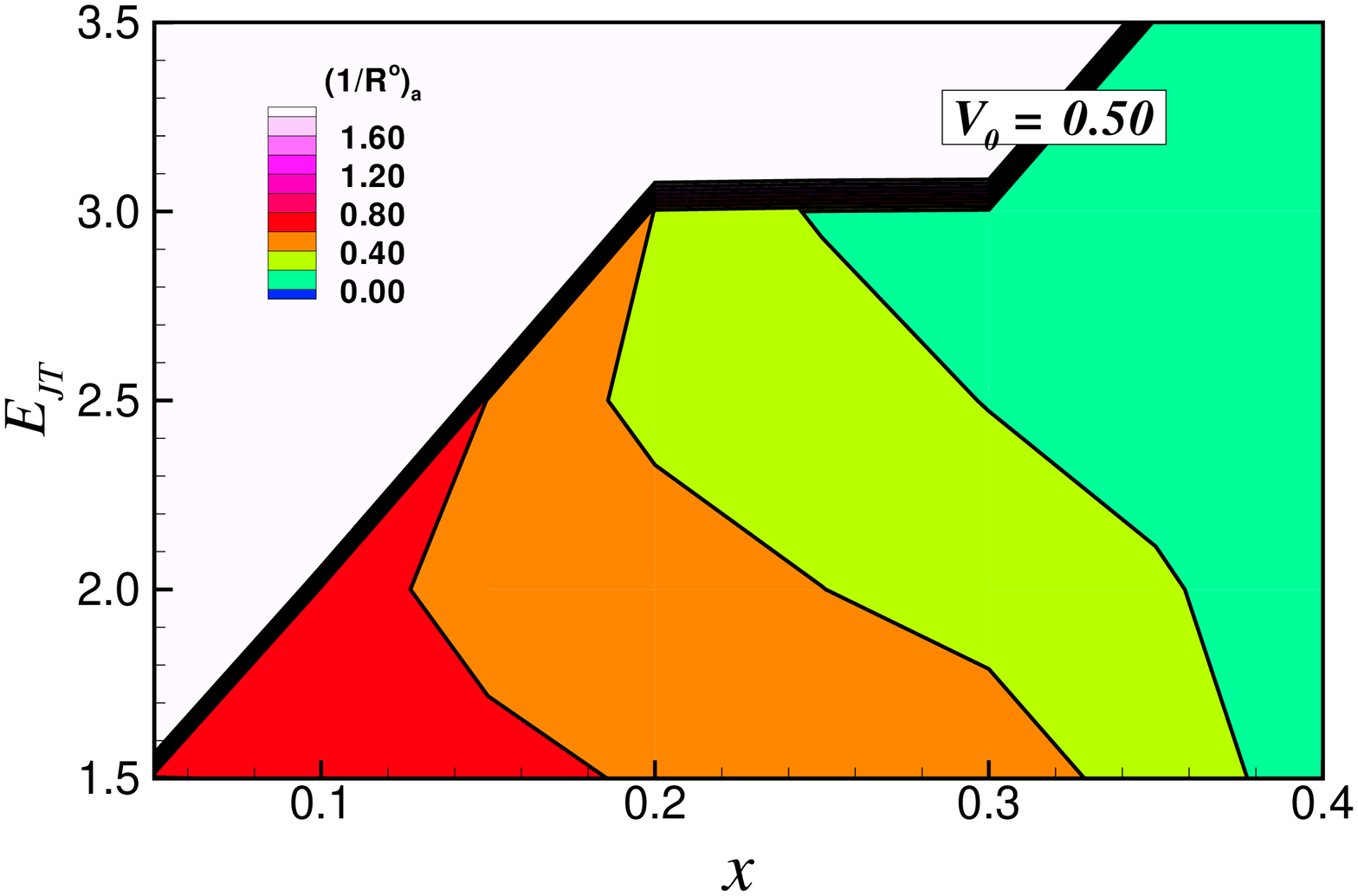}}
\centerline{(c)}
\caption{(color online) Average inverse size  $(1/R^o)_a$ of the  clumps with at least one $b$ state occupancy. Clump size $R$ is defined as the radius of gyration (see text) of the clump. For small $V_0$, the average inverse size at $x = 0.2$ and $\Ejt=1.0$ is between 0.44 and 0.88 -- thus corresponds to spherical puddle of containing 6 to 10 lattice sites. With increase in $V_0$, the sizes decreases and eventually there are no occupied clumps at $V_0=2.0$ (not shown). At larger $x$, there is one percolating clump for $V_0\lesssim1.0$. At larger $\Ejt$, again, there are no occupied clumps.
}
\label{rocave}
\end{figure*}

\begin{figure}
\centerline{\epsfxsize=\figwidth \epsfbox{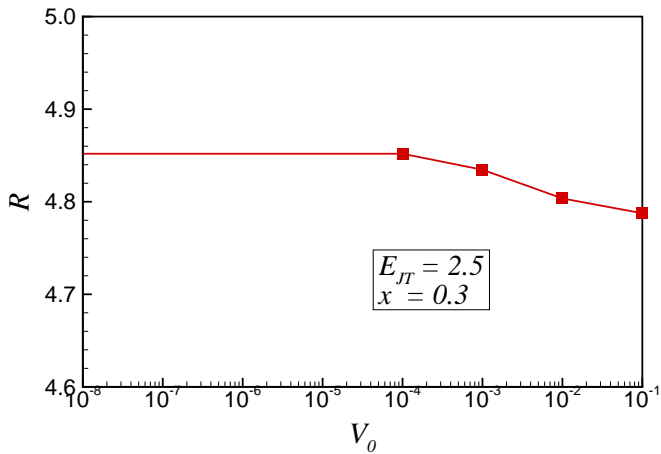}}
\caption{Clump size  as a function of the Coulomb interaction parameter $V_0$ (normalized by $t$)  for $E_{JT} = 2.5 t$, $x = 0.3$ obtained from simulations.  }
\label{Vdat}
\end{figure}

\begin{figure*}
\centerline{\epsfxsize=\figtwowidth \epsfbox{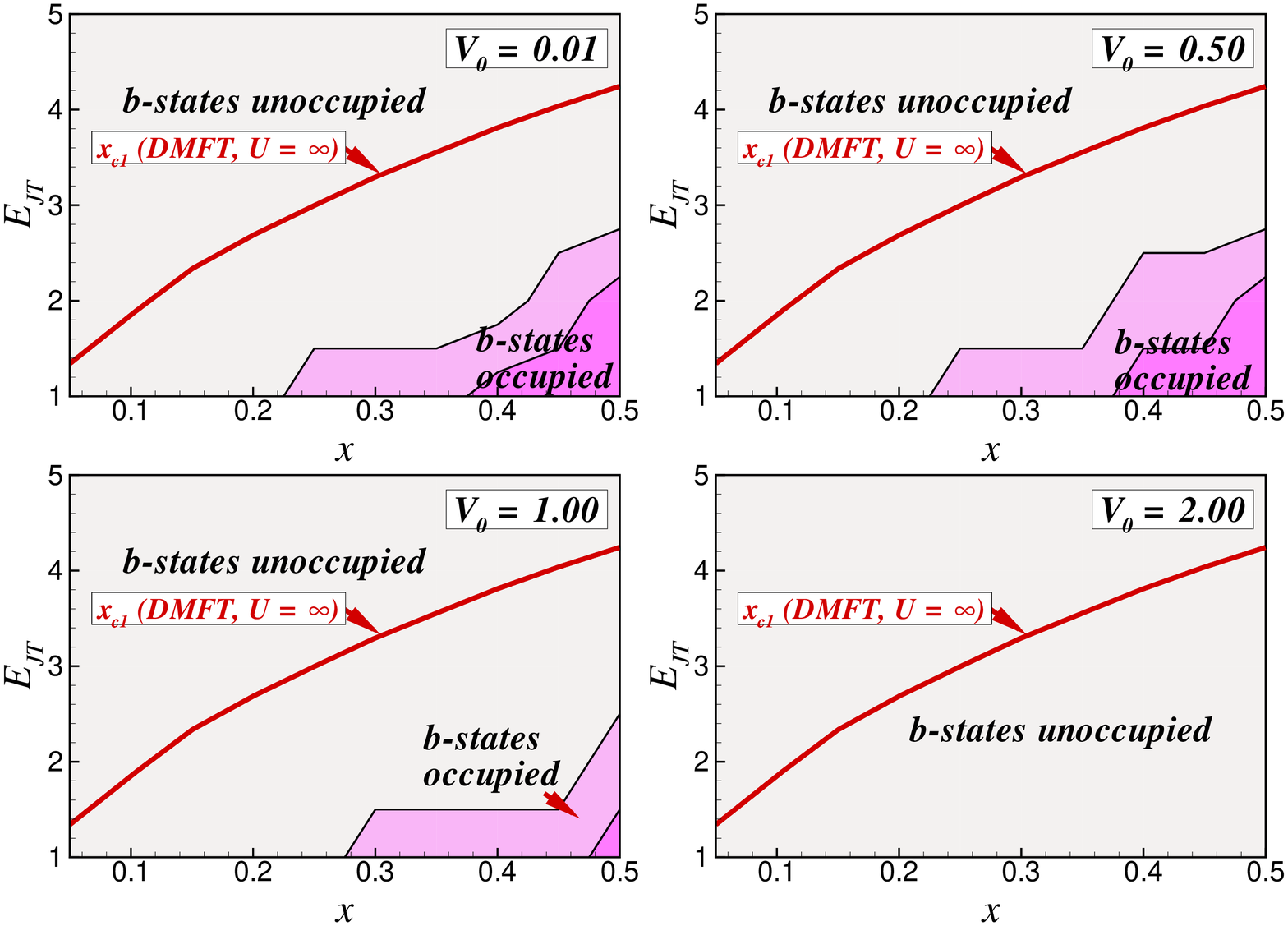}}
\caption{(color online) Critical doping levels $x_{c1}$ and $x_{c2}$ obtained from simulations with uniform distribution of Ak ions. The contour scheme is same as that of \fig{btype}. These results are for one initial realization of the holes (no average over initial conditions).}
\label{btypeu}
\end{figure*}

Finally, we discuss the clump size distribution. The size of a clump
is defined in terms of a radius of gyration $R$ calculated as $
\displaystyle{R = \sqrt{\left(\frac{1}{n} \sum_{i = 1}^n (\br(i) -
\bar{\br})^2\right)}}$ where $\bar{\br}$ is the center of mass of the
clump and $i$ runs over all the holes in the clump. We investigate
the distribution of the quantity $(1/R)$ as a function of the system parameters for clumps
with $b$ states occupied. For a percolating clump the quantity $1/R
\rightarrow 0$ and for the smallest clump with two holes $1/R = 2.0$.
Figures \ref{clzV}, \ref{clzx} and \ref{clzEjt} show the concentration
(number per lattice site) of the {\em occupied clumps} as a function of their inverse
size. For a given doping $x$ and $\Ejt$, there are many more larger clumps for
smaller $V_0$  (\Fig{clzV}). At high enough $V_0$
($V_0=2.0$) there are no $b$ electron puddles.  The
effect of doping $x$ is also as expected. At small doping, there
are clumps of various sizes, while at larger doping, there
is only one percolating clump (see \Fig{clzx}). The effect of $\Ejt$
is more interesting (\Fig{clzEjt}). For $\Ejt = 1.50$ there is a percolating clump
together with isolated small puddles. On increase of $\Ejt$ ($\Ejt =2.0$),
the isolated electron puddles are larger in size, and on further
increase of $\Ejt$ ($\Ejt = 3.0$), there is essentially a single large
percolating puddle. This is due to the fact that occupancy of a clump
requires that the gain in kinetic energy offsets the
loss in the Jahn-Teller energy $\Ejt$ at the least, thus larger $\Ejt$ requires larger
clumps to be occupied for sufficient gain in kinetic energy.
\Fig{rocave} shows the dependence of the average inverse size
$(1/R^{oc})_{ave}$ on the parameters; the results are as expected, in
that the average size of clumps is about a few lattice spacings for small
doping $(x < 0.25)$, followed by a regime of percolating clumps. A
result of particular interest is the dependence of the average clump
size on the long range Coulomb parameter $V_0$. This result is plotted
in \fig{Vdat}; we see that the clump size is essentially insensitive
to $V_0$ until $V_0$ approaches $t$. This may be contrasted with the
simple analytical result obtained earlier (see \Fig{clumpsize}). It is
evident that the the real system, any finite $V_0$ will prevent
``phase separation'', but the final details of the ground state is
strongly determined by the {\it random} distribution of the Ak ions,
which also determines the size scale of the clumps (and hence
their insensitivity to $V_0$ for small values of $V_0$).  We have
investigated the size dependence of the clump size on the size of the
simulation cube and found that there is no significant size
dependence.

Our study therefore indicates that for small doping $(x <
0.25)$, electronic inhomogeneities in the the form of  localized $b$
electron ``puddles'' exist in a background of polarons (see \Fig{clump})
while for larger doping, the system consists of intermingled states of
percolating, delocalized $b$ electron puddles coexisting with regions of
localized $\ell$ polarons. The scale of the inhomogeneities is
that of the local Ak ion disorder and not much larger. The study
therefore confirms that long range electrostatic interactions gives
rise to only nanometer (or lattice scale) electronic inhomogeneities
as was anticipated in earlier literature.\cite{Dagotto2003} The key
point is that this nanometric scale arises not out of phase
competition, but due to strong correlation effects of two types of
states that appear at the atomic scale, whose spatial arrangement is
then controlled by the long range Coulomb interaction and the random
distribution of the Ak ion disorder.

\begin{figure*}
\centerline{\epsfxsize=\figtwowidth \epsfbox{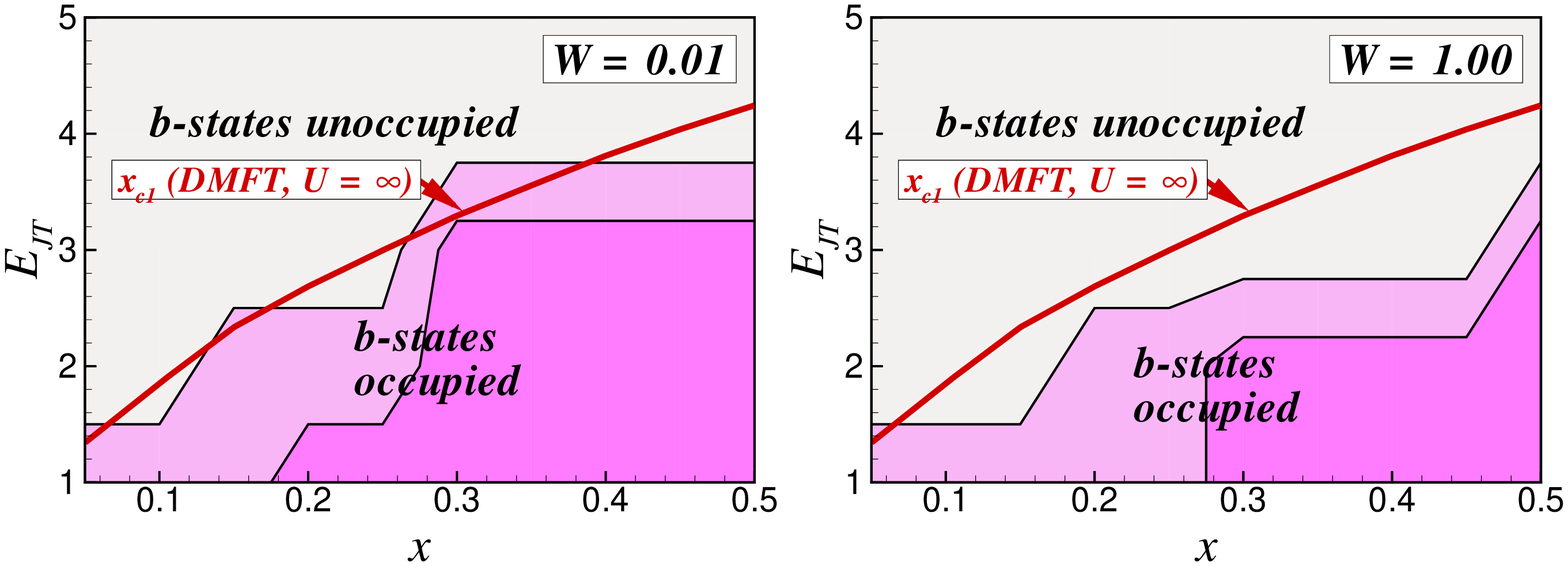}}
\caption{Critical doping levels $x_{c1}$ and $x_{c2}$ obtained from simulations with on-site disorder which varies from $-W$ to $+W$. The contour scheme is same as that of \fig{btype}. These results are for one realization of the disorder.}
\label{btypew}
\end{figure*}

\section{Discussion}\label{Disc}

Our study of the extended $\ell b$ model has established that long
ranged Coulomb interaction gives rise to nanoscale electronic
inhomogeneities. A key point is that even a very small $V_0$ completely
eliminates the macroscopic phase separation tendency of the
Falicov-Kimball like $\ell b$ model (compare \Fig{mphsep} with
\Fig{clump}). By using a dielectric constant of about 20, we
estimate an upper bound for $V_0$ in manganites to be about 0.02 eV and
thus well within the small $V_0$ regime; for $t \approx 0.2$ eV, the
dimensionless $V_0$ is about 0.1. Furthermore, for a given set of
energy parameters ($V_0$ and $\Ejt$), we have shown the existence of
two important doping thresholds: $x_{c1}$ which corresponding to the occupancy of
the $b$ states, and $x_{c2} > x_{c1}$ at which there is an
insulator to metal transition. Our results for $x_{c1}$ are close
those of the earlier DMFT prediction ; however, the existence of $x_{c2}$
is a new aspect of this work (in DMFT $x_{c2} = x_{c1}$).
The key physics behind the  agreement of $x_{c1}$ is {\em the
large $U$ condition} (no simultaneous occupancy of $\ell$ polaron and
$b$ electron on one site), and {\em the random distribution of the Ak
ions}. As we have shown, the random distribution of Ak ions produces local
clustering of holes around them which eventually causes electron
puddles and subsequent percolating clumps at higher doping. The
main consequence is that the electronic inhomogeneities are all at the
nanoscale and the material appears homogeneous at microscale.

How important is the random distribution of Ak ions? This question was
answered by distributing the charge of the Ak ions equally among all
the Ak sites. The physics of the problem is changed completely by this
step. First, the clustering tendency of the holes is completely
suppressed. This is evident from \Fig{gofr} (right column) where it is
seen that the probability of finding a hole neighboring a hole is
nearly reduced to zero. In such a case, for a given set of
energetic parameters, much larger doping levels would be required for $b$
state occupancy, and for the insulator to metal transition. Indeed,
these observations are borne out in the full scale simulations, see
\Fig{btypeu}. Thus the physically more realistic random distribution
of Ak ions is key to the agreement with the DMFT calculation.
In other words, the earlier DMFT calculation, although it
did not take into account the long range Coulomb interactions, premised a
homogeneous state (on a ``macroscopic'' scale), thereby effectively
incorporating the key effect of the long ranged Coulomb interaction.

It may be argued that effects of the random distribution of Ak ions (that help
the clustering of holes) is equivalent to a local disorder potential.
The results of the $\ell b$ Hamiltonian \prn{Hellb} with an
additional disorder potential (but no long range Coulomb
interactions),
\bea
{\cal H}_{dis} = \sum_i w_i \dagg{\ell}_i \ell + \sum_j w_j \dagg{b}_j b_j
\eea
where $w_j$ are distributed uniformly between $-W$ and $W$,
should be roughly similar to those of the extended $\ell b$ Hamiltonian \prn{Hellb}. The
premise is that since $\ell$ electrons will occupy sites of low $w_i$,
holes are likely to clump with a probability proportional to $x^n$
where $n$ is the size of the clump. This is similar to the probability
of finding clustering together of Ak ions. Thus the clumps that appear
in the disorder only model may be expected to have similar nature as
that of the case with random distribution of Ak ions.
The results of the disorder-only Hamiltonian are shown in \Fig{btypew}. Clearly, the
result of $x_{c1}$ is close to that of the DMFT  and the
extended $\ell b$ results. Also, the clump sizes etc.~are also quite
similar as expected. There is, however, a crucial difference -- the
density of states of the $\ell$ electrons does not have a Coulomb type
gap in the purely disordered model just discussed. In the real system
both  disorder and long ranged Coulomb interactions are present, and
the extended $\ell - b $ model includes both in a realistic way,
in contrast to the disorder only model.

We now discuss further predictions and inferences from the model. Our
model is particularly suited to study the low bandwidth manganites
such as Pr based compounds which have a large ferro-insulating regime
as a function of doping. Most of the inferences made hinge on the
coexistence of localized and delocalized states -- the key physical
idea of the $\ell b$ model. It may be inferred from the extended $\ell
b$ model that the low temperature conductivity in the low temperature
regime for doping $x < x_{c1}$ will be governed by the polaron Coulomb
glass. Thus, the low temperature conductivity is expected to be that
predicted by for the Coulomb glass\cite{Efros1975}, i.~e., $\sigma(T)
\sim e^{-a/ \sqrt{T}}$ where $a$ is a constant. Further, above $x_{c2}
> x > x_{c1}$, several excitations contribute to the
conductivity. These will include polaron hopping, variable range
hopping of the $b$-state electrons from one puddle to the other, even
two step processes where a $b$-state in an intermediate for an
$\ell$-polaron to hop from one site to another. Transport in this
regime is therefore expected to be involved and further investigation
is necessary to uncover the temperature dependence. We note that
transport measurements\cite{Sudheendra2003} on doped manganites do
show the features that we deduce from our model.  For $x > x_{c2}$,
the low temperature conductivity will be metallic with large residual
resistivity. An important contribution to this resistivity arises from
the Couloumb potentials of the Ak ions as has been noted
earlier\cite{Pickett1997}. At higher temperatures, scattering from the
thermally disordered $t_{2g}$ spins will cause a decrease in the
effective bandwidth of the $b$ states and can lead to the opening up
of a gap between the $\ell$ states and the $b$ states above the
ferromagnetic transition temperature. In this high temperature regime
the conductivity will have contributions both from polaron hopping and
thermal excitations of the $\ell$ polarons to the $b$ states similar
to that of a semiconductor.

The present model does not explicitly include the $t_{2g}$ core spins
as degrees of freedom. A simple minded approach of including these
degrees of freedom will lead to intractable computational
complexities. Novel approaches to treat both Coulomb interactions,
$b$ state quantum effects, and core spins need to be developed to
attain a final understanding of the manganite puzzle.

\begin{figure}
\centerline{\epsfxsize=\figwidth \epsfbox[133 342 491 729]{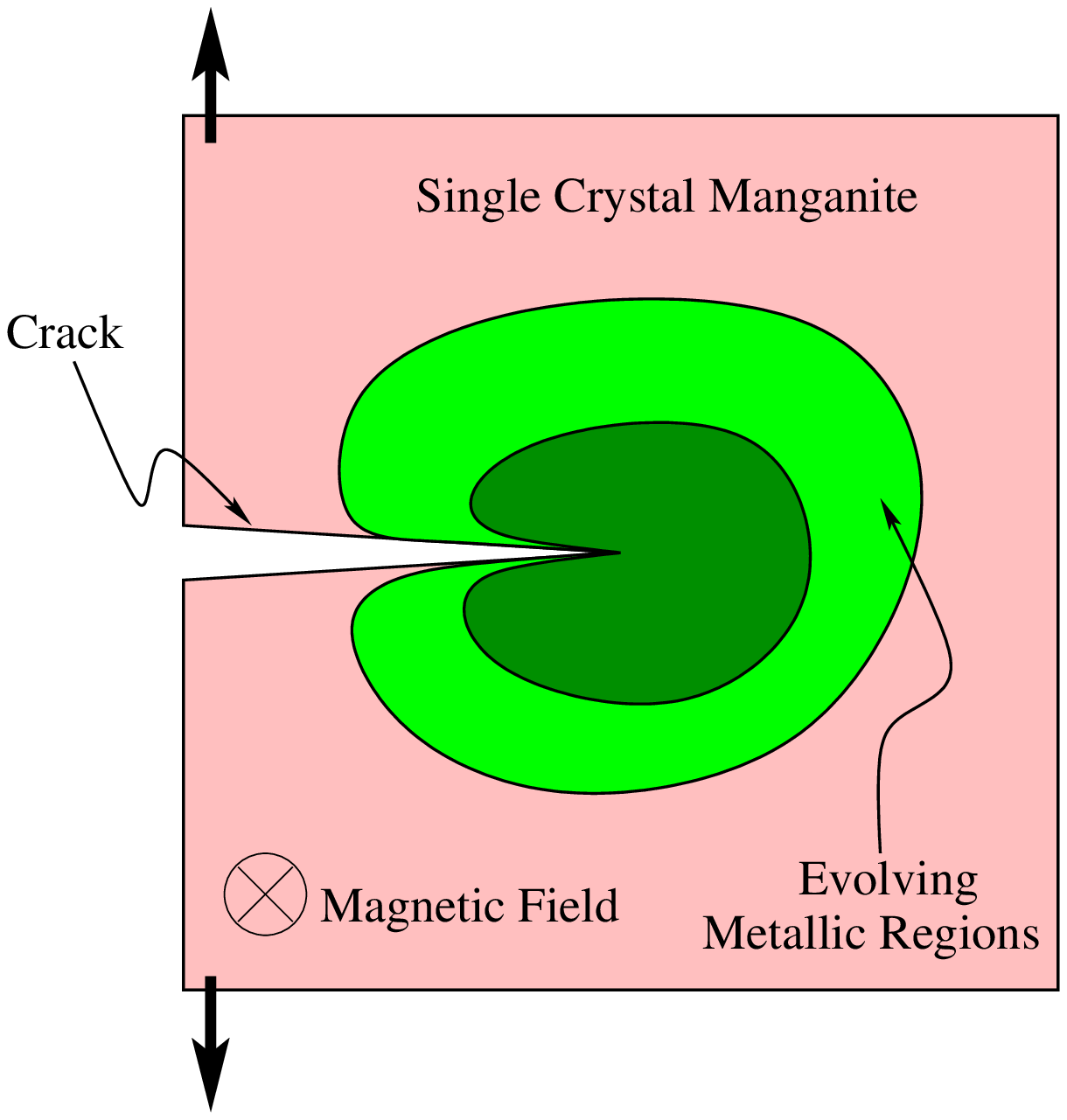}}
\caption{Schematic of a suggested experiment to study the influence of long range strain effects on mesoscale inhomogeneities in manganites. }
\label{crexp}
\end{figure}

We conclude the paper with a discussion of the important issue of
``phase separation'' and electronic inhomogeneities. As noted in the
introductory section several groups have implied that the presence of
electronic inhomogeneities is {\em essential} for the occurrence of the
colossal magnetoresistance. Our work suggests that inhomogeneities are
only present at the nanometer scale and a theory that {\em averages
over these} without explicit treatment of these inhomogeneities can and does
reproduce colossal magnetoresistance.\cite{Pai2003,Ramakrishnan2003,Ramakrishnan2004}
Indeed, very recent experimental
work\cite{Mathieu2004} shows that materials with no detectable ``phase
separation'' show colossal magnetoresistance.  Further, we find
that nanoscale inhomogeneities are a direct result of long range
Coulomb interaction frustrating ``phase separation'' induced by {\it
strong correlation} -- nanoscale electronic inhomogeneities are not a
result of phase competition in our model. Although similar mechanisms
based on long range coulomb interactions has been discussed earlier\cite{Dagotto2003,Nagaev1983,Emery1990,Emery1993,Lorenzana2001},
we believe that this is the first detailed quantitative treatment
of a realistic model any correlated oxide.

The large length scale (micrometer sized) inhomogeneities seen in
experiments remain to be explained in the present framework.  To the
best of our knowledge, the coexistence of metallic and insulating
regions have all been seen only in surface probe measurements or
measurements with thin films (for electron microscopy). Creation of
a surface introduces defects such as cracks and steps all of which
have long ranged elastic fields. As is well known, the phase of
manganites is strongly influenced by pressure
(stresses).\cite{Postorino2003} Thus the large scale inhomogeneities
are likely to be a result of pre-existing strain sources as is
indicated by recent photo-emission experiments\cite{Sarma2004}. It is
possible to explicitly test this in an experiment (see \fig{crexp})
with a pre-cracked manganite sample -- on loading the cracked sample
the motion of the metal-insulator boundary is expected to be
observed. Similar experiments\cite{Paranjape2003} (without cracks
etc.) do indeed suggest the strong effects of strains.

A second possibility for the existence of micron-scale clusters could
be due to ``kinetic arrest'' as seen recently in some rare-earth
compounds\cite{Chattopadhyay2005}. Thus the patches that appear could
arise to an ``incomplete phase transition'', and likely to show
``glassy'' behavior. Indeed many manganites are known to show glassy
behavior\cite{Dagotto2005a}, and this is possibly another important
line of further investigation.

\noindent
{\bf Acknowledgment} VBS wishes to acknowledge generous support from the
DST, India, through a Ramanujan grant.
H. R. K. acknowledges support from the Department of Science and Technology,
India as a J. C. Bose National Fellow. TVR acknowledges support from DST via the Ramanna Fellowship.We thank Pinaki
Majumdar, D.~D.~Sarma, Dinesh Topwal, A.~K.~Raychaudhuri for useful
discussions. A special note of thanks to P.~Sanyal for
discussions and data analysis.

\bibliography{phsep}
\end{document}